\documentclass[11pt]{article}
\usepackage{amsmath,amssymb,epsfig,color,bbold}
\renewcommand{\arraystretch}{1.2}

\usepackage{psfrag}
\usepackage{graphicx}
\usepackage{bm}
\usepackage{feynmf}

\newcommand{\be}{\begin{equation}}  
\newcommand{\ee}{\end{equation}} 
\def\slash#1{#1\!\!\!/\!\,\,}  
\newcommand{\Dslash}{D\!\!\!\!/\,}
\newcommand{\nl}{\nonumber \\ }

\newcommand{\order}{ {\cal O} }

\allowdisplaybreaks

\begin{document}

\begin{titlepage}

\begin{flushright}
EFI Preprint 14-25 \\
September 29, 2014

\end{flushright}

\vspace{0.7cm}
\begin{center}
\Large\bf 
Standard Model anatomy of WIMP dark matter direct detection 
\nl
II: QCD analysis and hadronic matrix elements
\end{center}

\vspace{0.8cm}
\begin{center}
{\sc   Richard J. Hill$^{1}$ and Mikhail P. Solon$^{1,2}$}\\
\vspace{0.4cm}
{\it $^{1}$Enrico Fermi Institute and Department of Physics \\
The University of Chicago, Chicago, Illinois, 60637, USA
\\[5mm]
$^{2}$Berkeley Center for Theoretical Physics, Department of Physics\\
and Theoretical Physics Group, Lawrence Berkeley National Laboratory\\
University of California, Berkeley, CA 94270, USA
}
\end{center}
\vspace{1.0cm}
\begin{abstract}
  \vspace{0.2cm}

\noindent 
Models of Weakly Interacting Massive Particles (WIMPs)
specified at the electroweak scale are systematically matched to  effective
theories at hadronic scales where WIMP-nucleus scattering observables
are evaluated.  Anomalous dimensions and heavy quark threshold
matching conditions are computed for the complete basis of
lowest-dimension effective operators involving quarks and gluons.  The
resulting QCD renormalization group evolution equations are solved.
The status of relevant hadronic matrix elements is reviewed and
phenomenological illustrations are given, including details for the
computation of the universal limit  of nucleon scattering with heavy
$SU(2)_W\times U(1)_Y$ charged WIMPs.  Several cases of previously
underestimated hadronic uncertainties are isolated.
The results connect arbitrary models specified at the
electroweak scale to a basis of $n_f=3$ flavor QCD operators.  The
complete basis of operators and Lorentz invariance constraints
through order $v^2/c^2$ in the nonrelativistic nucleon effective
theory are derived. 

\end{abstract}
\vfil

\end{titlepage}


\section{Introduction} 

In the search for Weakly Interacting Massive Particles (WIMPs),
experiments in the present decade and beyond will explore a broad
range of processes~\cite{Cushman:2013zza}, such as dark matter (DM)
production  at colliders, DM annihilation at the galactic center and
DM scattering from nuclear targets.    Given the multitude of WIMP
candidates and search strategies,  it is imperative to develop
theoretical formalism to delineate the possible interactions of DM
with known particles, making clear which uncertainties are inherently
model dependent and which can, at least in principle, be improved by
further Standard Model (SM) analysis. 

Particularly in the case of direct detection via nuclear scattering,
determining the relation between an underlying  particle physics model
and the experimental observable (i.e., scattering cross section)
demands analysis at multiple energy scales involving both perturbative
and  nonperturbative QCD.   Relating physics at (and above) the weak
scale to an  effective theory in which hadronic observables are
evaluated is a problem  that has received significant attention in 
other arenas, such as flavor transitions involving heavy mesons~\cite{Buchalla:1995vs}  and
electric dipole moment searches~\cite{Engel:2013lsa}.  The analogous
problem in dark matter direct detection contains a unique set of
challenges whose study is the focus of the
preceding~\cite{Hill:2014yka} and present paper.  In
\cite{Hill:2014yka} we described the steps involved in  matching an
ultraviolet completion DM model, consisting of some number  of SM
gauge multiplets, onto an effective theory renormalized at the weak
scale.   Here we develop the framework for the systematic treatment of
QCD effects when passing from a theory renormalized at the weak scale
to a low-energy theory of quarks and gluons.  We also identify pieces
of the framework whose further development significantly impacts
our knowledge of WIMP-nucleus scattering cross sections, including
heavy quark decoupling relations in perturbative QCD,  and
nonperturbative scalar quark matrix elements of the nucleon in $n_f=3$
or $n_f=4$ flavor QCD.  The results of this analysis can be
used as the basis for detailed nuclear modeling.   To this
end we derive the nucleon-level effective theory  and matching
conditions through two-derivative order in the one-nucleon sector.

While the theoretical formalism is general,  for the purposes of
illustration we focus our phenomenological illustrations on the
analysis of heavy $SU(2)_W\times U(1)_Y$ charged WIMPs.   We do so for
three reasons.   Firstly, this scenario is highly predictive: in  the
limit of large WIMP mass, the WIMP-nucleon scattering amplitude is
completely determined solely by SM parameters; secondly, this regime
provides an important illustration  of QCD effects, since generic
cancellations between subamplitudes enhance sensitivity
to subleading corrections; and thirdly, the hitherto absence of
significant deviations between observations and SM predictions at the
Large Hadron Collider  and elsewhere may actually indicate a new
physics scale lying somewhat above $m_W,\, m_Z \sim 100\,{\rm GeV}$.  

The unknown particle nature of dark matter is the source of great
intrigue but also complicates any analysis wishing to draw unambiguous
conclusions.   The separation of energy scales, formalized by a
sequence of effective theories, provides several choices for starting
point when  constraining potential WIMP interactions with SM fields,
or in relating constraints or  potential signals between observational
methods.    Each effective theory takes as input matching conditions
computed in a higher scale theory; alternatively, giving up the
connection  to the high scale theory, one may start at any point  in
the sequence by taking effective operator coefficients as free
parameters to be constrained by experiment.   

At the highest scales, a UV complete model may be
specified~\cite{Jungman:1995df,Feng:2013pwa,Fowlie:2013oua},  but may
involve many poorly constrained parameters with degenerate effects on
low energy observables.  Restriction to  a small number of postulated
fields~\cite{Bai:2013iqa,Bai:2014osa,Chang:2014tea,Agrawal:2014ufa,Chang:2013oia},
reduces the parameter number, but typically without justification for
choice of  field content.    A sparse distribution of pure gauge
states (measured in units of $m_W$) becomes generic when  masses of
particles beyond the SM (BSM) become large compared to $m_W$.  Here
the heavy WIMP expansion maintains theoretical control in the absence
of a specified UV completion and dramatically
simplifies loop integral computations that become numerically dominant
in  this regime~\cite{Hill:2011be,Hill:2013hoa,Hill:2014yka}.   In the case of an assumed large
mass scale for BSM particles mediating interactions with the SM,  a
basis of contact interactions can be
investigated~\cite{Goodman:2010ku,Matsumoto:2014rxa},  although the
connection between such contact interactions and UV completions may be
unclear.  As illustrated below in Section~\ref{sec:contact}, care
should also be taken to account for renormalization scale and scheme
dependence when relating high scale constraints to low scale
observables.  
The effects of renormalization group running in theories above the 
weak scale have been investigated in \cite{Crivellin:2014qxa,Crivellin:2014gpa}. 

Under the assumption that BSM particles (except perhaps the DM itself)
have mass at or above the weak scale,  the remaining analysis is
independent of which of the above approaches is taken to physics above
the weak scale.  The focus of the present paper is on the task of
relating the resulting effective theory specified at the weak scale to
the effective theory defined at low energy where hadronic matrix
elements are evaluated.   This $n_f=3$ flavor QCD theory is the
natural handoff point from particle to nuclear physics.      
Here again, there are several choices for the starting point of a nuclear
physics effective theory analysis. 
To the
extent that nuclear matrix elements are determined by single nucleon
matrix elements, an alternative  would be to take the coefficients of
single-nucleon operators as unknowns to be constrained by direct
detection
observables~\cite{Fitzpatrick:2012ix,Fitzpatrick:2012ib,Anand:2013yka,Catena:2014uqa},
generalizing the canonical spin-independent and spin-dependent
DM-nucleon interactions, cf. (\ref{eq:PT}) below, commonly considered
in  the presentation of DM direct detection
limits~\cite{Goodman:1984dc,Cushman:2013zza}.    As discussed below in
Section~\ref{sec:lorentz}, when including $v^2/c^2$ 
effects 
it is important to properly enforce Lorentz versus
Galilean invariance in the effective Lagrangian.    To the extent that
multi-nucleon effects are relevant, the complete nuclear response
cannot be derived from  information contained 
solely 
in the single nucleon
matching, requiring an extension of the effective theory to include
such effects and/or additional information from quark-level
matching~\cite{Cirigliano:2012pq}.  A heavy particle effective theory
may be constructed for an entire nucleus~\cite{Fan:2010gt} but 
requires further analysis in order to directly compare experiments using different nuclei~\cite{Peter:2013aha}.%
\footnote{
Other recent studies of general WIMP-nucleon interactions include \cite{Menendez:2012tm,Klos:2013rwa,Gardner:2013ama,DelNobile:2013sia}.
For a review including further references to early work, see \cite{Jungman:1995df}. 
}

The remainder of the paper is structured as follows.    In
Section~\ref{sec:eft} we discuss the construction of  complete
operator bases for DM-SM interactions after integrating out weak scale
particles.    We
present leading order weak scale matching conditions onto the lowest
dimension  effective theory operators for an illustrative UV completion
involving gauge-singlet  DM (the case of $SU(2)_W\times U(1)_Y$
charged dark matter was considered in  \cite{Hill:2014yka}).   In
Section~\ref{sec:weak} we compute the relevant operator
renormalization factors and anomalous dimensions. This
section also  presents the renormalization group evolution of
effective operators and coefficients,  and matching conditions at
heavy quark thresholds.     Section~\ref{sec:hadron} reviews the
relevant hadronic matrix elements  and Section~\ref{sec:nucleon}
describes the associated nucleon-level effective theory.
Section~\ref{sec:pheno} gives phenomenological illustrations of QCD
effects in DM direct detection. For example, the results for renormalization factors in Sec.~\ref{sec:weak} are combined with the computation in \cite{Hill:2014yka} to obtain
renormalized ($\overline{\rm MS}$) matching coefficients at the weak
scale for electroweak-charged self-conjugate heavy WIMPs.  
 Section~\ref{sec:summary} concludes
with a summary  and outlook. 
Appendices provide details of renormalization constants, and higher order
nucleon matrix elements discussed in the main text. 

\section{Effective theory below the electroweak scale \label{sec:eft}}

The tabulation of operators at a given mass dimension involving SM
fields and a finite collection of DM fields of given SM quantum
numbers is a straightforward task, but requires some care to ensure a
complete basis while avoiding redundant operators.  We construct
operator bases appropriate to energies below the weak scale, enforcing
$SU(3)_c \times U(1)_{\rm e.m.} \times U(1)_{\rm DM}$ or  $SU(3)_c
\times U(1)_{\rm e.m.} \times Z_2$  invariance, assuming a $U(1)_{\rm DM}$ or
$Z_2$ symmetry to stabilize the DM particle.  The massive electroweak
gauge bosons, $W^\pm, Z^0$, the top quark, $t$, and the physical Higgs
field, $h$, are integrated out, and we consider higher dimension
operators suppressed by the weak scale, for definiteness taken to be
$m_W$.  We focus on two cases:  firstly, the case $M\gtrsim m_W$ for
SM interactions  with the (assumed electrically neutral) lightest
state of a BSM sector;   and secondly, the case $M\ll m_W$ for SM
interactions with a gauge singlet scalar or fermion. 
These cases cover a large space of models, and illustrate principles
in any more general analysis.

\subsection{Standard Model building blocks}

For the SM degrees of freedom, we focus on the quark and gluon fields
of $n_f=5$ flavor QCD, and consider the photon field only in the case
of dimension five electric and magnetic dipole operators, i.e., when
operators containing  the photon are of lower dimension than quark and
gluon operators.  Operators with leptons may be constructed similarly
to quark operators.  The flavor diagonal, Hermitian, gauge-invariant
SM building blocks through dimension four are
\begin{align}\label{eq:SMblocks}  F^{\mu \nu} \,, \quad  {\bar q}
\big[ \gamma^\mu \,, \gamma^\mu \gamma_5 \big] \big[ 1\, ,  i D_-^\rho
\big] q \,, \quad G^A_{\mu \nu}G^A_{\rho \sigma} \,.  
\end{align}
We will perform Fierz rearrangements to the basis without spinor
contractions between SM and DM fields,  hence only free vector indices
appear in (\ref{eq:SMblocks}).   Collected within square brackets, $[
\  \ ]$, are the different structures that may be applied to the same
field bilinear. Total derivatives of building blocks are not listed
above but must be considered in the construction of the effective
lagrangian.  We use the shorthand $D_{\pm}^\mu \equiv D^\mu \pm
\overleftarrow{D}^\mu$, where $D_\mu = \partial_\mu - igA_\mu^A T^A -
i eQ A_\mu$  is the $SU(3)_c\times U(1)_{e.m.}$ covariant derivative
and $\overleftarrow{D}_\mu = \overleftarrow{\partial}_\mu + igA_\mu^A
T^A + ie Q A_\mu$ with $\overleftarrow{\partial}$ denoting a
derivative acting to the left.   Here $Q$ denotes the electric
charge in units of the proton electric charge $e$. 

In writing (\ref{eq:SMblocks}) we have considered only quark flavor
diagonal operators  and imposed global chiral symmetries $q_{L,R} \to
e^{i \epsilon_{L,R}} q_{L,R}$  when quark masses vanish.  These constraints can be formally
justified by restricting to ultraviolet completions for  which a
$U(3)_L \times U(3)_R^u \times U(3)_R^d$ symmetry (``minimal flavor
violation")  can be defined in the electroweak-symmetric theory. 
Additional operators through dimension four consistent with these
requirements are 
\begin{equation}\label{eq:redun} m_q {\bar q} \big[1 \, ,  i \gamma_5 ,
\, \sigma^{\mu \nu}   \big]  q \,. 
\end{equation}
However Lagrangian interactions containing  (\ref{eq:redun}) can be
shown to be  redundant by field redefinitions, leaving
(\ref{eq:SMblocks}) as  a complete basis of independent SM operators.
It is straightforward to extend the  building blocks in
(\ref{eq:SMblocks}) to consider more general flavor structure.

\subsection{Dark matter building blocks} 

\begin{table}[t]
\begin{center}
\small
\begin{minipage}[t]{6.8cm}
\renewcommand{\arraystretch}{1.7}
\begin{tabular}[t]{c|c}
$d$ &  Fermion \\
\hline
$3$  & $\bar{\psi}\big[1\,, \ i\gamma_5\,,  \ \gamma^\mu \gamma_5 \, , \ \{ \gamma^\mu  \, , \ \sigma^{\mu \nu} \} \big]  \psi  $\\
$4$  & $\bar{\psi} \big[ \{ 1\,, \ i\gamma_5\,,  \ \gamma^\mu \gamma_5 \} \, , \ \gamma^\mu \, , \  \sigma^{\mu \nu} \big] i\partial_-^\rho \psi$ 
\end{tabular}
\end{minipage}
\begin{minipage}[t]{2.9cm}
\renewcommand{\arraystretch}{1.7}
\begin{tabular}[t]{c|c}
$d$ & Scalar \\
\hline
$2$       & $|\phi|^2$ \\
$3$       & $\{ \phi^* i\partial^\mu_- \phi \} $ 
\end{tabular}
\end{minipage}
\begin{minipage}[t]{4.2cm}
\renewcommand{\arraystretch}{1.7}
\begin{tabular}[t]{c|c}
$d$ & Heavy particle \\
\hline
$3$          & $\bar{\chi}_v\big[1 \, , \ \{ \sigma^{\mu \nu}_\perp \}\big]  \chi_v  $\\
$4$                & $\bar{\chi}_v\big[\{ 1 \} \, , \ \sigma^{\mu \nu}_\perp \big]  i \partial_{\perp - }^\rho \chi_v  $
\end{tabular}
\end{minipage}
\end{center}
\caption{\label{tab:DMblocks}
Gauge-invariant DM operator building blocks of indicated dimension for
a relativistic fermion and scalar, and a heavy-particle fermion. For
the relativistic case, building blocks within curly brackets, $\{ \ \
\}$, vanish for self-conjugate fields such as a Majorana fermion or a
real scalar. For the heavy-particle case, building blocks within curly
brackets, $\{ \ \ \}$, are odd under the parity in
Eq.~(\ref{eq:vparity}). The list for a heavy-particle scalar (of mass
dimension $3/2$) is obtained by omitting building blocks with the spin
structure $\sigma_\perp^{\mu \nu}$ above. 
}
\end{table}

For the dark sector, we focus on operators involving the lightest
$SU(3)_c \times U(1)_{\rm e.m.}$-singlet WIMP state.  We collect in
the first two columns of Table~\ref{tab:DMblocks} the  lowest
dimension Hermitian, gauge-invariant DM bilinears for 
relativistic scalar and fermion fields, denoted respectively by a complex 
valued $\phi$ and a four-component spinor $\psi$.  We consider both the case where there
is a conserved global $U(1)_{\rm DM}$ DM particle number, i.e., a Dirac fermion
or complex scalar, and the case  where the DM particle is
self-conjugate and odd under an exact $Z_2$ symmetry,  i.e., a
Majorana fermion ($\psi =\psi^c$) or a real scalar ($\phi = \phi^*$).
As for the SM building blocks, we ignore total derivatives of DM bilinears, 
which must be considered when constructing lagrangian interactions. 

In the regime where the DM has mass comparable to or heavier than the
electroweak scale particles, $M \gtrsim m_W$, the scale separation $M
\gg m_b$ allows us to employ the heavy-particle building blocks listed
in the final column of Table~\ref{tab:DMblocks}. We list the building blocks appropriate for a spin 1/2 or spin 0 heavy particle; effective theories for
higher-spin particles may be similarly constructed.
Lorentz
transformations of the heavy particle field are governed by the little
group for massive particles defined by the time-like unit vector
$v^\mu$.  A heavy fermion has two degrees of freedom which may be
embedded in a Dirac spinor, $\chi_v$, with constraint $\slash{v}
\chi_v = \chi_v$ (see, e.g., Ref.~\cite{Heinonen:2012km} and Sec.~2 of
Ref.~\cite{Hill:2014yka} for more details). In writing the
heavy-particle building blocks in Table~\ref{tab:DMblocks} we assume
field redefinitions that eliminate operators with timelike derivatives
$v\cdot D$ acting on $\chi_v$, and hence only perpendicular components
of derivatives, $\partial_\perp^\mu$, appear.  In a standard notation
we define spacelike (with respect to the timelike unit vector $v^\mu$)
``perpendicular'' components using $g_\perp^{\mu \nu} \equiv g^{\mu
\nu} - v^\mu v^\nu$.  In particular, we have $\partial_\perp^\mu
\equiv \partial_\alpha g_\perp^{\alpha \mu} =  \partial^\mu - v^\mu v
\cdot \partial $ and $\sigma_\perp^{\mu \nu}  \equiv \sigma_{\alpha
\beta} g_\perp^{\alpha \mu} g_\perp^{\beta \nu}$.

For lagrangians containing heavy fields describing self-conjugate
particles  such as Majorana fermions or real scalars, we may
furthermore impose invariance under the self-conjugate parity,
enforced formally by the simultaneous
operations~\cite{Kopp:2011gg,Hill:2011be}%
\footnote{Here ${\cal C}$ is the charge conjugation matrix acting on the spinor index of 
$\chi_v$.  It is symmetric and unitary and satisfies ${\cal C}^\dagger \gamma_\mu {\cal C} = -\gamma_\mu^*$. 
For the extension to arbitrary spin see Ref.~\cite{Heinonen:2012km}. 
}
\begin{equation}\label{eq:vparity} v^\mu \to -v^\mu \, , \quad \chi_v
\to  \chi_v^c  = {\cal C} \chi_v^*\, .  
\end{equation}  
Equivalently we may impose $CPT$ invariance, applying the usual $CPT$
transformations for relativistic fields, but employing a modified
version of $CPT$ for the heavy-particle, under which
\footnote{The phases $\xi$, $\eta$ and $\zeta$ under $C$, $P$ and $T$ do not affect scattering observables.}
\begin{equation}
\label{eq:CPT}  
C: \chi(t,\bm{x}) \to \xi \, \chi(t,\bm{x}) \, , \quad   
P: \chi(t,\bm{x}) \to \eta \, \chi (t,-\bm{x}) \, , \quad   
T: \chi(t,\bm{x}) \to \zeta \, S \, \chi(-t,\bm{x}) \,,   
\end{equation}
where $S =  i\sigma_2$ for fermions and $S=1$ for scalars~\cite{Heinonen:2012km}. 
In this formulation of the self-conjugate parity, the action of
discrete symmetries transforms fields, but leaves the reference vector
$v^\mu$ unchanged. Hence, it may be readily employed even when the
reference vector is fixed, e.g., to $v^\mu=(1,\bm{0})$ in the rest
frame of the heavy particle.

\subsection{Operator basis}

Upon combining the SM building blocks in (\ref{eq:SMblocks}) with the
DM building blocks in Table~\ref{tab:DMblocks}, and performing field
redefinitions to eliminate redundant operators, we obtain the 
effective lagrangian for DM interactions below the
weak scale.   

For the relativistic scalar case we have the following interactions,
\begin{align}\label{eq:scalarL}
{\cal L}_{\phi,{\rm SM}} &=
\sum_{q=u,d,s,c,b} \Bigg\{ {c_{\phi1,q} \over m_W^2 } |\phi|^2 m_q {\bar q} q + 
{c_{\phi2,q} \over m_W^2 } |\phi|^2 m_q {\bar q} i \gamma_5 q + 
{c_{\phi 3,q} \over m_W^2} \phi^* i\partial_-^\mu \phi {\bar q} \gamma_\mu q 
\nl
& \quad 
+ 
{c_{\phi 4,q} \over m_W^2} \phi^* i\partial_-^\mu \phi {\bar q} \gamma_\mu \gamma_5 q \Bigg\}
+ 
{c_{\phi5} \over m_W^2 } |\phi|^2 G_{\alpha \beta}^A G^{A \alpha \beta} 
+ 
{c_{\phi6} \over m_W^2 } |\phi|^2 G_{\alpha \beta}^A {\tilde G}^{A \alpha \beta} +\dots \,. 
\end{align}
For antisymmetric tensors we define the shorthand notation 
$\tilde{T}^{\mu\nu} = \epsilon^{\mu\nu\rho\sigma} T_{\rho\sigma}/2$ (we use the convention
$\epsilon^{0123} = +1$).   The ellipsis in (\ref{eq:scalarL}) denotes
operators of dimension six and higher involving the photon, and
operators of dimension seven and higher involving quarks and gluons.
For a real scalar the coefficients $c_{\phi n}$ vanish for $n=3,4$. 

For the relativistic fermion case we have the following interactions,
\begin{align}\label{eq:fermionL}
{\cal L}_{\psi,{\rm SM}} &= 
{c_{\psi1} \over m_W} {\bar \psi} \sigma^{\mu \nu} \psi F_{\mu \nu} +
{c_{\psi2} \over m_W} {\bar \psi} {\sigma}^{\mu \nu} \psi \tilde{F}_{\mu \nu} +
\sum_{q=u,d,s,c,b} \Bigg\{
{c_{\psi3,q} \over m_W^2} {\bar \psi} \gamma^\mu \gamma_5 \psi {\bar q} \gamma_\mu q
+ {c_{\psi4,q} \over m_W^2} {\bar \psi} \gamma^\mu \gamma_5 \psi {\bar q} \gamma_\mu \gamma_5 q
\nl
&\quad 
+ {c_{\psi5,q} \over m_W^2} {\bar \psi} \gamma^\mu \psi {\bar q} \gamma_\mu q 
+ {c_{\psi6,q} \over m_W^2} {\bar \psi} \gamma^\mu \psi {\bar q} \gamma_\mu \gamma_5 q
+ {c_{\psi7,q} \over m_W^3} {\bar \psi } \psi m_q {\bar q} q
+ {c_{\psi8,q} \over m_W^3} {\bar \psi } i \gamma_5 \psi m_q {\bar q} q 
 \nl
&\quad 
+ {c_{\psi9,q} \over m_W^3} {\bar \psi } \psi m_q {\bar q} i \gamma_5 q 
+ {c_{\psi10,q} \over m_W^3} {\bar \psi } i \gamma_5 \psi m_q {\bar q} i \gamma_5 q  
+ {c_{\psi11,q} \over m_W^3} {\bar \psi} i \partial_-^\mu \psi {\bar q} \gamma_\mu q 
  \nl
 &\quad
+ {c_{\psi12,q} \over m_W^3} {\bar \psi} \gamma_5 \partial_-^\mu \psi {\bar q} \gamma_\mu q 
+ {c_{\psi13,q} \over m_W^3} {\bar \psi} i \partial_-^\mu \psi {\bar q} \gamma_\mu \gamma_5 q
 + {c_{\psi14,q} \over m_W^3} {\bar \psi} \gamma_5 \partial_-^\mu \psi {\bar q} \gamma_\mu \gamma_5 q  
\nl
&\quad 
+ {c_{\psi15,q} \over m_W^3}  {\bar \psi} \sigma_{\mu \nu} \psi m_q {\bar q} \sigma^{\mu \nu} q 
+ {c_{\psi16,q} \over m_W^3} \epsilon_{\mu\nu\rho\sigma} {\bar \psi} {\sigma}^{\mu \nu} \psi m_q {\bar q} \sigma^{\rho\sigma} q \Bigg\} 
 + {c_{\psi17} \over m_W^3} {\bar \psi} \psi G_{\alpha \beta}^A G^{A \alpha \beta} 
   \nl
 &\quad
 + {c_{\psi18} \over m_W^3} {\bar \psi} i \gamma_5 \psi G_{\alpha \beta}^A G^{A \alpha \beta} 
 + {c_{\psi19} \over m_W^3} {\bar \psi} \psi G_{\alpha \beta}^A {\tilde G}^{A \alpha \beta}
 + {c_{\psi20} \over m_W^3} {\bar \psi} i \gamma_5 \psi G_{\alpha \beta}^A {\tilde G}^{A \alpha \beta} 
+ \dots \, ,
\end{align}
where the ellipsis denotes operators of dimension six and higher
involving the photon,  and operators of dimension eight and higher
involving quarks and gluons.    
For a Majorana fermion the coefficients $c_{\psi n}$ with
$n=1,2,5,6,11,12,13,14,15,16$ vanish, leaving ten types of operators through 
dimension seven as considered in Ref.~\cite{Goodman:2010ku}. 

For the case of DM with mass $M \gtrsim m_W$, we have the following interactions,%
\footnote{
It is convenient to notice the identities, 
$G^{A\mu\alpha} \tilde{G}^{A\nu}_{\quad\alpha} = 
g^{\mu\nu}G^{A\alpha\beta}\tilde{G}^{A}_{\alpha\beta}/4$ and  
$v_\mu v_\nu G^{A \mu}_{\quad [\alpha} \tilde{G}^{A \nu}_{\quad \beta]} = -\epsilon_{\alpha\beta}^{\quad\mu\nu} v_\mu v^\rho 
G^A_{\nu\sigma} G^{A\, \sigma}_{\rho}/2$. 
}
\begin{align}\label{eq:heavyL}
{\cal L}_{\chi_v,{\rm SM}} &= 
{c_{\chi1} \over m_W} {\bar \chi}_v \sigma_\perp^{\mu \nu} \chi_v F_{\mu \nu} + 
{c_{\chi2} \over m_W} {\bar \chi}_v {\sigma}_\perp^{\mu \nu} \chi_v {\tilde F}_{\mu \nu} +
\sum_{q=u,d,s,c,b} \Bigg\{
{c_{\chi3,q} \over m_W^2} \epsilon_{\mu\nu\rho\sigma} v^\mu {\bar \chi}_v {\sigma}_\perp^{\nu\rho}  \chi_v {\bar q} \gamma^\sigma q 
\nl
&\quad 
+ {c_{\chi4,q} \over m_W^2}  \epsilon_{\mu\nu\rho\sigma} v^\mu {\bar \chi}_v  {\sigma}_\perp^{\nu\rho}  \chi_v {\bar q} \gamma^\sigma \gamma_5 q
+ {c_{\chi5,q} \over m_W^2} {\bar \chi}_v \chi_v {\bar q} \slash{v} q  
+ {c_{\chi6,q} \over m_W^2}  {\bar \chi}_v \chi_v {\bar q} \slash{v} \gamma_5 q  
+ {c_{\chi7,q} \over m_W^3} {\bar \chi}_v \chi_v m_q {\bar q} q 
\nl
&\quad 
+ {c_{\chi8,q} \over m_W^3} {\bar \chi}_v \chi_v  {\bar q} \slash{v} iv \cdot D_- q 
+ {c_{\chi9,q} \over m_W^3} {\bar \chi}_v \chi_v m_q {\bar q} i \gamma_5 q 
+ {c_{\chi10,q} \over m_W^3} {\bar \chi}_v \chi_v {\bar q} \slash{v} \gamma_5 i v \cdot D_- q 
\nl
&\quad 
 + {c_{\chi11,q} \over m_W^3} {\bar \chi}_v  {\sigma}_\perp^{\mu \nu} i\partial_{-\mu}^{\perp} \chi_v {\bar q} \gamma_\nu  q 
 + {c_{\chi12,q} \over m_W^3}  \epsilon_{\mu\nu\rho\sigma} {\bar \chi}_v  {\sigma}_\perp^{\mu \nu} i\partial_-^{\perp \rho} \chi_v {\bar q} \gamma^\sigma  q 
 + {c_{\chi13,q} \over m_W^3}  {\bar \chi}_v  { \sigma}_\perp^{\mu \nu} i\partial_{-\mu}^{\perp} \chi_v {\bar q} \gamma_\nu \gamma_5  q 
\nl
&\quad 
+ {c_{\chi14,q} \over m_W^3} \epsilon_{\mu\nu\rho\sigma} {\bar \chi}_v  {\sigma}_\perp^{\mu \nu} i\partial_-^{\perp \rho} \chi_v {\bar q} \gamma^\sigma \gamma_5  q  
+ {c_{\chi15,q} \over m_W^3} \epsilon_{\mu\nu\rho\sigma} v^\mu {\bar \chi}_v \sigma_\perp^{\nu\rho} \chi_v 
\, {\bar q} ( \slash{v} iD_-^\sigma + \gamma^\sigma iv\cdot D_- ) q 
\nl
&\quad 
+ {c_{\chi16,q} \over m_W^3} \epsilon_{\mu\nu\rho \sigma} v^\mu {\bar \chi}_v \sigma_\perp^{\nu\rho} \chi_v 
{\bar q} ( \slash{v} iD_-^\sigma + \gamma^\sigma iv\cdot D_-) \gamma_5 q 
+ {c_{\chi17,q} \over m_W^3} {\bar \chi}_v i \partial_-^{\perp \mu} \chi_v {\bar q} \gamma_\mu q  
\nl
&\quad 
+ {c_{\chi18,q} \over m_W^3} {\bar \chi}_v \sigma_\perp^{\mu \nu} \partial_{+\mu}^\perp \chi_v  {\bar q} \gamma_\nu q  
+ {c_{\chi18,q} \over m_W^3} \epsilon_{\mu\nu\rho\sigma} {\bar \chi}_v  {\sigma}_\perp^{\mu \nu} \partial_{+}^{\perp\rho} \chi_v {\bar q} \gamma^\sigma  q 
+ {c_{\chi20,q} \over m_W^3} {\bar \chi}_v i \partial_-^{\perp \mu} \chi_v {\bar q} \gamma_\mu \gamma_5 q 
\nl
&\quad 
+ {c_{\chi21,q} \over m_W^3} {\bar \chi}_v \sigma_\perp^{\mu \nu} \partial_{+\mu}^{\perp} \chi_v {\bar q} \gamma_\nu \gamma_5 q 
+ {c_{\chi22,q} \over m_W^3} \epsilon_{\mu\nu\rho\sigma} {\bar \chi}_v  {\sigma}_\perp^{\mu \nu} \partial_+^{\perp \rho} \chi_v {\bar q} \gamma^\sigma \gamma_5  q 
+ {c_{\chi23,q} \over m_W^3} {\bar \chi}_v \sigma_\perp^{\mu \nu} \chi_v m_q {\bar q} \sigma_{\mu \nu} q 
  \nl
&\quad 
+ {c_{\chi24,q} \over m_W^3} \epsilon_{\mu\nu\rho\sigma} {\bar \chi}_v {\sigma}_\perp^{\mu \nu} \chi_v m_q {\bar q} \sigma^{\rho\sigma} q 
\Bigg\}
+ {c_{\chi25} \over m_W^3} {\bar \chi}_v \chi_v G_{\alpha \beta}^A G^{A \alpha \beta} 
+ { c_{\chi26} \over m_W^3} {\bar \chi}_v \chi_v G_{\alpha \beta}^A {\tilde G}^{A \alpha \beta} 
  \nl
&\quad 
+ {c_{\chi27} \over m_W^3} {\bar \chi}_v \chi_v v_\mu v_\nu G_{\,\,\,\,\,\,\alpha}^{A \mu} G^{A \nu\alpha}   
+ {c_{\chi28} \over m_W^3} {\bar \chi}_v \sigma_\perp^{\mu \nu} \chi_v 
\epsilon_{\mu\nu\alpha\beta} v^\alpha v^\gamma G^{A \beta \delta} G^{A}_{\gamma\delta} 
+ \dots \, ,
\end{align}
where the ellipsis denotes operators of dimension six and higher
involving the photon,  and operators of dimension eight and higher
involving quarks and gluons.    
In each of (\ref{eq:scalarL}), (\ref{eq:fermionL}) and (\ref{eq:heavyL})
we have employed field redefinitions and chosen a basis of Hermitian QCD operators as 
in the following Section~\ref{sec:QCDbasis}.%
\footnote{
For the dimension four QCD operators, field redefinitions implement the equations of motion
$m_q \bar{q} \sigma^{\mu\nu} q = \partial^{[\mu} \bar{q} \gamma^{\nu]} q 
+ \frac12 \epsilon^{\mu\nu\alpha\beta} \bar{q} \gamma_\mu iD_{-\nu} \gamma_5 q$ and 
$\bar{q}\gamma^{[\mu} iD_-^{\nu]} q = \frac12 \epsilon^{\mu\nu\alpha\beta} \partial_\alpha( \bar{q}\gamma_\sigma\gamma_5 q )$. 
}
Lorentz-invariance constraints on the coefficients in
Eq.~(\ref{eq:heavyL}) may be derived by performing an infinitesimal
boost,
\be\label{eq:boost} 
{\cal B}(q)^\mu_{\ \nu} = g^{\mu}_{\ \nu} 
+ {v^\mu q_\nu - q^\mu v_\nu \over M} + \order(q^2) \,.
\ee 
Relativistic fields transform in the usual way, while  
the heavy field $\chi_v$ transforms as~\cite{Luke:1992cs,Heinonen:2012km}
\begin{align}\label{eq:boostvar}
\chi_v(x) \to e^{iq\cdot x} \left[ 1 + {i q \cdot D_\perp \over 2M^2} + {1\over 4M^2}\sigma_{\alpha\beta} q^\alpha 
D_\perp^\beta 
\dots \right] \chi_v({\cal B}^{-1} x) \, , 
\end{align}
where the ellipsis denotes terms higher order in $1/M$. 
Working through ${\cal O}(M^{-1})$ for photon operators and ${\cal O}(M^{-3})$ 
for quark and gluon operators, we find that the variation of 
Eq.~(\ref{eq:heavyL}) under the boost transformation vanishes upon enforcing the constraints
\be
  {m_W\over M} c_{\chi 3} + 2 c_{\chi 12}  =  {m_W\over M}c_{\chi 4} + 2 c_{\chi 14}  
= {m_W\over M} c_{\chi 5} - 2 c_{\chi 17}  =  {m_W\over M}c_{\chi 6} - 2 c_{\chi 20} 
= c_{\chi 11} = c_{\chi 13} = 0 \, ,
\ee
where the subscript $q$ on coefficients of quark operators is suppressed.
This leaves sixteen independent quark operators (for each quark flavor) through dimension seven, 
which reduce, upon imposing parity and time-reversal symmetry, 
to the seven operators describing nucleon-lepton 
interactions in NRQED~\cite{Hill:2012rh}.

The basis for a heavy scalar is obtained by omitting in Eq.~(\ref{eq:heavyL})
 operators containing the spin structure $\sigma_\perp^{\mu \nu}$. 
The basis for a self-conjugate heavy particle 
is obtained by imposing invariance under Eq.~(\ref{eq:vparity}) 
or Eq.~(\ref{eq:CPT}); in particular we find that the coefficients 
$c_{\chi n}$ vanish for $n$=1, 2, 5, 6, 15, 16, 17, 18, 19, 20, 21, 22, 23, 24. 

\subsection{Weak scale matching\label{sec:weakmatch}}

Above the weak scale, the theory for the WIMP, symmetric under
$SU(3)_c \times SU(2)_W \times U(1)_Y$, may be specified in terms of a
renormalizable UV completion (e.g., a supersymmetric extension), a
basis of contact operators in the case of a heavy mediator, or heavy
particle effective theory in the case of a heavy WIMP.  By performing
a matching calculation between the theories above and below the weak
scale, thereby integrating out the weak scale particles including
$W^\pm, Z^0, t, h$, we obtain a solution for the coefficients $c_i$ of
the low-energy effective theories in Eqs.~(\ref{eq:scalarL}),
(\ref{eq:fermionL}), or (\ref{eq:heavyL}) in terms of parameters in
the high-energy theory.

As a simple illustration, let us consider the case of a Majorana
fermion  electroweak singlet.  The lowest dimension operators
involving SM interactions are given in the electroweak symmetric
theory by
\begin{align} \label{eq:psiHH}
{\cal L}_{\psi,{\rm SM}} &= 
\frac12 \bar{\psi} \left( i\slash{\partial} - M^\prime \right) \psi 
- {1\over \Lambda} \bar{\psi} \left( c^\prime_{\psi 1} + i c^\prime_{\psi 2} \gamma_5 \right)\psi H^\dagger H  + \dots \, , 
\end{align}
where the ellipsis denotes terms suppressed by higher powers of
$\Lambda$,  the scale associated with a heavy mediator. Let us further
assume  $\psi$ to have mass parameter $M^\prime \ll m_W$, and hence organize the
matching by a  power counting employing a scale separation $M^\prime \ll m_W
\ll \Lambda$. 

Upon integrating out the physical Higgs field $h$ and the top quark
$t$, and performing the field redefinition,
\be
\psi \to e^{-i\phi\gamma_5} \psi \,, 
\quad
\tan{2\phi} =  { c^\prime_{\psi 2} v^2 \over c^\prime_{\psi1} v^2 + M^\prime \Lambda } \,,
\ee
to retain a positive real mass convention for $\psi$, we obtain the
effective  lagrangian below the weak scale 
\begin{align} \label{eq:phi}
{\cal L}_{\psi,{\rm SM}} &= 
 \frac12 \bar{\psi} \left( i\slash{\partial} - M \right) \psi 
+ {1 \over m_W^3}\bigg[ 
 \bar{\psi}\left( c_{\psi 7} + i c_{\psi 8} \gamma_5 \right)\psi   
\sum_{q} m_q \bar{q} q  
+ 
 \bar{\psi} \left( c_{\psi 17} + i c_{\psi 18} \gamma_5 \right) \psi \, G^A_{\mu\nu} G^{A {\mu\nu}}
\bigg] 
+ \dots \,,
\end{align}
where the sum runs over the active quark mass eigenstates
$q=u,d,s,c,b$,  and the ellipsis denotes higher-order perturbative and
power corrections.  The physical DM mass and the effective couplings
in the low energy theory are given at leading order by 
\begin{align}\label{eq:majoranaparams}
M &= \sqrt{\left(M^\prime +   {c^\prime_{\psi 1} v^2  \over \Lambda} \right)^2 + \left( c^\prime_{\psi 2} v^2 \over \Lambda \right)^2} \, , \nl
\{ c_{\psi7} \, , c_{\psi 8} \} &= { m_W^3 M^\prime \over  m_h^2 \Lambda M}
 \left\{ c^\prime_{\psi1} + {v^2 \over M^\prime \Lambda} \big[ c^{\prime 2}_{\psi1} + c^{\prime 2}_{\psi2} \big] \, , c^\prime_{\psi 2}\right\}\,, 
\quad 
 \{ c_{\psi17} \, , c_{\psi18} \} 
= -{\alpha_s(m_W) \over 12 \pi} \{ c_{\psi 7} \, , c_{\psi 8} \} \, .
\end{align}
Note that a vanishing $c^\prime_{\psi 1}$ does not imply  a
velocity-suppressed spin-independent cross section for WIMP nucleon scattering,   
since a nonvanishing $c_{\psi 8} \sim (v^2/ M^\prime \Lambda)
c_{\psi 2}^{\prime 2}$ is  induced in the low energy theory.%
\footnote{ This observation has been employed in
\cite{Fedderke:2014wda,Matsumoto:2014rxa}.    }
While we do not pursue a detailed phenomenology of the model
(\ref{eq:phi}),  this example illustrates some generic features of
weak scale matching. 
Firstly, particular UV completions may have nontrivial  correlations
and suppression factors amongst coefficients;  e.g., $c_{3,4,9,10}$
are suppressed by loop ($\sim g^2$) or power ($\sim 1/\Lambda$)
corrections.   Secondly, effects that are naively absent from the high
scale lagrangian are  nonetheless present once a complete analysis is
performed.  It is essential to include a complete basis that is closed
under renormalization and contains all operators not forbidden by symmetry. 

Weak scale matching for an electroweak singlet Dirac fermion or (real or complex) 
scalar can be similarly performed. Weak scale matching for the case of electroweak 
charged dark matter, requires a more intricate analysis as detailed in Ref.~\cite{Hill:2014yka}.

\section{Operator renormalization, scale evolution and matching at heavy quark thresholds\label{sec:weak}} 

Having determined the basis of effective operators and their coefficients 
at the weak scale, 
we may proceed to map onto a theory valid at lower energy scales. We 
identify the relevant QCD operators and compute their anomalous dimensions. 
We then solve the corresponding renormalization group evolution equations and 
enforce matching conditions at heavy quark thresholds, passing from $n_f=5$ renormalized 
at $\mu\sim m_W$ to $n_f=3$ (or $n_f=4$) renormalized below the charm (or bottom) threshold.  

\subsection{QCD operator basis \label{sec:QCDbasis}}

\begin{table}[t]
\begin{center}
\small
\renewcommand{\arraystretch}{1.7}
\begin{tabular}[t]{c|c}
$d$ & QCD operator basis\\
\hline
$3$  & $ V_q^\mu = \bar{q} \gamma^\mu q $\\
& $A_q^\mu = \bar{q} \gamma^\mu \gamma_5 q $ \\
\hline
$4$  & $T_q^{\mu \nu} = i m_q {\bar q} \sigma^{\mu \nu} \gamma_5 q$ \\
& $O_q^{(0)} = m_q {\bar q} q \, , \quad O_g^{(0)} = G^A_{\mu \nu} G^{A \mu \nu}$  \\
& $O^{(0)}_{5q}  = m_q \bar{q} i \gamma_5 q \,,
\quad 
O^{(0)}_{5g} = \epsilon^{\mu\nu\rho\sigma} G^A_{\mu\nu} {G}^{A}_{\rho\sigma}$ \\
& $O^{(2)\mu\nu}_{q} = \frac12 \bar{q}\left( \gamma^{\{\mu} iD_-^{\nu\}}  
- {g^{\mu\nu} \over 4} i\Dslash_- \right) q \, , \quad O^{(2)\mu\nu}_g = -G^{A \mu\lambda} G^{A \nu}_{\phantom{A \nu} \lambda} + { g^{\mu\nu} \over 4} (G^A_{\alpha\beta})^2$\\
& $O^{(2)\mu\nu}_{5q} = \frac12 \bar{q} \gamma^{\{\mu} iD_-^{\nu\}} \gamma_5 q$ \\
\end{tabular}
\end{center}
\caption{\label{tab:QCDoperators}
The seven operator classes: vector $\big( V_q\big) $, axial-vector $\big( A_q\big) $, tensor $\big( T_q \big) $, scalar $ \big( O^{(0)}_{q} \, , O^{(0)}_{g}  \big) $, pseudoscalar $\big( O^{(0)}_{5q} \, , O^{(0)}_{5g} \big) $,
 $C$-even spin-2 $\big( O^{(2)}_{q} \, , O^{(2)}_{g} \big) $ and $C$-odd spin-2 $\big( O^{(2)}_{5q} \big)$. 
Here $A^{[\mu}B^{\nu]} \equiv (A^\mu B^\nu - A^\nu B^\mu)/2$ and $A^{\{\mu}B^{\nu\}} \equiv (A^\mu B^\nu + A^\nu B^\mu)/2$ respectively denote antisymmetrization and symmetrization, 
and the subscript $q$ denotes an active quark flavor. The antisymmetric tensor current $T_q$ and the quark pseudoscalar operator $O^{(0)}_{5q}$ both include a conventional quark mass prefactor.
}
\end{table}

Inspection of the low-energy SM building
blocks in (\ref{eq:SMblocks}) shows that, up to field
redefinitions,
the strong interaction matrix elements relevant for 
WIMP-SM interactions through dimension seven 
involve seven QCD operator classes collected in Table~\ref{tab:QCDoperators}: at dimension three we have the vector and axial-vector currents; at dimension four we have the antisymmetric tensor currents, the scalar operators, the pseudoscalar operators, the $C$-even spin-2 operators and the $C$-odd spin-2 operators. Each of these classes transforms irreducibly under continuous and discrete Lorentz transformations, and is separately closed under renormalization. 

\subsection{Renormalization constants}

\begin{table}[t]
\begin{center}
\small
\renewcommand{\arraystretch}{1.7}
\begin{tabular}[t]{c|c}
Operator & Renormalization constant \\
\hline
$V_q$  & $ Z_{V} = 1$\\
\hline
$A_q$ & $Z_{A}^{({\rm singlet})} = 1 + {\alpha_s \over 4\pi} {16 \over 3} - \big({\alpha_s \over 4\pi} \big)^2 {1 \over \epsilon} \Big(  {20 \over 9} n_f + {88 \over 3}  \Big)  + \order(\alpha_s^3) \, ,$ \\
 & $Z_{A}^{({\rm non-singlet})} = 1 + {\alpha_s \over 4\pi} {16 \over 3} + \big({\alpha_s \over 4\pi} \big)^2 {1 \over \epsilon} \Big( {16 \over 9} n_f - {88 \over 3}  \Big)  + \order(\alpha_s^3) $ \\
 \hline
 $T_q$ & $Z_{T} = 1 - {\alpha_s \over 4\pi} {1 \over \epsilon} {16 \over 3} + \order(\alpha_s^2) $ \\
 \hline
$O_q^{(0)}\, , O_g^{(0)}$  & $Z^{(0)}_{qq} =1 \, , \quad  Z^{(0)}_{qg} =0\, ,$\\
&$ Z^{(0)}_{gq} = {2\gamma_m  \over \epsilon} 
\, , \quad Z^{(0)}_{gg} = 1 - {{\tilde \beta} \over \epsilon}  
$\\
\hline
$O_{5q}^{(0)}\, , O_{5g}^{(0)}$  & $Z^{(0)}_{5,qq} = 1 + {\alpha_s \over 4\pi} {32 \over 3} + \order(\alpha_s^2) \, , \quad   Z^{(0)}_{5,qg} =0 + \order(\alpha_s^2) \, ,$ \\
&$ Z^{(0)}_{5,gq}  =  {\alpha_s \over 4\pi} {1 \over \epsilon} {16 }  + \order(\alpha_s^2) \, , \quad Z^{(0)}_{5,gg} = 1 + {\alpha_s \over 4\pi} {1 \over \epsilon} \beta_0 + \order(\alpha_s^2) $\\
\hline
$O_{q}^{(2)}\, , O_{g}^{(2)}$  & $Z^{(2)}_{qq} = 1 - {\alpha_s \over 4\pi} {1 \over \epsilon} {32 \over 9} + \order(\alpha_s^2) \, , \quad   Z^{(2)}_{qg} = {\alpha_s \over 4\pi} {1 \over \epsilon} {\frac23 } + \order(\alpha_s^2)  \, ,$ \\
&$ Z^{(2)}_{gq}  =  {\alpha_s \over 4\pi} {1 \over \epsilon} {32 \over 9 }  + \order(\alpha_s^2) \, , \quad Z^{(2)}_{gg} = 1 - {\alpha_s \over 4\pi} {1 \over \epsilon} {2 n_f \over 3} + \order(\alpha_s^2) $\\
\hline
$O_{5q}^{(2)}$ & $Z_{5}^{(2)} = 1 - {\alpha_s \over 4\pi} {1 \over \epsilon}{32 \over 9} + \order(\alpha_s^2)$
\end{tabular}
\end{center}
\caption{\label{tab:QCDZ}
Renormalization constants for each of the seven operator classes arising in the low-energy effective theory for the DM particle. Here $n_f$ is the number of active quark flavors and $\beta_0 = 11-2n_f/3$.
}
\end{table}

Let us denote by $O_i$ a generic operator with coefficient $c_i$ belonging to
one of the seven operator classes closed under renormalization.  The relations
between bare and renormalized operators and coefficients are given by
\be\label{eq:Oren}
O^{\rm bare}_i  = Z_{ij}(\mu) O^{\rm ren}_j(\mu) \,, \quad c^{\rm ren}_i(\mu) =  Z_{ji}(\mu) c_j^{\rm bare} \, ,
\ee  
with an implicit sum over repeated indices.
We define the operator renormalization constants $Z_{ij}$ in the
$\overline{\rm MS}$ scheme, except for the axial-vector and
pseudoscalar operators where we consider an additional finite
renormalization to retain a conventional axial current divergence and the scale independence of the quark pseudoscalar matrix elements.

For vector currents, axial-vector currents, tensor currents and $C$-odd spin-two operators, the renormalization constants are quark flavor diagonal, and have the form $Z_{ij} = Z \delta_{ij}$, with $Z$ listed in Table~\ref{tab:QCDZ}. For scalar, pseudoscalar and $C$-even spin-two operators, the renormalization constants, in the basis $(u,d,s,\dots | g)$, have the form
\renewcommand{\arraystretch}{1.4}
\begin{align} \label{eq:Zmatrix}
{Z} &= 
\left( \begin{array}{ccc|c} 
Z_{qq} & & & Z_{qg} \\
& \ddots & & \vdots \\
&& Z_{qq} & Z_{qg} \\
\hline
Z_{gq} & \cdots & Z_{gq} & Z_{gg}
\end{array} 
\right) \, ,
\end{align}
with elements $Z_{ij}$ listed in Table~\ref{tab:QCDZ}.
 
The vector currents, representing conserved quark number,
$\partial_\mu V_q^{ \mu} = 0$, evolve trivially under QCD
renormalization. 
For the axial-vector currents, we consider separately the quark-flavor singlet and non-singlet combinations (see Eq.~(\ref{eq:def380})), and work in the 't~Hooft-Veltman scheme with the convention $\epsilon^{0123} = +1$,
\be\label{eq:tHV}
\gamma_5 = i \gamma^0 \gamma^1 \gamma^2 \gamma^3 
= -{i\over 4!} \epsilon^{\mu\nu\rho\sigma} \gamma_\mu \gamma_\nu \gamma_\rho \gamma_\sigma \,. 
\ee
The renormalization constants $Z_A^{(\rm singlet)}$ and $Z_A^{(\rm non-singlet)}$ include a finite correction in addition to the $\overline{\rm MS}$ scheme~\cite{Larin:1993tq} (see Appendix~\ref{sec:constants} for details), which retains the one-loop anomaly condition,
\be\label{eq:anomaly}
\sum_q 
\partial_\mu A_q^{\mu} = \sum_q 2im_q \bar{q} \gamma_5 q - {g^2 n_f \over 32\pi^2} \epsilon^{\mu\nu\rho\sigma} G^a_{\mu\nu} G^a_{\rho\sigma} \,,
\ee
for the singlet combination, and imposes a vanishing anomalous dimension for the non-singlet combination.
Terms contributing to the one-loop matching and two-loop anomalous dimension have been retained in both $Z_A^{(\rm non-singlet)}$ and $Z_A^{(\rm singlet)}$. Corrections through three-loop order are also available~\cite{Larin:1993tq}. 

For the tensor current, the renormalization constant includes the contribution $Z_m$ (given in Appendix~\ref{sec:constants}) from the quark mass appearing in the definition of $T_q$. Two loop corrections to $Z_T$ are also available~\cite{Kumano:1997qp, Hayashigaki:1997dn, Vogelsang:1997ak}. 
For the scalar operators, the all-orders expression for the coefficient of the
$1/\epsilon$ term of ${Z}^{(0)}$ is specified in terms of coupling
and mass renormalization functions,%
\footnote{A typo appears in the expression after equation (24) of \cite{Hill:2011be}, 
which should read $g^{-1}\beta = g^{-1}dg/d\log\mu \approx -\beta_0 \alpha_s/4\pi$.
}
\be\label{eq:Z0functions}
{\tilde \beta} = \beta/g \, , \quad \beta = { dg \over d \log \mu} \,, \quad \gamma_m = { d \log m_q \over d \log \mu}\, ,
\ee
which are given explicitly in Appendix~\ref{sec:constants}. 

For the pseudoscalar operators, we employ the $\gamma_5$ scheme in Eq.~(\ref{eq:tHV}), and have included the contribution $Z_m$ from the quark mass appearing in the definition of $O_{5q}^{(0)}$. The renormalization constant $Z^{(0)}_5$ also includes an additional finite renormalization constant that ensures nonrenormalization of the pseudoscalar quark operators~\cite{Larin:1993tq} (see Appendix~\ref{sec:constants} for details). 
Terms contributing to the one-loop matching and two-loop anomalous dimension have been retained in $Z^{(0)}_5$.
For the $C$-even spin-two operators, three-loop corrections to the renormalization constant are available from Refs.~\cite{Vogt:2004mw,Moch:2004pa}.
For the $C$-odd spin-two operators, the two-loop anomalous dimension may be obtained from Ref.~\cite{Mertig:1995ny}.

\subsection{Anomalous dimensions and renormalization group evolution \label{sec:anomRG}}

\begin{table}[t]
\begin{center}
\small
\renewcommand{\arraystretch}{1.7}
\begin{tabular}[t]{c|c}
Operator & Anomalous dimension \\
\hline
$V_q$  & $ \gamma_{V} = 0$\\
\hline
$A_q$ & $\gamma_{A}^{({\rm singlet})} =  \big({\alpha_s \over 4\pi} \big)^2 16 n_f + \order(\alpha_s^3) \, , $ \\
 & $\gamma_{A}^{({\rm non-singlet})} = 0$ \\
 \hline
 $T_q$ & $\gamma_{T} =  - {\alpha_s \over 4\pi} {32 \over 3} + \order(\alpha_s^2)  \, ,$ \\
 \hline
$O_q^{(0)}\, , O_g^{(0)}$  & $\gamma^{(0)}_{qq} =0 \, , \quad  \gamma^{(0)}_{qg} =0\, ,$\\
&$ \gamma^{(0)}_{gq} = - 2\gamma^\prime_m  
\, , \quad \gamma^{(0)}_{gg} = {\tilde \beta}^\prime  $\\
\hline
$O_{5q}^{(0)}\, , O_{5g}^{(0)}$  & $\gamma^{(0)}_{5,qq} =0 \, , \quad   \gamma^{(0)}_{5,qg} =0  \, ,$ \\
&$ \gamma^{(0)}_{5,gq}  =  - {\alpha_s \over 4\pi} 32 + \order(\alpha_s^2)  \, , \quad \gamma^{(0)}_{5,gg} = - {\alpha_s \over 4\pi} 2 \beta_0  + \order(\alpha_s^2) $\\
\hline
$O_{q}^{(2)}\, , O_{g}^{(2)}$  & $\gamma^{(2)}_{qq} =  {\alpha_s \over 4\pi}  {64 \over 9}  + \order(\alpha_s^2)  \, , \quad   \gamma^{(2)}_{qg} = - {\alpha_s \over 4\pi}  {\frac43 } + \order(\alpha_s^2)  \, ,$ \\
&$ \gamma^{(2)}_{gq}  = -  {\alpha_s \over 4\pi}  {64 \over 9 }  + \order(\alpha_s^2) \, , \quad \gamma^{(2)}_{gg} =  {\alpha_s \over 4\pi}  {4 n_f \over 3} + \order(\alpha_s^2) $\\
\hline
$O_{5q}^{(2)}$ & $\gamma_{5}^{(2)} = {\alpha_s \over 4\pi} {64 \over 9} + \order(\alpha_s^2)$
\end{tabular}
\end{center}
\caption{\label{tab:QCDgamma}
Anomalous dimensions for the seven operator classes arising in the low-energy effective theory for the DM particle. Here we denote ${X^\prime} \equiv g {\partial \over \partial g } X $.
}
\end{table}

From the relations between bare and renormalized quantities
in Eq.~(\ref{eq:Oren}), we obtain the scale evolution equations
\be\label{eq:evolve}
{d\over d\log\mu} O_i = - \gamma_{ij} O_j \,, \quad
{d\over d\log\mu} c_i = \gamma_{ji} c_j \,, \quad 
\gamma_{ij} \equiv Z_{ik}^{-1} {d\over d\log\mu} Z_{kj}  \, ,
\ee
where the scale dependence and superscript ``${\rm ren}$" on
renormalized quantities in (\ref{eq:Oren}) 
have been suppressed,  and we have defined the
anomalous dimension matrix $\gamma_{ij}$.  In the $\overline{\rm MS}$
scheme the anomalous dimension is given to all orders in $\alpha_s$ in
terms of the coefficient of $1/\epsilon$ in $Z_{ij}$,
\be\label{eq:gammaMSbar}
\gamma_{ij} = -g{\partial\over \partial g} Z_{(1)ij} \,, \quad 
Z_{ij} = \delta_{ij} + \sum_{n=1}^\infty { Z_{(n)ij} \over \epsilon^n} \,. 
\ee
The renormalization constants for axial-vector currents and pseudoscalar operators include a finite contribution beyond ${\overline{\rm MS}}$, and hence we employ the general definition in (\ref{eq:evolve}) to determine their anomalous dimensions. 

For vector currents, axial-vector currents, tensor currents and $C$-odd spin-two operators, the anomalous dimensions have the form $\gamma_{ij} = \gamma \delta_{ij}$, with $\gamma$ listed in Table~\ref{tab:QCDgamma}. For scalar, pseudoscalar and $C$-even spin-two operators, the anomalous dimensions, in the basis $(u,d,s,\dots | g)$, have the form
\renewcommand{\arraystretch}{1.4}
\begin{align}\label{eq:anom}
{\gamma}
&=  \left( \begin{array}{ccc|c} 
\gamma_{qq} & & & \gamma_{qg} \\
& \ddots & & \vdots \\
&& \gamma_{qq} & \gamma_{qg} \\
\hline
\gamma_{gq} & \cdots & \gamma_{gq} & \gamma_{gg}
\end{array} 
\right) \,.
\end{align}
with elements $\gamma_{ij}$ listed in Table~\ref{tab:QCDgamma}.

It is straightforward to solve for the evolution of coefficients from a
high scale $\mu_h$ down to a low scale $\mu_l$, employing the anomalous dimension for each of the seven operator classes. Let us express the solutions as
\be\label{eq:Rdef}
c_i(\mu_l) = R_{ij}(\mu_l,\mu_h) c_j(\mu_h) \, .
\ee
For vector currents, axial-vector currents, tensor currents and $C$-odd spin-two operators, the solutions have the form $R_{ij} = R \delta_{ij}$, with $R$ listed in Table~\ref{tab:QCDrunning}. For scalar, pseudoscalar operators and $C$-even spin-two operators, the solutions in the basis $(u,d,s,\dots | g)$ have the form
\renewcommand{\arraystretch}{1.4}
\begin{align}\label{eq:Rsol1}
R
&=  \left( \begin{array}{ccc|c} 
 & & & R_{qg} \\
&  \mathbb{1} (R_{qq} - R_{qq^\prime} ) + \mathbb{J}  R_{qq^\prime} & & \vdots \\
&&  & R_{qg} \\
\hline
R_{gq} & \cdots & R_{gq} & R_{gg}
\end{array} 
\right) \,,
\end{align}
where the $n_f \times n_f$ matrices $\mathbb{1}$ and $\mathbb{J}$ are respectively the identity matrix and the matrix with all elements equal to unity. For the scalar and pseudoscalar operators $R_{qq^\prime} =0$. The elements $R_{ij}$  are specified in Table~\ref{tab:QCDrunning}, where the results for the $C$-even spin-two operators involve the function
\be\label{eq:roft}
r(t) = \left(\alpha_s(\mu_l) \over \alpha_s(\mu_h) \right)^{ -{1\over 2\beta_0} \left( {64\over 9}  + \frac43 t \right) }\,.
\ee

\begin{table}[t]
\begin{center}
\small
\renewcommand{\arraystretch}{2}
\begin{tabular}[t]{c|c}
Operator & Solution to coefficient running \\
\hline
$V_q$  & $ R_{V} = 1$\\
\hline
$A_q$ & $R_A^{({\rm singlet})} = \exp\bigg\{ {2 n_f \over \pi \beta_0} \big[ \alpha_s(\mu_h) - \alpha_s(\mu_l) \big] 
+ \order(\alpha_s^2) \bigg\}  \, , $ \\
 & $R_A^{({\rm non-singlet})} = 1$ \\
 \hline
 $T_q$ & $R_T =\left( \alpha_s(\mu_l) \over \alpha_s(\mu_h) \right)^{-{16 \over 3\beta_0} } \big[ 1+ \order(\alpha_s)  \big]  $ \\
 \hline
$O_q^{(0)}\, , O_g^{(0)}$  & $R^{(0)}_{qq} =1 \, , \quad  R^{(0)}_{qg} = 2 [\gamma_m (\mu_h) - \gamma_m(\mu_l)]/{\tilde \beta}(\mu_h) \, ,$\\
&$ R^{(0)}_{gq} = 0  
\, , \quad R^{(0)}_{gg} = {\tilde \beta}(\mu_l)/ {\tilde \beta}(\mu_h)  $\\
\hline
$O_{5q}^{(0)}\, , O_{5g}^{(0)}$  & $R^{(0)}_{5,qq} =1 \, , \quad   R^{(0)}_{5,qg} = {16\over\beta_0}\left({\alpha_s(\mu_l)\over \alpha_s(\mu_h)} - 1 \right) + \order(\alpha_s)  \, ,$ \\
&$R^{(0)}_{5,gq}  =0  \, , \quad R^{(0)}_{5,gg} =  {\alpha_s(\mu_l)\over \alpha_s(\mu_h)} + \order(\alpha_s)  $\\
\hline
$O_{q}^{(2)}\, , O_{g}^{(2)}$  & $ R^{(2)}_{qq} - R^{(2)}_{qq^\prime} = r(0) + \order(\alpha_s) \,, \quad R^{(2)}_{qq^\prime} =  {1 \over n_f} \Big[ {16 r(n_f) +3 n_f \over 16 + 3n_f} - r(0) \Big] + \order(\alpha_s) \, ,$ \\
&$ R^{(2)}_{qg} = {16[1-r(n_f)] \over 16 + 3n_f} + \order(\alpha_s) \, ,$ \\
&$ R^{(2)}_{gq}  = {3[1-r(n_f)] \over 16 + 3n_f} + \order(\alpha_s)\, , \quad R^{(2)}_{gg} =  {16+3n_f r(n_f) \over 16 + 3n_f}+ \order(\alpha_s)$\\
\hline
$O_{5q}^{(2)}$ & $R_{5}^{(2)} = \left( \alpha_s(\mu_l) \over \alpha_s(\mu_h) \right)^{-{ 32 \over 9 \beta_0} } \big[ 1+ \order(\alpha_s)  \big]$
\end{tabular}
\end{center}
\caption{\label{tab:QCDrunning}
Solutions to coefficient running for each of the seven operator classes arising in the low-energy effective theory for the DM particle. The coefficient running for $C$-even spin-two operators are given in terms of the function $r(t)$ defined in Eq.~(\ref{eq:roft}). 
}
\end{table}

The vector and non-singlet axial-vector currents have vanishing anomalous dimension, and hence trivial scale evolution. For the singlet axial-vector current, non-trivial renormalization begins at two-loop. 
For the tensor current and $C$-odd spin-two operator we have presented the leading logarithmic order solutions. The chosen renormalization prescription ensures scale invariance of the quark pseudoscalar operators to all orders.

For most phenomenological applications we may simply evaluate the matrix elements of the $C$-even spin-two operators in terms of parton distribution functions (PDFs) at the weak scale $\mu_h \sim m_W$. This avoids the need 
for renormalization group analysis (apart from matching to a convenient
scale to evaluate matrix elements) and heavy-quark threshold matching
conditions. Nonetheless, we include the above results for future analyses 
which may require an evaluation
of tensor matrix elements at low scales, such as in considering
multi-nucleon contributions to matrix elements~\cite{Prezeau:2003sv,Cirigliano:2012pq,Beane:2013kca}, 
or in investigating
the power-suppressed mixing between scalar and tensor operators.

\subsection{Heavy quark threshold matching \label{sec:threshold}}

\begin{table}[t]
\begin{center}
\small
\renewcommand{\arraystretch}{2}
\begin{tabular}[t]{c|c}
Operator & Solution to matching condition \\
\hline
$V_q$  & $ M_V = 1$\\
\hline
$A_q$ & $M_A =  1 + \order(\alpha_s^2) $  \\
 \hline
 $T_q$ & $M_T =  1 + \order(\alpha_s^2)  $ \\
 \hline
$O_q^{(0)}\, , O_g^{(0)}$  & $M^{(0)}_{gQ} = -{\alpha_s^{\prime}(\mu_Q)\over 12\pi} 
\Big\{ 1 + {\alpha_s^{\prime}(\mu_Q) \over 4\pi} \left[  11 - \frac43 \log {\mu_Q \over m_Q} \right]  +\order(\alpha_s^2) \Big\} \, , $ \\
& $ M^{(0)}_{gg} =1 - {\alpha_s^{\prime}(\mu_Q) \over 3\pi} \log {\mu_Q \over m_Q}   +\order(\alpha_s^2) $ \\
\hline
$O_{5q}^{(0)}\, , O_{5g}^{(0)}$  & $M^{(0)}_{5,gQ} ={ \alpha_s^{\prime}(\mu_Q) \over 8 \pi}+\order(\alpha_s^2) \, , \quad  M^{(0)}_{5,gg} =1 +\order(\alpha_s) $ \\
\hline
$O_{q}^{(2)}\, , O_{g}^{(2)}$  & $M^{(2)}_{gQ} =  {\alpha_s^{\prime} \over 3\pi}  \log {\mu_Q \over m_Q}  + \order(\alpha_s^2)  \, , \quad   M^{(2)}_{gg} = 1 + \order(\alpha_s)  $ \\
\hline
$O_{5q}^{(2)}$ & $M_{5}^{(2)} = 1 + \order(\alpha_s^2)$
\end{tabular}
\end{center}
\caption{\label{tab:HQmatching}
Heavy quark threshold matching relations for the seven operator classes.  The strong coupling in the $(n_f+1)$-flavor theory is denoted $\alpha_s^{\prime}$.}
\end{table}

After evolving to the scale $\mu_Q \sim m_Q$, we integrate out the heavy quark, i.e., 
the bottom or charm quark, of mass $m_Q$. The coefficients in the $n_f$- 
and $(n_f+1)$-flavor theories are related by matching physical matrix elements. In terms 
of renormalized coefficients and operators the matching condition is
\be
c^\prime_i \langle O^\prime_i \rangle = c_i \langle O_i \rangle  + \order(1/m_Q) \, ,
\ee
where primed and unprimed quantities are in the $(n_f+1)$- and $n_f$-flavor theories, respectively.\footnote{For example,
the matching condition for scalar operators,  
between physical matrix elements in the 
$5$- and $4$-flavor theories, is given by
$c_g^{(0)\prime} \langle O_g^{(0)\prime} \rangle  
+ \sum_{q=u,d,s,c,b} c_{q}^{(0)\prime} \langle O_{q}^{(0)\prime} \rangle
= c^{(0)}_g \langle O^{(0)}_g \rangle 
+ \sum_{q=u,d,s,c} c^{(0)}_{q} \langle O^{(0)}_{q} \rangle
 + \order(1/m_b) \,,$
where primed and unprimed quantities are in the $5$- and $4$-flavor theories, 
respectively, and the scale dependence is implicit.} Let us express the solution to the matching condition as
\begin{align}\label{eq:Mdef}
c_i(\mu_Q) = {M}_{ij}(\mu_Q) c^{\prime}_j (\mu_Q) \,.
\end{align}
 
The vector currents have trivial matching conditions up to power corrections, while the axial-vector currents, tensor currents and $C$-odd spin-two operators receive threshold matching corrections beginning at $\order(\alpha_s^2)$. Since the latter operator classes have nuclear spin-dependent and/or velocity-suppressed matrix elements in physical WIMP-nucleon processes at small relative velocity, we restrict attention to the leading effects of renormalization scale evolution as detailed in the previous section, and neglect heavy quark threshold matching conditions which are suppressed in each case by a further power of $\alpha_s$.\footnote{For explicit results at two and three loop order see \cite{Grozin:1998kf,Grozin:2006xm}.}
In terms of Eq.~(\ref{eq:Mdef}), we express these solutions in the basis $(u,d,s, \dots |Q)$ as the $n_f \times (n_f +1)$ matrix $M_{ij} = M \delta_{ij}$, with $i=u,d,s, \dots$ and $j = u,d,s, \dots,Q$. The constants $M$ are collected in Table~\ref{tab:HQmatching}.

For the scalar, pseudoscalar and $C$-even spin-two operators, threshold matching involving gluon operators begins at $\order(\alpha_s)$, and the solution to the matching condition may be expressed in terms of an $(n_f+1) \times (n_f+2)$ matrix in the basis $(u,d,s, \dots |Q|g)$ as
\renewcommand{\arraystretch}{1.2}
\begin{align}\label{eq:Mlead}
M = \left( \begin{array}{ccc|c|c} 
 1 &  & & 0 & 0 \\
&  \ddots  & &\vdots & \vdots \\
&  &  1 & 0 & 0 \\
\hline
0 & \cdots & 0 & M_{gQ} & M_{gg} 
\end{array} 
\right) \, .
\end{align}
This parameterization is sufficient for matching at NLO for scalar operators~\cite{Inami:1982xt} and at LO for pseudoscalar and $C$-even spin-two operators.\footnote{In the next section we generalize the parameterization of $M_{ij}$ for higher-order matching in the case of scalar operators.} The elements $M_{ij}$ are given in Table~\ref{tab:HQmatching}.
Scheme dependence for 
the heavy quark mass (e.g. pole versus $\overline{\rm MS}$) appears
at higher order. 

Due to the lightness of 
the charm quark, and correspondingly poorly convergent $\alpha_s(m_c)$
expansion, WIMP-nucleon cross sections can depend sensitively on threshold corrections for the scalar operator. Contributions from matrix elements of the heavy
quark operator, i.e., the column vector $M_{i (n_f + 1)}^{(0)}$, are
known through $\order(\alpha_s^3)$~\cite{Chetyrkin:1997un}. 
In the next section, we employ a sum rule
for matrix elements of scalar operators, derived from the QCD energy
momentum tensor, to obtain new relations amongst the elements of $M^{(0)}$, 
thus extending the available results at higher-orders. 

\subsection{Sum rule constraints on scale evolution and heavy quark threshold matching}\label{sec:sumrule}

The equivalence of physical matrix elements determined in theories
defined at different scales or with different numbers of active quark
flavors, together with the solutions for coefficient evolution and
matching at heavy quark thresholds given in Eqs.~(\ref{eq:Rdef}) and~(\ref{eq:Mdef}), imply relations between operator matrix elements:
\be\label{eq:operatorRM}
\langle O^{\prime (S)}_i \rangle (\mu_h) = R^{(S)}_{ji}(\mu, \mu_h) \langle O^{(S)}_j \rangle (\mu)\, , \quad 
\langle O^{\prime(S)}_i \rangle (\mu_b)= M^{(S)}_{ji}(\mu_b) \langle O^{(S) }_j \rangle (\mu_b) +\order(1/m_b) \, ,
\ee
where $\langle \ \cdot \ \rangle \equiv \langle N | \ \cdot \ | N \rangle $ 
denotes a physical matrix element (for definiteness we consider the matrix element in 
a nucleon state $|N\rangle$). 
The first relation links operator matrix elements at different scales
but with the same number of active quarks, while the second relation
links operator matrix elements at the same scale (here taken to be the
bottom threshold for definiteness) but with $n_f+1$ (primed) and $n_f$
(unprimed) active flavors.

The 
matrix elements $\langle O_i^{(S)} \rangle$ are not independent
but linked by sum rules derived from the trace and traceless part of
the (symmetric and conserved)
QCD energy momentum tensor $\theta^{\mu\nu}$. Let us focus on the scalar case, $S=0$, where
the sum rule for $n_f$ flavors is given by the trace part as
\be\label{eq:scalarsumrule}
\langle \theta^\mu_\mu \rangle = m_N = (1 - \gamma_m) \sum_{q=u,d,s, \dots }^{n_f} \langle O_q^{(0)} \rangle + {{\tilde \beta} \over 2} \langle O_g^{(0)} \rangle \, .
\ee
The sum rule relating matrix elements $\langle O_i^{\prime (S)} \rangle$ in a theory with $n_f+1$ flavors has the analogous form.

Consistency between Eqs.~(\ref{eq:operatorRM})
and~(\ref{eq:scalarsumrule}) yields a system of equations which
imposes constraints on the matrices $R^{(0)}$ and $M^{(0)}$. In the following, we drop the superscript ${(0)}$ for brevity. In the
case of scale evolution, the sum rule 
determines $R$. Starting from the general form,
\renewcommand{\arraystretch}{1.4}
\begin{align}
R(\mu,\mu_h) = \left( \begin{array}{ccc|c} 
1 & & & R_{qg}  \\
& \ddots & & \vdots \\
&& 1 & R_{qg} \\
\hline
0 & \cdots & 0 & R_{gg}
\end{array} 
\right) \, ,
\end{align}
which follows from the scale invariance of $\langle O_q^{(0)}
\rangle$, the functions $R_{qg}$ and $R_{gg}$ are
determined by the system of equations derived from
Eqs.~(\ref{eq:operatorRM}) and~(\ref{eq:scalarsumrule}):
\be
{2 \over {\tilde \beta}(\mu) } R_{gg} = {2 \over {\tilde \beta}(\mu_h) }\, , \quad 
R_{qg} - {2 \over {\tilde \beta}(\mu) } \big[ 1 - \gamma_m(\mu) \big] R_{gg} 
= - {2 \over {\tilde \beta}(\mu_h) } \big[ 1 - \gamma_m(\mu_h) \big] \, .
\ee
This yields the results given in Table~\ref{tab:QCDrunning}.

In the case of heavy quark threshold matching, relations between
elements of the matrix $M$ can be similarly derived.  Consider
the general form,
\renewcommand{\arraystretch}{1.2}
\begin{align}\label{eq:M0muQ}
M(\mu_Q) = \left( \begin{array}{ccc|c|c} 
 &  & & M_{qQ} & M_{qg} \\
&  \mathbb{1}(M_{qq} - M_{qq^\prime}) + \mathbb{J}M_{qq^\prime}   & &\vdots & \vdots \\
&  &  & M_{qQ} & M_{qg} \\
\hline
M_{gq} & \cdots & M_{gq} & M_{gQ} & M_{gg} 
\end{array} 
\right) \, ,
\end{align}
where the $n_f \times n_f$ matrices $\mathbb{1}$ and $\mathbb{J}$  are
respectively the identity matrix and the matrix with all elements
equal to unity. The system of equations derived from
Eqs.~(\ref{eq:operatorRM}) and~(\ref{eq:scalarsumrule}) yield the
following relations
\begin{align}\label{eq:Mrelations}
0 &=  {\tilde \beta}^{( n_f)} - {\tilde \beta}^{( n_f+1)} M_{gg} - 2 \big[ 1-\gamma_m^{(n_f+1)} \big] ( M_{gQ}  + n_f M_{gq})     \, , \nl
0 &=2  
\Big\{     1 - \gamma_m^{(n_f)}  - \big[ 1 - \gamma_m^{(n_f+1)} \big]  ( M_{qQ} + M_{qq} + (n_f-1) M_{qq^\prime} ) \Big\} - {\tilde \beta} ^{(n_f+1)} M_{qg}  \, ,
\end{align}
where the superscripts on $\gamma_m$ and ${\tilde \beta}$ denote 
the $n_f$ dependence, while the $\mu_Q$ dependence is implicit.

We may further simplify the matrix (\ref{eq:M0muQ}). 
By dimensional analysis, the gauge invariant operator $m_q \bar{q}q$
matches onto $(G^A_{\mu\nu})^2$ with power suppression, $\sim m_q/m_Q$, and hence 
$M_{gq}\equiv 0$.  Conserved global chiral symmetries, $q_{L,R} \to e^{i\epsilon_{L,R}}q_{L,R}$ when $m_q\to 0$, 
imply that integrating out the heavy quark $Q$ in the presence of $m_q {\bar q} q$ does 
not induce $m_{q^\prime} {\bar q}^\prime q^\prime$ for $q^\prime \ne q$, i.e., $M_{qq^\prime} \equiv 0$.%
\footnote{
We are free to assume here an anticommuting $\gamma_5$ prescription, since $\gamma_5$ does
not enter the QCD analysis of the scalar operators.  The assumption of diagonal quark 
matching underlies the light quark mass decoupling analysis~\cite{Chetyrkin:1997un,Grozin:2011nk}.  
For an explicit comparison of decoupling relations for pseudoscalar and axial currents 
using different $\gamma_5$ prescriptions, see \cite{Grozin:2006xm}. 
}
Finally, since the quark masses in the $n_f$ and $n_f-1$ flavor theories are defined 
to include the induced effects of the heavy quark, we have simply $M_{qq}\equiv 1$.  
These arguments imply from (\ref{eq:M0muQ}) a solution for all elements in terms of
$M_{gQ}$ and $M_{qQ}$: 
\begin{align}
M_{qq} &\equiv 1\,, \quad M_{qq^\prime} \equiv 0 \,, \quad M_{gq} \equiv 0 \,, 
\nl
M_{gg} &= {\tilde{\beta}^{(n_f)} \over \tilde{\beta}^{(n_f+1)}} 
- {2 \over \tilde{\beta}^{(n_f+1)}} \big[ 1 - \gamma_m^{(n_f+1)} \big] M_{gQ} \,,
\nl
M_{gq} &= {2 \over \tilde{\beta}^{(n_f+1)}} \big[ \gamma^{(n_f+1)}_m - \gamma_m^{(n_f)} \big] 
- {2\over \tilde{\beta}^{(n_f+1)}} \big[ 1 - \gamma_m^{(n_f+1)} \big] M_{qQ} \,.
\end{align} 

Let us consider solutions for the elements of $M^{(0)}$ expanded in powers of $\alpha_s$,
\be
M = \sum_{n=0}^\infty \left( { \alpha_s^{(n_f+1)}(\mu_Q)  \over \pi  }  \right)^n M^{(n)}\, ,
\ee
where the superscript signifies that the strong coupling constant is defined in the
$(n_f+1)$-flavor theory.  
Employing this $\alpha_s$ counting and the
$\order(\alpha_s^4)$ results for $M_{gQ}$ and $M_{qQ}$ from
Ref.~\cite{Chetyrkin:1997un}, we may solve the relations in
Eq.~(\ref{eq:Mrelations}) order by order.%
\footnote{In the notation of Ref.~\cite{Chetyrkin:1997un}, $M_{gQ} = C_1$ and $M_{qQ} = C_2 -1$. 
Scheme dependence of $C_{1}$ and $C_{2}$ enters at $\order(\alpha_s^3)$.}
Let us work in the $\overline{\rm MS}$ scheme, employing results for $M_{gQ}$ and $M_{qQ}$, 
as well as for the nontrivial matching condition between $\alpha_s^{(n_f)}(\mu_Q)$ and
$\alpha_s^{(n_f+1)}(\mu_Q)$ found in
Ref.~\cite{Chetyrkin:1997un}, expressed in terms of the heavy quark mass $m_Q$ defined in this scheme. 
Working through NLO, we recover the result in Table~\ref{tab:HQmatching}.
At NNLO, we find
\begin{align}\label{eq:MsolutionsNNLO}
M_{gg}^{(2)} &=  {11 \over 36} 
- {11 \over 6} \log {\mu_Q \over m_Q}  + {1 \over 9} \log^2 {\mu_Q \over m_Q}  \,. 
\end{align}
At NNNLO, we find
\begin{align}\label{eq:MsolutionsNNNLO}
M_{gg}^{(3)}&=
 { 564731 \over 41472 } 
- { 2821 \over 288} \log {\mu_Q \over m_Q } + {3 \over 16} \log^2 {\mu_Q \over m_Q} 
- {1 \over 27} \log^3 {\mu_Q \over m_Q} - {82043 \over 9216} \zeta(3) 
\nl
&\quad 
+n_f \Bigg[ -{2633 \over 10368} + {67 \over 96}  \log {\mu_Q \over m_Q}  - \frac13 \log^2 {\mu_Q \over m_Q} \Bigg] \, ,\nl
M_{qg}^{(2)} &= 
- {89 \over 54} + {20 \over 9} \log {\mu_Q \over m_Q} - {8 \over 3} \log^2 {\mu_Q \over m_Q}  \, .
\end{align}

Conversely, if $M$ is known, the relation in
Eq.~(\ref{eq:operatorRM}) determines quark matrix elements in the
$(n_f+1)$-flavor theory in terms of those in the $n_f$-flavor theory,
up to power corrections. Employing the results for $M_{gQ}$ and
$M_{qQ}$ from Ref.~\cite{Chetyrkin:1997un}, the matrix element for the
heavy quark in the $(n_f+1)$-flavor theory is given by
\begin{align}\label{eq:heavyME}
\langle O_Q^{\prime (0)} \rangle /m_N &= M_{qQ}  \lambda   + M_{gQ}  {2 \over {\tilde \beta}^{(n_f)} } [1 - (1- \gamma_m^{(n_f)})  \lambda ]  \nl
&=  {1 \over 3 \beta_0^{(n_f)}} \Bigg\{ 2-2\lambda \Bigg\} + { \alpha_s^{(n_f+1)}(\mu_Q) \over \pi} \left( {1 \over 3 \beta_0^{(n_f)}} \right)^2 \Bigg\{  {57 \over 2} - {321 \lambda \over 2} + 8 n_f \Bigg\} \nl
&\quad + \left( { \alpha_s^{(n_f+1)}(\mu_Q) \over \pi} \right)^2 \left( {1 \over 3 \beta_0^{(n_f)}} \right)^3 \Bigg\{{9145 \over 8} - {90985 \lambda \over 8} + {19437 \over 4} \log {\mu_Q \over m_Q} 
- {109461 \lambda \over 4}  \log {\mu_Q \over m_Q} \nl
&\quad +  n_f \Bigg[ {374 \over 3}  +{1420 \lambda \over 3} + 756 \log {\mu_Q \over m_Q} + 3424 \lambda \log {\mu_Q \over m_Q}  \Bigg] 
+ n_f^2 \Bigg[ {7661 \over 144 } -{7469 \lambda \over 144 } \nl
&\quad 
- {455 \over 3} \log {\mu_Q \over m_Q} 
 -107 \lambda \log {\mu_Q \over m_Q}  \Bigg] + n_f^3 \Bigg[ - {77 \over 72 } + {77 \lambda \over 72} 
+{16 \over 3} \log {\mu_Q \over m_Q} \Bigg]
\Bigg\} \nl
&\quad + \left( { \alpha_s^{(n_f+1)}(\mu_Q) \over \pi} \right)^3 \left( {1 \over 3 \beta_0^{(n_f)}} \right)^4 \langle O_Q^{ \prime (0)} \rangle_4 + \order(\alpha_s^4) \, ,
\end{align}
where the scale independent quantity $\lambda \equiv \sum_{q=u,d,s,\dots} \langle O_q^{(0)} \rangle/m_N$ is the sum of light quark scalar matrix elements in the $n_f$-flavor
theory. 
The result for $\langle O_Q^{ \prime (0)} \rangle_4$ can be found in
Appendix~\ref{sec:MEappendix}. 
The functions $M_{gQ}$, $M_{qQ}$ and the relation between $\alpha_s^{(n_f)}(\mu_Q)$ and
$\alpha_s^{(n_f+1)}(\mu_Q)$ are also given in Ref.~\cite{Chetyrkin:1997un} in terms of the pole mass $m_Q^{({\rm pole})}$, and we check that the resulting matrix element $\langle O_Q^{\prime (0)} \rangle$ is consistent with the relation between $m_Q$ and $m_Q^{({\rm pole})}$ given in Ref.~\cite{Chetyrkin:2000yt}.

In Sec.~\ref{sec:hadron}, we employ
this solution to determine the charm scalar matrix element in the $4$-flavor theory in terms of light quark scalar matrix elements measured in 3-flavor lattice QCD. We note that the solutions for $M_{qq}$, $M_{qq^\prime}$ and $M_{gq}$, 
imply the equality of light quark scalar nucleon matrix elements 
in $n_f$ and $n_f+1$ flavor theories, up to power corrections, 
\be\label{eq:lightME}
\langle O^{\prime (0)}_q \rangle = \langle O^{ (0)}_q \rangle  + \order(1/m_Q) \,. 
\ee
Further iteration of these solutions determine scalar matrix elements for the bottom and top quarks.

Our result in Eq.~(\ref{eq:heavyME}) disagrees with the result given in Eq.~(B9) in Appendix B of Ref.~\cite{Vecchi:2013iza}. In particular, the expression for $\langle O_Q^{\prime (0)} \rangle$ given there implies results for $M_{gQ}$ and $M_{qQ}$ that do not agree with those of Ref.~\cite{Chetyrkin:1997un} beyond leading order. Moreover, employing the result of Ref.~\cite{Vecchi:2013iza} in (\ref{eq:Mrelations}) yields the NLO result for arbitrary $\mu_Q$, $M_{gg}=1+\order(\alpha_s^2)$, in disagreement with Ref.~\cite{Inami:1982xt}.
A complete comparison cannot be made since Ref.~\cite{Vecchi:2013iza} does not specify a scheme choice for the heavy quark mass, however $M_{gQ}$ at $\order(\alpha_s^2)$, $M_{qQ}$ at $\order(\alpha_s^3)$ and $M_{gg}$ at $\order(\alpha_s)$ are independent of scheme choice.
In terms of the matrix element $\langle O_Q^{\prime (0)} \rangle$, the $\order(\alpha_s)$ piece differs by terms proportional to $\log{\mu_Q \over m_Q}$, while the $\order(\alpha_s^2)$,  $\order(\alpha_s^3)$ and $\order(\alpha_s^4)$ pieces disagree even at $\mu_Q =m_Q$. The scalar matrix element for a heavy quark was also determined in Ref.~\cite{Kryjevski:2003mh}, however a clear comparison is not straightforward given the details presented there.\footnote{The result in Ref.~\cite{Kryjevski:2003mh} has the scaling $\langle O_Q^{(0)} \rangle \propto (1-\lambda)$, which does not agree with Eq.~(\ref{eq:heavyME}) and Ref.~\cite{Vecchi:2013iza}.} 

\subsection{Low-energy coefficients}

To summarize, the matrices ${R}$ given in Table~\ref{tab:QCDrunning} of Sec.~\ref{sec:anomRG}
and ${M}$ given in Table~\ref{tab:HQmatching} of Secs.~\ref{sec:threshold} and~\ref{sec:sumrule} 
completely specify the mapping of coefficients down to low
energies. For example, coefficients $c_i(\mu_t)$ defined in the
five-flavor theory at scale $\mu_t$ are mapped onto coefficients
$c_i(\mu_0)$ defined in the 3-flavor theory at scale $\mu_0$ as
\be
c_j (\mu_0) = R_{jk} (\mu_0, \mu_c)M_{kl}(\mu_c) R_{lm}(\mu_c, \mu_b) M_{mn} (\mu_b) R_{ni}(\mu_b,\mu_t) c_i (\mu_t) \, .
\ee
Having determined these coefficients, we proceed to analyze the relevant 
nucleon matrix elements. 

\section{Hadronic matrix elements\label{sec:hadron}}

Having determined the structure of the effective theory in terms of quark and gluon degrees of freedom in $n_f=3$ (or $n_f=4$) flavor QCD, we may evaluate the resulting nuclear matrix elements at a renormalization scale $\mu \sim 1-2\,{\rm GeV}$. As a natural handoff point to nuclear modeling, the subsequent section identifies these matrix elements with
matching coefficients of a nucleon-level effective theory. 

In this section, we use nonrelativistic normalization $\bar{u}(k) u(k) = m_N/E_{\bm{k}}$ for nucleon spinors. For the matrix elements of the vector, axial-vector, $C$-even spin-two and $C$-odd spin-two operators, we employ approximate isospin symmetry, neglecting small corrections proportional to $m_u-m_d$ and $\alpha$, to relate proton and neutron matrix elements as
\be\label{eq:PNiso}
\langle p | O_{u} | p \rangle = \langle n | O_{d} | n \rangle \, , \quad \langle p | O_{d} | p \rangle = \langle n | O_{u} | n \rangle \, , \quad \langle p | O_{s} | p \rangle = \langle n | O_{s} | n \rangle \, .
\ee
The proton and neutron tensor charges $t_{q,N}$ defined in Eqs.~(\ref{eq:tensor}) and~(\ref{eq:tensorcharge}) are also related by (\ref{eq:PNiso}), while the matrix element of the tensor current $T_q$ itself requires the appropriate quark mass factor. For the scalar and pseudoscalar matrix elements, we tabulate both the proton and neutron form factors. The corrections to zero momentum transfer ($q^2 \to 0$) are suppressed in the nonrelativistic regime of typical WIMP-nucleon scattering processes. We discuss the 
these corrections in Appendix~\ref{sec:MEappendix}.

\subsection{Vector current matrix elements} 

\begin{table}[t]
\begin{center}
\begin{tabular}{c|ccc}
 $q  $ & $F_1^{(p,q)}(0) $ & $F_2^{(p,q)}(0)$  & $F_2^{(p,q)}(0)$ 
\\
\hline 
$u$ & 2 & $1.62(2)$& $1.65(7)$
\\
$d$ & 1 & $- 2.08(2)$ & $ - 2.05(7)$ 
\\
$s$ & 0 & $- 0.046(19) $& $ - 0.017(74)$
\end{tabular} 
\end{center}
\caption{\label{tab:vec}
Scale independent vector form factors for the proton at $q^2=0$ for light quark flavors $u,d,s$. For $F_2^{(p,q)}(0)$ we present values in the second and third column employing $\mu_s$ from Refs.~\cite{Leinweber:2004tc} and~\cite{Doi:2009sq}, respectively.
The uncertainties are combined in quadrature and symmetrized. The vector form factors for the neutron follow from approximate isospin symmetry expressed in (\ref{eq:PNiso}).
} 
\end{table}

For vector currents we parametrize matrix elements as
\begin{align}
\langle N(k^\prime)| V^{(q)}_\mu | N(k) \rangle 
&\equiv \bar{u}(k^\prime) \left[ F^{(N,q)}_{1}(q^2) \gamma_\mu + {i\over 2 m_N} F^{(N,q)}_{2}(q^2) \sigma_{\mu\nu} q^\nu \right]
u(k) \,, 
\end{align}
where  $q\equiv k^\prime-k$ and $N$ denotes a proton ($p$) or neutron ($n$).
The Dirac $F_1^{(N,q)}$ form factors are normalized according to quark content. 
The Pauli form factors
$F_2^{(N,q)}(0)$ give the contribution of quark flavor $q$ to the nucleon anomalous magnetic moment $a_N$, 
\begin{align}\label{eq:aNucleon}
a_p &\equiv F_2^{(p)}(0) = \frac23 F_2^{(p,u)}(0) -\frac13 F_2^{(p,d)}(0) -\frac13 F_2^{(p,s)}(0)  \,,
\nl
a_n &\equiv F_2^{(n)}(0) = \frac23 F_2^{(n,u)}(0) -\frac13 F_2^{(n,d)}(0) -\frac13 F_2^{(n,s)}(0) \,, 
\end{align}  
where $a_p\approx 1.79$ and $a_n\approx -1.91$.
A phenomenological analysis employing lattice data~\cite{Leinweber:2004tc}
and a direct lattice simulation with $n_f=2+1$ dynamical quarks~\cite{Doi:2009sq} 
support a small value for the strange contribution to 
the proton magnetic moment~\cite{Armstrong:2012bi},
\be\label{eq:mustrange}
F_2^{(p,s)}(0) \equiv \mu_s = \left\{ 
\begin{array}{lc} 
-0.046(19) & \cite{Leinweber:2004tc} 
\\
-0.017(25)(70) & \cite{Doi:2009sq} 
\end{array} \right. \,.   
\ee
Equations~(\ref{eq:aNucleon}) and~(\ref{eq:mustrange}), together with the approximate isospin symmetry expressed in~(\ref{eq:PNiso}), yield $F_2^{(p,u)}(0) = 2a_p + a_n + \mu_s$ and $F_2^{(p,d)}(0) =  a_p +2 a_n + \mu_s$. Numerical values for the proton form factors are collected in Table~\ref{tab:vec}. The $q^2$ dependence of $F_1^{(p,q)}(q^2)$ is described in Appendix~\ref{sec:MEappendix}. Following from (\ref{eq:PNiso}), the neutron form factors for $i=1,2$ are
\be
F_i^{(n,d)} = F_i^{(p,u)} \, , \quad F_i^{(n,u)} = F_i^{(p,d)} \, , \quad  F_i^{(n,s)} = F_i^{(p,s)} \, .
\ee

\subsection{Axial-vector current matrix elements}

\begin{table}[t]
\begin{center}
\begin{tabular}{c|cccc}
 $\mu\ ({\rm GeV}) $ & $F_A^{(p,u)} (0)$ & $F_A^{(p,d)} (0)$ & $F_A^{(p,s)} (0)$ &Ref
\\
\hline 
1-2 & 0.75(8) & -0.51(8) & -0.15(8) & \cite{Kaplan:1988ku}
\\
1 & 0.80(3) & -0.46(4) & -0.12(8) & \cite{Nocera:2014gqa}
\\
2 & 0.79(5) & -0.46(5) & -0.13(10) & \cite{Nocera:2014gqa} \\
\end{tabular} 
\end{center}
\caption{\label{tab:axial}
Axial-vector form factors for the proton at $q^2=0$ for light quark flavors $u,d,s$.
The form factors in the first line are extracted from the non-singlet and singlet form factors in Eq.~(\ref{eq:FA380}), while the form factors in the second and third lines are from the NNPDF parameterization~\cite{Nocera:2014gqa} at indicated values of $\mu$. The axial-vector form factors for the neutron follow from approximate isospin symmetry expressed in (\ref{eq:PNiso}).
} 
\end{table}

For the axial-vector currents we parametrize matrix elements as 
\begin{align}\label{eq:axial}
\langle N(k^\prime)| A^{(q)}_\mu | N(k) \rangle 
&\equiv \bar{u}^{(N)}(k^\prime) 
\left[ F^{(N,q)}_{A}(q^2) \gamma_\mu\gamma_5 + {1\over 2 m_N} F^{(N,q)}_{P^\prime}(q^2) \gamma_5 q_\mu \right] 
u^{(N)}(k) \,,
\end{align}
and it is convenient to consider flavor non-singlet ($A^{(3)}$, $A^{(8)}$) 
and flavor singlet ($A^{(0)}$) linear combinations, 
\begin{align}\label{eq:def380}
A^{(3)}_\mu &= \bar{Q} \gamma_\mu \gamma_5 T^3 Q  
= \frac12 \big[ \bar{u}\gamma_\mu \gamma_5 u - \bar{d}\gamma_\mu \gamma_5 d \big] 
\,, 
\nl
A^{(8)}_\mu &= \bar{Q} \gamma_\mu \gamma_5 T^8 Q 
= {1\over 2\sqrt{3}}\big[ \bar{u}\gamma_\mu \gamma_5 u + \bar{d}\gamma_\mu \gamma_5 d 
- 2 \bar{s} \gamma_\mu \gamma_5 s \big]
\,, 
\nl
A^{(0)}_\mu &= \frac13 \bar{Q} \gamma_\mu \gamma_5 Q 
= \frac13 \big[  \bar{u}\gamma_\mu \gamma_5 u + \bar{d}\gamma_\mu \gamma_5 d 
+ \bar{s} \gamma_\mu \gamma_5 s \big] \,. 
\end{align} 
In the limit of $SU(3)$ flavor symmetry, the $q^2=0$ limit for these matrix elements 
can be extracted from hyperon semileptonic decay and $\nu p$ scattering~\cite{Kaplan:1988ku},
\begin{align}\label{eq:FA380}
F_A^{(p,3)}(0) &= {(F+D) \over 2} = 0.63(2)\,,
\quad
F_A^{(p,8)}(0) = {( 3F - D ) \over 2 \sqrt{3}}  = 0.16(2) \,,
\quad 
F_A^{(p,0)}(0\, , \mu) = 0.03(8)\, ,
\end{align}
where $D=0.80(2)$ and $F=0.45(2)$. The non-singlet currents are scale independent but the flavor singlet current has weak scale dependence governed by the anomalous dimension $\gamma_A^{({\rm singlet})}$ in Table~\ref{tab:QCDgamma}, corresponding to the solution $R_A^{({\rm singlet})}$ in Table~\ref{tab:QCDrunning}. In particular, with $n_f=3$, running from $\mu=2 \, {\rm GeV}$ to $\mu=1 \, {\rm GeV}$ gives a factor of $R_A^{({\rm singlet})}(1 \, {\rm GeV} ,2\, {\rm GeV}) = 0.96$, and we may thus consider $F_A^{(p,0)}$ in Eq.~(\ref{eq:FA380}) to be evaluated at $\mu=1-2 \, {\rm GeV}$. The first line of Table~\ref{tab:axial} lists the matrix elements of definite quark flavor from solving~(\ref{eq:def380}) and employing numerical values in~(\ref{eq:FA380}).

The $q^2=0$ limit of these form factors may also be constrained by observables of polarized deep inelastic scattering, via 
\be
F_A^{(p,q)}(0) = \int_0^1 dx \, \Big[ \Delta q(x,\mu) + \Delta {\bar q} (x,\mu) \Big] \,,
\ee 
where $\Delta q(x,\mu)$ is the quark helicity distribution evaluated at scale $\mu$. Numerical values for these matrix elements extracted from the NNPDF collaboration's parameterization of $\Delta q$ in Ref.~\cite{Nocera:2014gqa} are listed in Table~\ref{tab:axial}, showing a negligible scale dependence.
Results from lattice calculations~\cite{Fukugita:1994fh,Dong:1994zs} are numerically similar. 
The $q^2$ dependence of $F_A^{(p,a)}$ is described in Appendix~\ref{sec:MEappendix}. Following from (\ref{eq:PNiso}), the neutron form factors are
\be
F_A^{(n,d)} = F_A^{(p,u)} \, , \quad F_A^{(n,u)} = F_A^{(p,d)} \, , \quad  F_A^{(n,s)} = F_A^{(p,s)} \, .
\ee

The terms parametrized by induced pseudoscalar form factors $F_{P^\prime}$ in (\ref{eq:axial}) 
are suppressed by two powers of $|\bm{k}|/m_N$, and lead to numerically small 
contributions in typical WIMP-nucleus scattering processes. For completeness
we describe the leading contributions to these form factors in Appendix~\ref{sec:MEappendix}.

\subsection{Antisymmetric tensor current matrix element} 

\begin{table}[t]
\begin{center}
\begin{tabular}{c|cccc}
 $\mu \ ({\rm GeV}) $ & $t_{u,p} (\mu)$ & $t_{d,p} (\mu)$ &  $t_{s,p} (\mu)$ & Ref \\
\hline 
- & 4/3 & -1/3 & 0 & -
\\
$1.0$ & 0.88(6) & -0.24(5) & -0.05(3) & -
\\
$1.4$ & 0.84(6) & -0.23(5) & -0.05(3) & \cite{Aoki:1996pi}
\\
$2.0$ & 0.81(6) & -0.22(5) & -0.05(3) & -
\end{tabular} 
\end{center}
\caption{\label{tab:tensor}
Tensor charges from a nonrelativistic quark model ($\mu$ unspecified) and the lattice measurement in Ref.~\cite{Aoki:1996pi} at $\mu \approx 1.4 \, {\rm GeV}$ for a proton. The values at $\mu=1,2 \, {\rm GeV}$ are obtained by  scale evolution of the tensor charges from $\mu = 1.4 \, {\rm GeV}$.
The tensor charges for the neutron follow from approximate isospin symmetry expressed in (\ref{eq:PNiso}).
} 
\end{table}

For the antisymmetric tensor currents, we
parametrize the matrix element as 
\be\label{eq:tensor}
{E_{\bm{k}}\over
m_N}  \langle N(k)| T^{(q)}_{\mu\nu} | N(k) \rangle
\equiv {2\over m_N} s^{[\mu} k^{\nu]} m_q(\mu) t_{q,N} (\mu) \,,  
\ee 
where
$s^\mu=-(E_{\bm{k}}/2m_N^2) \epsilon^{\mu\nu\rho\sigma} k_\nu
\bar{u}(k) \sigma_{\rho\sigma} u(k)$ is the covariant spin vector
satisfying $k^\mu s_\mu =0$ and $s^2 =-1$. In terms of structure
functions appearing in polarized deep inelastic scattering, the tensor charges are given as
\be \label{eq:tensorcharge}
t_{q,N} (\mu) = \int_{-1}^1 dx \, \delta q_N  (x,\mu) \,.  
\ee 
The functions $\delta q (x,\mu)$ are not yet well constrained
experimentally. Table~\ref{tab:tensor} lists values for the proton tensor
charges $t_{q,p}$ from a nonrelativistic quark model with $SU(6)$ spin flavor
symmetry and from a lattice measurement~\cite{Aoki:1996pi}. Other estimates of $t_{u,p}$, $t_{d,p}$ or $t_{u,p} - t_{d,p}$ have been obtained using
lattice QCD methods~\cite{Dolgov:2002zm, Gockeler:2005cj, Lin:2008uz}, QCD sum rules~\cite{He:1994gz},
modeling~\cite{Wakamatsu:2007nc, Cloet:2007em} and semi-inclusive deep
inelastic scattering data~\cite{Anselmino:2008jk}.

The tensor charges at $\mu =1,\, 2 \ {\rm GeV}$ in Table~\ref{tab:tensor} are obtained by scale evolution of the tensor charges at $\mu =1.4 \ {\rm GeV}$ using the anomalous dimension $\gamma_T - \gamma_m$ with $\gamma_T$ given in Table~\ref{tab:QCDgamma} and $\gamma_m$ the quark mass anomalous dimension given in Appendix~\ref{sec:constants}. Together with $m_q(\mu)$, e.g., 
taken from the PDG~\cite{Beringer:1900zz} or Ref.~\cite{Xing:2007fb}, the tensor charges in Table~\ref{tab:tensor} specify the matrix element of the antisymmetric tensor current $T^{\mu \nu}_q$. Following from (\ref{eq:PNiso}), the neutron tensor charges are
\be
t_{d,n} = t_{u,p} \, , \quad t_{u,n} = t_{d,p} \, , \quad  t_{s,n} = t_{s,p} \, .
\ee

\subsection{Scalar matrix elements} 

For the dimension four scalar
operators, we restrict attention to 
forward nucleon matrix elements.
Let us define
\begin{align} \label{eq:scalar}
{E_{\bm{k}}\over m_N} \langle N(k) | O_{q}^{(0)}| N(k) \rangle &\equiv m_N f_{q,N}^{(0)} \,,
\quad 
{- 9\alpha_s(\mu)\over 8\pi} 
{E_{\bm{k}}\over m_N} 
\langle N(k) | O_g^{(0)}(\mu) | N(k) \rangle 
\equiv m_N f_{g,N}^{(0)}(\mu) \,,
\end{align}
where the appearance of the numerical factor involving $\alpha_s(\mu)$ is 
purely conventional. The operator matrix elements are not independent, 
being linked by the sum rule in Eq.~(\ref{eq:scalarsumrule}) as
\begin{align}\label{eq:gluon}
m_N \bar{u}(k) u(k) 
= (1- \gamma_m) \sum_q \langle N(k) | m_q \bar{q} q | N(k) \rangle 
+ {\tilde{\beta}\over 2 } \langle N(k) | (G^a_{\mu\nu})^2 | N(k) \rangle  \,,
\end{align} 
ignoring $\order(1/m_N)$ power corrections. Combining~(\ref{eq:scalar}) and~(\ref{eq:gluon}) we have 
\be \label{eq:fglue}
f_{g,N}^{(0)} =    - { \alpha_s \over 4 \pi} { 9 \over {\tilde \beta}}   
\Big\{  1- \big( 1- \gamma_m \big) \lambda \Big\} = 1 - \lambda + \order(\alpha_s)  \,,
\ee
where $\lambda= \sum_{q=u,d,s} f_{q,N}^{(0)}$, the scale dependence is implicit, and the second equality is obtained by
neglecting $\gamma_m$ and $\order(\alpha_s^2)$ contributions to $\tilde{\beta}$.
In Sec.~\ref{sec:pheno}, we will see that corrections to the leading order relation are numerically important 
in the case of electroweak-charged WIMPs. 

We may extract the up and down quark scalar nucleon matrix elements from the scale-invariant combinations, 
\begin{align}\label{eq:sigmas}
\Sigma_{\pi N} &= {m_u+m_d\over 2} \langle N | (\bar{u} u + \bar{d} d) | N \rangle = 
44 (13) \,{\rm MeV}  \,,
\nl
\Sigma_{-} &= (m_d - m_u) \langle N | (\bar{u} u - \bar{d} d) | N \rangle =  \pm 2(2) \,{\rm MeV} \,,
\end{align}
where the upper (lower) sign in $\Sigma_-$ is for the proton (neutron)~\cite{Gasser:1982ap}.
The numerical value for the pion-nucleon sigma term $\Sigma_{\pi N} $ is the lattice result from Ref.~\cite{Durr:2011mp} with errors symmetrized.
For the strange scalar nucleon matrix element, we use the updated lattice result $m_N f^{(0)}_{s,N} = 40 \pm 20 \, {\rm MeV}$ from Ref.~\cite{Junnarkar:2013ac}, where we assume a conservative $50 \%$ uncertainty compared to their estimate of $25 \%$.

\begin{table}[t]
\begin{center}
\begin{tabular}{c|cc}
 $q  $ & $f^{(0)}_{q,p}  $ & $f^{(0)}_{q,n}  $   
\\
\hline 
$u$ & 0.016(5)(3)(1) & $0.014(5)({}^{+2}_{-3})(1)$
\\
$d$ & 0.029(9)(3)(2) & $0.034(9)({}^{+3}_{-2})(2)$ 
\\
$s$ & 0.043(21) & $0.043(21) $
\end{tabular} 
\end{center}
\caption{\label{tab:scalar}
Scale independent scalar form factors for the proton and neutron for light quark flavors $u,d,s$. The first, second and third uncertainties are from
$\Sigma_{\pi N}$, $m_u/m_d$ and $\Sigma_-$, respectively. As discussed below Eq.~(\ref{eq:massratios}), the parameterization in Eq.~(\ref{eq:fufd}) leads to highly correlated uncertainties in $f_{u,N}^{(0)}$ and $f_{d,N}^{(0)}$.} 
\end{table}

For models with identical couplings to up and down quarks, it is sufficient to take as input $m_N\big( f_{u,N}^{(0)} + f_{d,N}^{(0)}\big) = \Sigma_{\pi N}  - \Sigma_- /2 \approx  \Sigma_{\pi N}   $, neglecting the small contribution from $\Sigma_-$. For general applications requiring separately the up and down quark scalar matrix elements let us write
\begin{align}\label{eq:fufd}
f^{(0)}_{u,N} = {R_{ud} \over 1 + R_{ud}} \, {\Sigma_{\pi N} \over m_N} (1 + \xi) \, , \quad f^{(0)}_{d,N} = {1 \over 1 + R_{ud}} \, {\Sigma_{\pi N} \over m_N} (1 - \xi) \, , \quad \xi = {1 + R_{ud} \over 1 - R_{ud}}  \, { \Sigma_-  \over 2 \Sigma_{\pi N} } \, ,
\end{align}
where we employ the quark mass ratios adopted from PDG values~\cite{Beringer:1900zz} (symmetrizing errors),
\begin{align}\label{eq:massratios}
R_{ud} \equiv {m_u\over m_d} = 0.49 \pm 0.13 \,,
\quad 
R_{sd} \equiv {m_s \over m_d} = 19.5 \pm 2.5 \,.
\end{align}
The resulting numerical values for the light quark scalar matrix elements are collected in Table~\ref{tab:scalar}. The uncertainties in $f_{u,N}^{(0)}$ and $f_{d,N}^{(0)}$ are highly correlated, and for applications we use Eq.~(\ref{eq:fufd}), varying the inputs $\Sigma_{\pi N}$, $R_{ud}$ and $\Sigma_-$ whose uncertainties are taken as uncorrelated. 
For both proton and neutron, the gluon matrix element $f_{g,N}^{(0)}$ is obtained from the quark matrix elements via
the sum rule in Eq.~(\ref{eq:gluon}).

From the analysis of heavy quark matching conditions in Sec.~\ref{sec:sumrule}, we may determine the scalar matrix elements of heavy quark flavors. For definiteness, let us consider 4-flavor QCD with a heavy charm quark. Denoting quantities in the 4-flavor (3-flavor) theory with (without) a prime, the results in Eqs.~(\ref{eq:heavyME}) and~(\ref{eq:lightME}) yield
\begin{align}\label{eq:fcharm}
f^{(0)\prime}_{c,N}  &= 0.083 - 0.103 \lambda + \order(\alpha_s^4,1/m_c) = 0.073(3)+ \order(\alpha_s^4,1/m_c)\,, \nl
f_{q,N}^{(0) \prime} &= f_{q,N}^{(0)} + \order(1/m_c)\,,
\end{align}
where we use $\lambda \approx \Sigma_{\pi N}/m_N + f_{s,N}^{(0)} = 0.089(26)\, {\rm MeV}$, neglecting the small contribution from $\Sigma_-$. An expression for $f^{(0)\prime}_{c,N}$ in terms of $\alpha_s^\prime(\mu_c)$ is given in Appendix~\ref{sec:MEappendix}; in particular, the $\order(\alpha_s^3)$ term in $f_{c,N}^{(0)\prime}$ employs $\langle O_Q^{\prime (0)} \rangle_4$ derived in Sec.~\ref{sec:sumrule}. The uncertainty in $f_{c,N}^{(0)\prime}$ is presently dominated by hadronic inputs, and 
in (\ref{eq:fcharm})
we neglect the small uncertainty ($ < 1 \%$) from scale variation of $\mu_c$.
Recent lattice measurements of the charm
matrix element in Refs.~\cite{Freeman:2012ry}  and~\cite{Gong:2013vja}  have determined 
\be\label{eq:fcharmlattice}
{f}^{(0) \prime}_{c,N} = \left\{ 
\begin{array}{lc} 
0.10(3) & \cite{Freeman:2012ry} 
\\
0.07(3) & \cite{Gong:2013vja} 
\end{array} \right. \,,
\ee
which are consistent within large errors with (\ref{eq:fcharm}). As discussed below (\ref{eq:heavyME}), we find discrepancies with previous determinations of the heavy quark scalar matrix elements~\cite{Vecchi:2013iza,Kryjevski:2003mh}.\footnote{In Ref.~\cite{Junnarkar:2013ac}, the result of Ref.~\cite{Kryjevski:2003mh} was presented with updated inputs.} Nonetheless, due to a large $\order(30\%)$ uncertainty in $\lambda$, the resulting numerical values are consistent.
A nonperturbative determination of the charm and light quark matrix elements in 4-flavor lattice QCD would avoid uncertainties associated with the charm scale $\mu_c \sim m_c$, such as $\order(1/m_c)$ power corrections and $\order(\alpha_s )$ perturbative corrections. In Sec.~\ref{sec:pheno}, we investigate the 
evaluation of 
the spin-independent cross section for heavy electroweak-charged WIMPs in the 4-flavor theory.

\subsection{Pseudoscalar matrix elements}

\begin{table}[t]
\begin{center}
\begin{tabular}{c|cc|cc}
 $q  $ & $f^{(0)}_{5q,p}  $ &  Ref.~\cite{Cheng:2012qr}  & $f^{(0)}_{5q,n}  $ & Ref.~\cite{Cheng:2012qr} 
\\
\hline 
$u$ & 0.42(8)(1) & 0.43 &-0.41(8)(1)& -0.42
\\
$d$ & -0.84(8)(3) & -0.84 &0.85(8)(3)& 0.85
\\
$s$ & -0.48(8)(1)(3)  & -0.50 &-0.06(8)(1)(3)& -0.08
\end{tabular} 
\end{center}
\caption{\label{tab:pseudo}
Scale invariant quark pseudoscalar form factors evaluated at $\kappa(0,\mu) = 0$. We list numbers for the proton and neutron obtained from (\ref{eq:f5sol}) with inputs from (\ref{eq:massratios}) and (\ref{eq:FA380}), and compare to the values in Table II of Ref.~\cite{Cheng:2012qr}. The first, second and third uncertainties are respectively from $R_{ud}$, $F_A^{(p,3)}$ and $F_A^{(p,8)}$; negligible uncertainties are not shown.} 
\end{table}

For the quark and gluon pseudoscalar operators we parametrize the matrix elements as\begin{align}\label{eq:pseudoME}
{E_{\bm{k}}\over m_N}
\langle N(k^\prime)| O_{5q}^{(0)} | N(k) \rangle 
&\equiv m_N f_{5q,N}^{(0)}(q^2) \bar{u}(k^\prime)i\gamma_5 u(k) \,,
\nl
{E_{\bm{k}}\over m_N}
\langle N(k^\prime)| O_{5g}^{(0)} | N(k) \rangle
&\equiv m_N f_{5g,N}^{(0)}(q^2)  \bar{u}(k^\prime)i\gamma_5 u(k)  \, ,
\end{align}
where the quark pseudoscalar operators have been defined independent of renormalization scale, while the gluon operators have a weak scale dependence. The matrix elements in Eq.~(\ref{eq:pseudoME}) are related to the matrix elements of the axial vector current through the axial anomaly in Eq.~(\ref{eq:anomaly}). Employing the matrix elements for the non-singlet axial-vector currents in Eq.~(\ref{eq:FA380}), together with the additional definition,
\be\label{eq:singletpseudo}
\sum_{q=u,d,s} \langle N(k^\prime) | {\bar q} i \gamma_5 q  | N (k) \rangle
\equiv  \kappa(q^2,\mu) \bar{u}(k^\prime)i\gamma_5 u(k)  \,,
\ee
we find the following quark pseudoscalar form factors at $q^2=0$:
\begin{align}\label{eq:f5sol}
f_{5u,p}^{(0)} (0)
&= { R_{ud} \Big( \sqrt{3} F_A^{(p,8)}(0) + \big[ 1 + 2 R_{sd}\big] F_A^{(p,3)}(0)  \Big) \over  R_{sd} + R_{ud} +R_{sd} R_{ud}  } +\omega  \, , \nl
f_{5d,p}^{(0)} (0)
&= { \sqrt{3} F_A^{(p,8)}(0) R_{ud} - \big[ R_{ud} + 2 R_{sd}\big] F_A^{(p,3)}(0)  \over  R_{sd} + R_{ud} +R_{sd} R_{ud}  }  +\omega  \, , \nl
f_{5s,p}^{(0)} (0)
&= {R_{sd} \Big( - \sqrt{3}  \big[ 1 + R_{ud} \big] F_A^{(p,8)}(0) - \big[ 1 - R_{ud} \big] F_A^{(p,3)}(0)   \Big) \over  R_{sd} + R_{ud} +R_{sd} R_{ud}} 
+\omega  \, ,
\end{align} 
where the quark mass ratios $R_{qq^\prime} = m_q / m_{q^\prime}$ are given in (\ref{eq:massratios}) and 
$\omega$ is the scale independent quantity, 
\begin{align} \label{eq:omega}
 \omega &=  {  \kappa(0,\mu) m_d(\mu) R_{sd} R_{ud} \over m_N \Big( R_{sd} + R_{ud} + R_{sd} R_{ud} \Big) } \,.
\end{align}
In the absence of better information on the quantity $\kappa(q^2,\mu)$, we list numerical values for the quark form factors in 
Table~\ref{tab:pseudo} setting $\kappa(0,\mu)=0$, as motivated by large $N_c$ arguments~\cite{Cheng:2012qr}.
This standard ansatz should be revisited if observables are found to be sensitive to nonzero $\omega$. 
The matrix element for the pseudoscalar gluon operator may then be obtained through Eq.~(\ref{eq:anomaly}),
\be
f_{5g,N}^{(0)}(0) =  {16 \pi \over \alpha_s(\mu)} \left[ \frac13 \sum_q f_{5q,N}^{(0)}(0) -  F_A^{(p,0)}(0,\mu) \right].
\ee
As discussed below (\ref{eq:FA380}), the scale dependence from the singlet axial-vector form factor $F_A^{(p,0)}(0,\mu)$ is weak.
The neutron form factors presented in Table.~\ref{tab:pseudo} were obtained using approximate isospin symmetry for the axial-vector currents, i.e., taking $F_A^{(n,3)} = - F_A^{(p,3)}$ in (\ref{eq:f5sol}). 

\subsection{$C$-even spin-two matrix elements} 

\begin{table}[t]
\begin{center}
\begin{tabular}{c|cccccc}
 $\mu \ (\rm GeV) $ & $f_{u,p}^{(2)}(\mu)$ & $f_{d,p}^{(2)}(\mu)$ & $f_{s,p}^{(2)}(\mu)$ & $f_{c,p}^{(2)}(\mu)$ & $f_{b,p}^{(2)}(\mu)$ & $f_{g,p}^{(2)}(\mu)$ 
\\
\hline 
1 & 0.404(9) & 0.217(8) & 0.024(4) & - & - & 0.356(29)
\\
1.2 & 0.383(8) & 0.208(8) & 0.027(4) & - & - &0.381(25)
\\
1.4 & 0.370(8) & 0.202(7) & 0.030(4) & - & - & 0.398(23)
\\
2 & 0.346(7) & 0.192(6) & 0.034(3) & - & - & 0.419(19)
\\
$80.4/\sqrt{2}$ & 0.260(4) & 0.158(4) & 0.053(2) & 0.036(1) & 0.0219(4) & 0.470(8)
\\
100 & 0.253(4) & 0.156(4) & 0.055(2) & 0.038(1) & 0.0246(5) & 0.472(8)
\\
$172\sqrt{2}$ & 0.244(4) & 0.152(3) & 0.057(2) & 0.042(1) & 0.028(1) & 0.476(7)
\end{tabular} 
\end{center}
\caption{\label{tab:momfraction}
Form factors for $C$-even spin-two operators derived from MSTW analysis~\cite{Martin:2009iq} at different
values of $\mu$. The neutron form factors follow from approximate isospin symmetry expressed in (\ref{eq:PNiso}).
} 
\end{table}

For $C$-even spin-two operators, the forward matrix elements are parametrized as 
\begin{align}
{E_{\bm{k}}\over m_N} \langle N(k) | O_q^{(2)\mu\nu}(\mu) | N(k) \rangle 
&\equiv {1\over m_N} \left(k^\mu k^\nu - {g^{\mu\nu} \over 4} m_N^2 \right) f_{q,N}^{(2)}(\mu) \,,
\nl
{E_{\bm{k}}\over m_N} \langle N(k) | O_g^{(2)\mu\nu}(\mu) | N(k) \rangle 
&\equiv {1\over m_N} \left(k^\mu k^\nu - {g^{\mu\nu} \over 4} m_N^2 \right) f_{g,N}^{(2)}(\mu) \,,
\end{align}
and are identified as moments of parton distribution functions 
constrained in unpolarized deep inelastic scattering,  
\be
f_{q,p}^{(2)}(\mu) = \int_0^1 dx \, x [ q(x,\mu) + {\bar{q}}(x,\mu) ] \,,
\ee
where $q(x,\mu)$ is the parton distribution function evaluated at scale $\mu$. 
Neglecting power corrections, the sum of spin two operators is identified as the
traceless part of the QCD energy momentum tensor, 
\be
\sum_{q=u,d,s} f_{q,p}^{(2)}(\mu) + f_{g,p}^{(2)}(\mu) = 1 \,.
\ee
Table~\ref{tab:momfraction} lists coefficient values for
renormalization scales $\mu=1, 1.2, 1.4, 2, m_W/\sqrt{2}, 100, m_t \sqrt{2} \ {\rm GeV}$ 
using the parameterization and analysis of MSTW~\cite{Martin:2009iq}. 
Following from (\ref{eq:PNiso}), the neutron form factors are
\be
f_{u,n}^{(2)} = f_{d,p}^{(2)} \,, \quad
f_{d,n}^{(2)} = f_{u,p}^{(2)} \,, \quad
f_{s,n}^{(2)} = f_{s,p}^{(2)} \,.
\ee

\subsection{$C$ odd, spin two matrix elements} 

\begin{table}[h]
\begin{center}
\begin{tabular}
{c|ccc}
 $\mu \ (\rm GeV)$ & $f_{5u,p}^{(2)}(\mu)$ & $f_{5d,p}^{(2)}(\mu)$ & $f_{5s,p}^{(2)}(\mu)$ 
\\
\hline 
1.0 & 0.186(7) & -0.069(8) & -0.007(6) 
\\
1.2 & 0.175(6) & -0.065(7) & -0.006(6)  
\\
1.4 & 0.167(6) & -0.062(7) & -0.006(5) 
\\
2.0 & 0.154(5) & -0.056(6) & -0.005(5)
\end{tabular} 
\end{center}
\caption{\label{tab:polmomfraction}
Form factors for $C$-odd spin-two operators derived from NNPDF analysis~\cite{Nocera:2014gqa} at different
values of $\mu$. The neutron form factors follow from approximate isospin symmetry expressed in (\ref{eq:PNiso}).
} 
\end{table}

For $C$-odd spin-two operators, we parametrize the matrix elements as~\cite{Kodaira:1998jn}
\begin{align}
{E_{\bm{k}}\over m_N} 
\langle N(k) | O_{5q}^{(2)\mu\nu}(\mu) | N(k) \rangle 
&\equiv s^{\{ \mu} k^{\nu \} } f_{5q,N}^{(2)}(\mu) \,,
\end{align}
where $s^\mu$ is the nucleon spin defined below (\ref{eq:tensor}). 
The coefficients are moments of polarized structure functions 
\be
f_{5q,N}^{(2)}(\mu) = \int_0^1 dx \, x [ \Delta q(x,\mu) + \Delta {\bar{q}}(x,\mu) ] \,. 
\ee
Table~\ref{tab:polmomfraction} lists coefficient values for the proton, at renormalization scales $\mu = 1,1.2,1.4, 2 \ {\rm GeV}$ using the parameterization and analysis of NNPDF~\cite{Nocera:2014gqa}. 
Following from (\ref{eq:PNiso}), the neutron form factors are
\be
f_{5u,n}^{(2)} = f_{5d,p}^{(2)} \,, \quad
f_{5d,n}^{(2)} = f_{5u,p}^{(2)} \,, \quad
f_{5s,n}^{(2)} = f_{5s,p}^{(2)} \,.
\ee

\section{Nucleon level effective theory\label{sec:nucleon}} 

At energy scales much lower than $\Lambda_{\rm QCD}, m_\pi$, it is useful to employ an effective description in terms 
of nucleon degrees of freedom.   
We consider WIMP-hadron interactions given either through 
electromagnetic couplings, or by contact operators with contractions of 
Lorentz vector indices (perhaps including heavy-particle reference vectors $v_\mu$) 
and the QCD operators of the previous section.    
The heavy nucleon lagrangian is given by 
\be
{\cal L}_N = \bar{N}_u\bigg\{  iu\cdot D  - {D_\perp^2 \over 2 m_N} + \dots  \bigg\} N_u  \,,
\ee
where $D_\mu = \partial_\mu - ieQ A_\mu$ is the electromagnetic gauge covariant derivative, and 
we have introduced the timelike invariant
vector $u^\mu$ for the nucleon $N_u$, in addition to $v^\mu$ for the WIMP.   

\subsection{Matching conditions  in single nucleon sector}
We begin by constructing the heavy particle representation of the nucleon.  
For the SM current, at $d=2$ we require the representation for the photon $F_{\mu\nu}$, which is trivial. At $d=3$ we have the vector and axial-vector currents which match to
\begin{align}
u_\mu V_q^\mu &= 
\bigg[ F_1^{(q)}(0) \bigg] \bar{N}_u N_u 
+ {1\over m_N^2}  
\bigg\{
\bigg[ -\frac18 F_1^{(q)}(0) - m_N^2 F_1^{(q)\prime}(0) -\frac14 F_2^{(q)}(0) \bigg] 
\partial_\perp^2 \big( \bar{N}_u N_u \big) 
\nl
&\quad
+\bigg[ -\frac14 F_1^{(q)}(0)  -\frac12 F_2^{(q)}(0) \bigg]
i\bar{N}_u\partial_\perp^\mu \overleftarrow{\partial}_\perp^\nu \sigma_{\perp\mu\nu}N_u  
\bigg\}
+ \order(1/m_N^4) \,,
\nl
V^{\mu}_{q \perp} &= {1\over m_N} \bigg\{ 
\bigg[ \frac12 F_1^{(q)}(0) \bigg]
i \bar{N}_u \overleftrightarrow{\partial}_\perp^\mu N_u 
+ 
\bigg[ \frac12 F_1^{(q)}(0) + \frac12 F_2^{(q)}(0) \bigg] 
\partial_{\perp\nu} \big( \bar{N}_u \sigma_\perp^{\mu\nu} N_u \big)
\bigg\} 
+ \order(1/m_N^3)  \,,
\nl
u_\mu A_q^{\mu} &= {1\over m_N} \bigg\{ 
\bigg[ -\frac14 F_A^{(q)}(0) \bigg] i\epsilon^{\mu\nu\rho\sigma} u_\mu \bar{N}_u 
\overleftrightarrow{\partial}_{\perp\nu} \sigma_{\perp \rho\sigma} N_u 
\bigg\} + \order(1/m_N^3) \,,
\nl
A_{q\perp}^{\mu} &= \bigg[ -\frac12 F_A^{(q)}(0) \bigg]  \epsilon^{\mu\nu\rho\sigma} u_\nu \bar{N}_u\sigma_{\perp\rho\sigma} N_u 
\nl 
&\quad
+ {1\over m_N^2}\bigg\{ 
\bigg[ \frac18 F_A^{(q)}(0) + m_N^2 F_A^{(q)\prime}(0) \bigg] 
\epsilon^{\mu\nu\rho\sigma} u_\nu 
\bar{N}_u \overleftarrow{\partial}_{\perp}^\alpha \partial_{\perp\alpha} \sigma_{\perp\rho\sigma} N_u 
\nl
&\quad
+ \bigg[ -\frac{1}{16} F_A^{(q)}(0) + \frac12 m_N^2 F_A^{(q)\prime}(0) \bigg] 
\epsilon^{\mu\nu\rho\sigma} u_\nu 
\bar{N}_u \big( \overleftarrow{\partial}^2 +  \partial_{\perp}^2 \big) \sigma_{\perp\rho\sigma} N_u 
\nl
&\quad
+ \bigg[ -\frac18 F_{P^\prime}^{(q)}(0) \bigg] \epsilon_{\alpha\beta\gamma\delta} u^\gamma \bar{N}_u 
\big( \partial_\perp^\mu \partial_\perp^\alpha + \overleftarrow{\partial}_\perp^\mu \overleftarrow{\partial}_\perp^\alpha 
\big) \sigma_\perp^{\beta\delta} N_u 
\nl
&\quad 
+ \bigg[ -\frac18 F_A^{(q)}(0) -\frac18 F_{P^\prime}^{(q)}(0) \bigg] 
\epsilon_{\alpha\beta\gamma\delta} u^\gamma \bar{N}_u 
\big( \partial_\perp^\mu \overleftarrow{\partial}_\perp^\alpha + \overleftarrow{\partial}_\perp^\mu{\partial}_\perp^\alpha 
\big) \sigma_\perp^{\beta\delta} N_u 
\nl
&\quad
+ \bigg[ -\frac14 F_A^{(q)}(0) \bigg] 
i\epsilon^{\mu\nu\alpha\beta} u_\nu \bar{N}_u \partial_{\perp\alpha} \overleftarrow{\partial}_{\perp\beta} N_u 
\bigg\} + \order(1/m_N^4) \,,
\end{align}
where
we have expressed the matching coefficients (the quantities in square brackets) 
in terms of the form factors of the previous section, and decomposed the currents into components along and perpendicular to $u^\mu$. At $d=4$, we work through $\order(1/m_N)$, i.e., first derivative order, and have the antisymmetric tensor currents, the scalar and pseudoscalar operators, and the $C$-even and $C$-odd spin-two operators. Employing the notation in Table~\ref{tab:QCDoperators} and expressing results in terms of matrix elements of the previous section, the matching conditions are
\begin{align}
T_q^{\mu\nu} &= m_N \bigg[ 
 \left( m_q t_{q} \over m_N \right)  
\epsilon^{\alpha\beta\gamma[\mu} u^{\nu]} u_\alpha \bar{N} \sigma^\perp_{\beta\gamma} N 
 + \order( 1/m_N^2 ) \bigg] \,, 
 \nl
O_q^{(0)} &= m_N \bigg[ f_{q}^{(0)} \bar{N}_u N_u   + \order(1/m_N^2)  \bigg]\,, 
\nl
O_g^{(0)} &= m_N \bigg[ \left(- 8\pi \over 9\alpha_s \right) f_{g}^{(0)} \bar{N}_u N_u + \order(1/m_N^2) 
\bigg] \,, 
\nl
O_{5q,5g}^{(0)} &= \frac14 f_{5q,5g}^{(0)} 
\epsilon^{\mu\nu\rho\sigma} u_\mu \partial_{\perp\nu} ( \bar{N} \sigma^\perp_{\rho\sigma}N ) 
+ \order(1/m_N^2) \,,
\nl
u_\mu u_\nu O_{q,g}^{(2)\mu\nu} &= m_N \bigg[ \frac34 f_{q,g}^{(2)} \bar{N}_u N_u  + \order(1/m_N^2) \bigg]
\,,
\nl
O_{5q}^{(2) \mu \nu} &= m_N \bigg[ 
\frac12 f_{5q}^{(2)} \epsilon^{\alpha\beta\gamma\{\mu} u^{\nu\}} u_\alpha \bar{N}\sigma^\perp_{\beta\gamma}N 
+ \order(1/m_N^2) 
\bigg]
\,,
\end{align}
where the subscript label $N$ on form factors has been suppressed. 

\subsection{Nucleon effective theory for light mediators} 

The forgoing analysis, with additional matching onto multinucleon operators, 
provides a general framework for WIMP-nucleus scattering in the case where all new states in the dark sector 
have mass $\gg \Lambda_{\rm QCD}$, such that below this scale, a complete description is possible 
in terms of a systematic expansion of operators in 
$n_f=3$ flavor QCD.   Subsequent matching onto nucleon operators is given simply by 
evaluating the necessary form factors, whose low-$q^2$ behavior may be determined by lattice QCD, 
chiral perturbation theory or other nonperturbative methods.  

For completeness let us consider a more general situation allowing for light degrees
of freedom, with mass only assumed large compared to a typical WIMP-nucleon momentum transfer.%
\footnote{We are here also assuming that the considered momentum transfers are small enough that pions may be 
integrated out.}
We assume that all new states of the dark sector are integrated out, and consider the resulting
basis of operators in the one-nucleon sector. 
Specializing to the choice $v^\mu = u^\mu = (1,0,0,0)$, and neglecting electromagnetic interactions, 
the kinetic terms may be written, 
\begin{align} 
{\cal L}_N &= N^\dagger\bigg\{ i\partial_t + {\bm \partial^2 \over 2 m_N} + \dots \bigg\} N 
\,, 
\quad 
{\cal L}_\chi = \chi^\dagger\bigg\{ i\partial_t + {\bm \partial^2 \over 2 m_\chi} + \dots \bigg\} \chi \,,
\end{align}
where $N$ and $\chi$ denote the nonrelativistic nucleon and WIMP fields, respectively.
For interactions even under $P$ and $T$, we find through dimension eight the operators~\cite{Hill:2012rh,Brambilla:2008zg},
\begin{align}\label{eq:PT}
{\cal L}_{N\chi, PT} &= {1\over m_N^2} \bigg\{ 
{ d_1 } N^\dagger \sigma^i N  \ \chi^\dagger \sigma^i \chi
+ {d_2 } N^\dagger N \ \chi^\dagger \chi 
\bigg\} 
+ {1\over m_N^4} \bigg\{ 
{d_3} N^\dagger \partial_+^i N \ \chi^\dagger \partial_+^i \chi 
+ {d_4} N^\dagger \partial_-^i N \ \chi^\dagger \partial_-^i \chi 
\nl
&\quad
+ {d_5} N^\dagger (\bm{\partial}^2 + \overleftarrow{\bm{\partial}}^2 )N \ \chi^\dagger\chi 
+ {d_6} N^\dagger N \ \chi^\dagger (\bm{\partial}^2 + \overleftarrow{\bm{\partial}}^2 )\chi
+ {i d_8} \epsilon^{ijk} N^\dagger \sigma^i \partial_-^j N \ \chi^\dagger \partial_+^k \chi 
\nl
&\quad
+ {i d_9} \epsilon^{ijk} N^\dagger \sigma^i \partial_+^j N \ \chi^\dagger \partial_-^k \chi 
+ {i d_{11}} \epsilon^{ijk} N^\dagger \partial_+^k N \ \chi^\dagger \sigma^i \partial_-^j \chi 
+ {i d_{12}} \epsilon^{ijk} N^\dagger  \partial_-^k N \ \chi^\dagger \sigma^i  \partial_+^j \chi 
\nl
&\quad
+ {d_{13}} N^\dagger \sigma^i \partial_+^j N \ \chi^\dagger \sigma^i \partial_+^j \chi 
+ {d_{14}} N^\dagger \sigma^i \partial_-^j N \ \chi^\dagger \sigma^i{\partial}_-^j \chi 
+ {d_{15}} N^\dagger \bm{\sigma}\cdot \bm{\partial}_+ N \ \chi^\dagger \bm{\sigma}\cdot \bm{\partial}_+ \chi 
\nl
&\quad
+ {d_{16}} N^\dagger \bm{\sigma}\cdot \bm{\partial}_- N \ \chi^\dagger \bm{\sigma}\cdot \bm{\partial}_- \chi 
+ {d_{17}} N^\dagger \sigma^i \partial_-^j N \ \chi^\dagger \sigma^j \partial_-^i \chi 
\nl
&\quad
+ {d_{18}} N^\dagger \sigma^i ( \bm{\partial}^2 + \overleftarrow{\bm{\partial}}^2 ) N \ \chi^\dagger \sigma^i \chi
+ {d_{19}} N^\dagger \sigma^i ( \partial^i \partial^j + \overleftarrow{\partial}^j \overleftarrow{\partial}^i) N \ \chi^\dagger \sigma^j \chi 
\nl
&\quad
+ {d_{20}} N^\dagger \sigma^i N \ \chi^\dagger \sigma^i ( \bm{\partial}^2 + \overleftarrow{\bm{\partial}}^2 ) \chi
+ {d_{21}} N^\dagger \sigma^i N \ \chi^\dagger \sigma^j ( \partial^i \partial^j + \overleftarrow{\partial}^j \overleftarrow{\partial}^i) \chi  
\bigg\} 
+ \order(1/m_N^6) 
\,,
\end{align} 
where the naming scheme for Wilson coefficients is from Ref.~\cite{Hill:2012rh}. (Note in particular that
$d_i$ for $i=7,10$ are absent in (\ref{eq:PT}), since these operators are proportional to electromagnetic field strength.) 
Lorentz symmetry is imposed by enforcing invariance under the infinitesimal boost $\bm{\eta}$~\cite{Heinonen:2012km,Hill:2012rh}
\begin{align}\label{eq:Lorentz}
N &\to e^{i m_N \bm{\eta}\cdot\bm{x}} 
\bigg[ 1 - {i\bm{\eta}\cdot \bm{\partial} \over 2 m_N} 
+ {\bm{\sigma}\times \bm{\eta}\cdot \bm{\partial} \over 4 m_N} + \dots \bigg] N  \,,
\nl
\chi &\to e^{i m_\chi \bm{\eta}\cdot\bm{x}} 
\bigg[ 1 - {i\bm{\eta}\cdot \bm{\partial} \over 2 m_\chi} 
+ {\bm{\sigma}\times \bm{\eta}\cdot \bm{\partial} \over 4 m_\chi} + \dots \bigg] \chi  \,,
\nl
\partial_t &\to \partial_t - \bm{\eta}\cdot\bm{\partial} \,, \quad
\bm{\partial} \to \bm{\partial} -\bm{\eta} \partial_t \,. 
\end{align}
This implies the constraints, 
\begin{align}\label{eq:fourrelations}
  rd_4 +d_5  &= {d_2\over 4} \, , \quad  
d_5  = r^2 d_6 \,, \quad
8r(d_8 + rd_9) = -rd_2 + d_1  \,, \quad 
8r ( rd_{11}+d_{12} ) = -d_2  + rd_1 \,, 
\nl
rd_{14} + d_{18} &= {d_1\over 4} \, , \quad  
d_{18}  = r^2 d_{20}  \,, \quad 
2rd_{16} + d_{19} = {d_1\over 4} \,, \quad 
r(d_{16}+d_{17})+d_{19}=0 \,, \quad
d_{19} = r^2 d_{21} \,,
\end{align}
where $r=m_\chi/m_N$. With these constraints in place there are ten independent $P$ and $T$ conserving 
four-fermion operators through dimension eight, including two operators at dimension six. 

Operators even under $T$ but odd under $P$ are 
\begin{align}
{\cal L}_{N\chi, \slash{P}} &= {1\over m_N^3} \bigg\{ 
d_1^\prime i N^\dagger \bm{\sigma}\cdot\bm{\partial}_- N \chi^\dagger\chi 
+ d_2^\prime i N^\dagger \sigma^i N \chi^\dagger \partial_-^i \chi 
+ d_3^\prime i N^\dagger \partial_-^i N \chi^\dagger \sigma^i \chi 
+ d_4^\prime i N^\dagger N \chi^\dagger \bm{\sigma}\cdot \bm{\partial}_- \chi
\nl
&\quad
+ 
d_5^\prime \epsilon^{ijk} N^\dagger \sigma^i \partial_+^j N \chi^\dagger \sigma^k \chi 
\bigg\}  + \order(1/m_N^5) \,.
\end{align}
Relativistic invariance enforces the constraints
\be
d_1^\prime + r d_2^\prime = d_3^\prime + r d_4^\prime = 0 \,,
\ee
leaving three independent operators. 
Operators odd under both $P$ and $T$ are 
\begin{align} 
{\cal L}_{N\chi, \slash{P}\slash{T}} &= 
{1\over m_N^3} \bigg\{ 
f_1^\prime N^\dagger \bm{\sigma} \cdot \bm{\partial}_+ N \chi^\dagger \chi 
+ f_2^\prime N^\dagger N \chi^\dagger \bm{\sigma} \cdot \bm{\partial}_+ \chi 
+ f_3^\prime i \epsilon^{ijk} N^\dagger \sigma^i \partial_-^j N \chi^\dagger \sigma^k \chi 
\nl
&\quad
+ f_4^\prime i\epsilon^{ijk} N^\dagger \sigma^i N \chi^\dagger \sigma^j \partial_-^k \chi 
\bigg\} + \order(1/m_N^5) \,.
\end{align}
Relativistic invariance enforces the constraints
\be
f_3^\prime = r f_4^\prime \,,
\ee
leaving three independent operators. 
Operators even under $P$ and odd under $T$ are 
\begin{align}
{\cal L}_{N \chi, \slash{T}} &= {1\over m_N^4} \bigg\{ 
i f_1  N^\dagger \bm{\partial}_+\cdot \bm{\partial}_- N \chi^\dagger\chi 
+ i f_2  N^\dagger N \chi^\dagger \bm{\partial}_+ \cdot  \bm{\partial}_- \chi
+ f_3 \epsilon^{ijk} N^\dagger \sigma^i \partial_-^j N \chi^\dagger \partial_-^k\chi  
\nl
&\quad
+ f_4 \epsilon^{ijk} N^\dagger \partial_-^i N \chi^\dagger \sigma^j \partial_-^k \chi 
+ i f_5 N^\dagger \bm{\partial}_+\cdot \bm{\partial}_-  \sigma^i N \chi^\dagger \sigma^i \chi 
+i f_{6} N^\dagger \bm{\sigma} \cdot \bm{\partial}_+ \partial_-^i N \chi^\dagger \sigma^i \chi
\nl
&\quad
+ i f_7 N^\dagger   \bm{\sigma} \cdot \bm{\partial}_- N \chi^\dagger \bm{\sigma} \cdot  \bm{\partial}_+ \chi 
+ i f_8  N^\dagger \sigma^i  N \chi^\dagger \sigma^i \bm{\partial}_+ \cdot \bm{\partial}_- \chi 
+ i f_9 N^\dagger \sigma^i  N \chi^\dagger \bm{\sigma} \cdot \bm{\partial}_+ \partial_-^i \chi 
\nl
&\quad
+ i f_{10} N^\dagger  \bm{\sigma}  \cdot \bm{\partial}_+ N \chi^\dagger \bm{\sigma} \cdot \bm{\partial}_- \chi 
\bigg\} 
+ \order(1/m_N^6)  \,. 
\end{align}
Relativistic invariance enforces the constraints
\be 
f_1 + r f_2 = f_5 + r f_8 = f_7 + r f_9 = f_6 + r f_{10}= f_3 = f_4 = 0 \,
\ee
leaving four independent operators.

\subsubsection{Lorentz versus Galilean invariance \label{sec:lorentz}} 

We remark that the basis of operators in Eq.~(\ref{eq:PT}) under the 
constraints in (\ref{eq:fourrelations}) is Lorentz invariant.  
If in place of the transformations in (\ref{eq:Lorentz}) we instead enforced Galilean symmetry~\cite{Fitzpatrick:2012ix}, 
defined by 
\begin{align}\label{eq:galilean}
N &\to e^{i m_N \bm{\eta}\cdot\bm{x}} N \,,  \quad
\chi \to e^{i m_\chi \bm{\eta}\cdot\bm{x}} \chi \,, \quad
\partial_t \to \partial_t - \bm{\eta}\cdot\bm{\partial} \,, \quad
\bm{\partial} \to \bm{\partial} \,, 
\end{align} 
we would obtain constraints on dimension eight operators different from (\ref{eq:fourrelations}).\footnote{Galilean constraints would be given 
by the formal limit $d_1=d_2=0$ in (\ref{eq:fourrelations}).}
These constraints would imply that all Hermitian operators are constructed from 
the combinations of derivatives corresponding to  
\be
\bm{v}_{\rm rel} \equiv \frac12 \bigg[ {\bm{p} + \bm{p}^\prime \over m_N} 
- {\bm{k} + \bm{k}^\prime \over m_\chi} \bigg] \,, \quad
\bm{q} \equiv \bm{p}^\prime - \bm{p} = \bm{k} - \bm{k}^\prime \,,
\ee
where $p$ and $k$ ($p^\prime$ and $k^\prime$) are the incoming (outgoing) momenta
of $N$ and $\chi$ respectively.    
In particular, 
the violation of Lorentz symmetry obtained by using (\ref{eq:galilean}) in 
place of (\ref{eq:Lorentz}) 
would manifest itself as the absence of 
operators coupling to total momentum $\bm{P}$, 
\be
\bm{P}  \equiv \bm{p} + \bm{k} = \bm{p}^\prime + \bm{k}^\prime \,. 
\ee
Note that Lorentz symmetry links a leading order nucleon spin-dependent operator 
($d_1$) to subleading nucleon spin-independent operators.   
The phenomenological impact of such terms remains to be investigated.  
Note that Lorentz symmetry cannot be obtained by enforcing additional 
constraints on operators present in the Galilean invariant theory.   

\section{Phenomenological illustrations \label{sec:pheno}} 

The forgoing analysis provides a framework to systematically evolve
coefficients defined at the weak scale, to obtain the effective low-energy theory where nuclear matrix elements are evaluated.   
As illustration we focus attention on two cases: firstly the specification of contact interactions at or 
above the weak scale, and secondly, the specification of the complete basis of 
coefficients at the weak scale by the leading order of the heavy WIMP expansion. 

\subsection{Contact interactions \label{sec:contact}} 

\begin{figure}[top]
\centering
\includegraphics[scale=0.5]{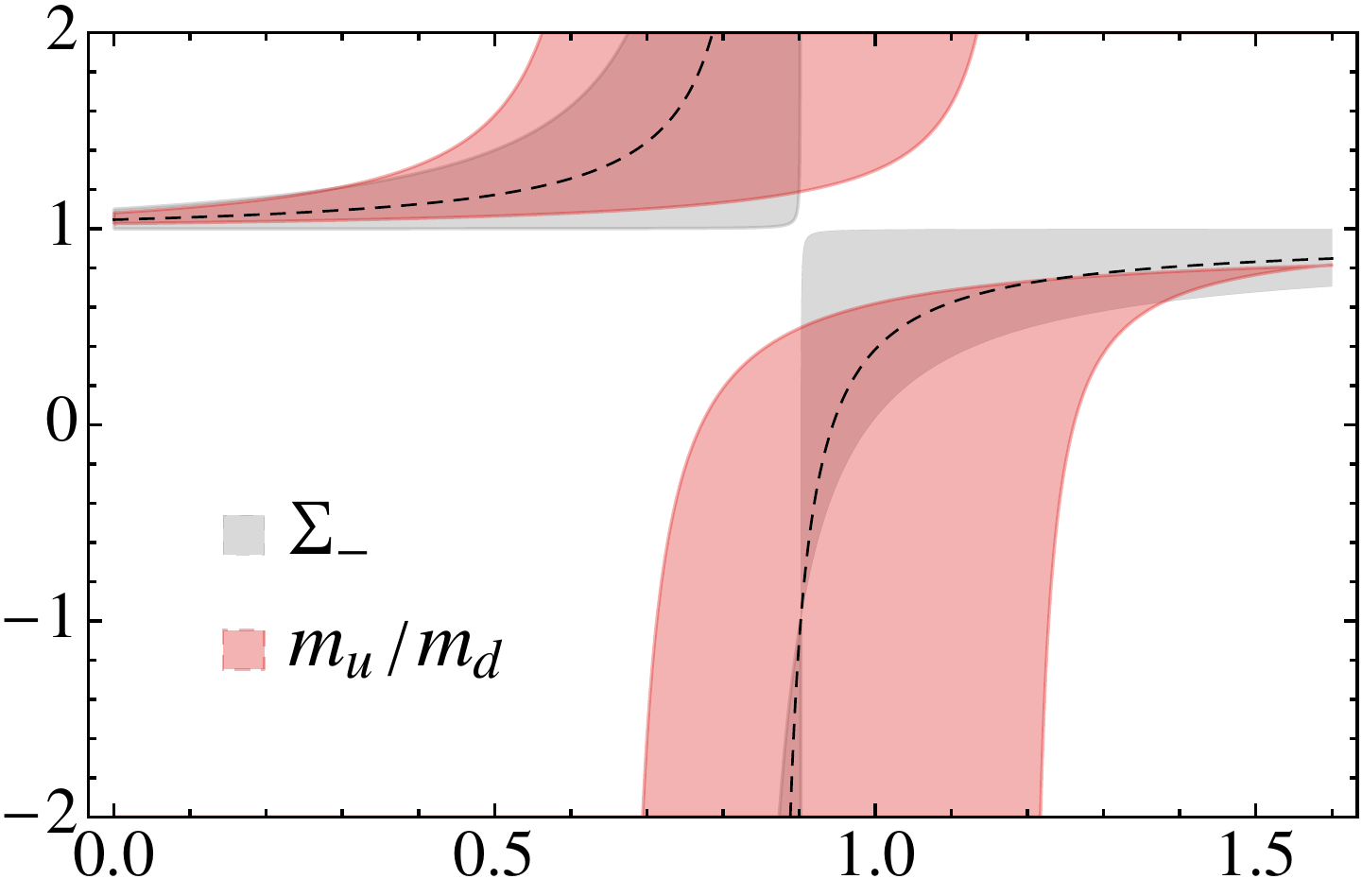}
\hspace*{10mm}\includegraphics[scale=0.5]{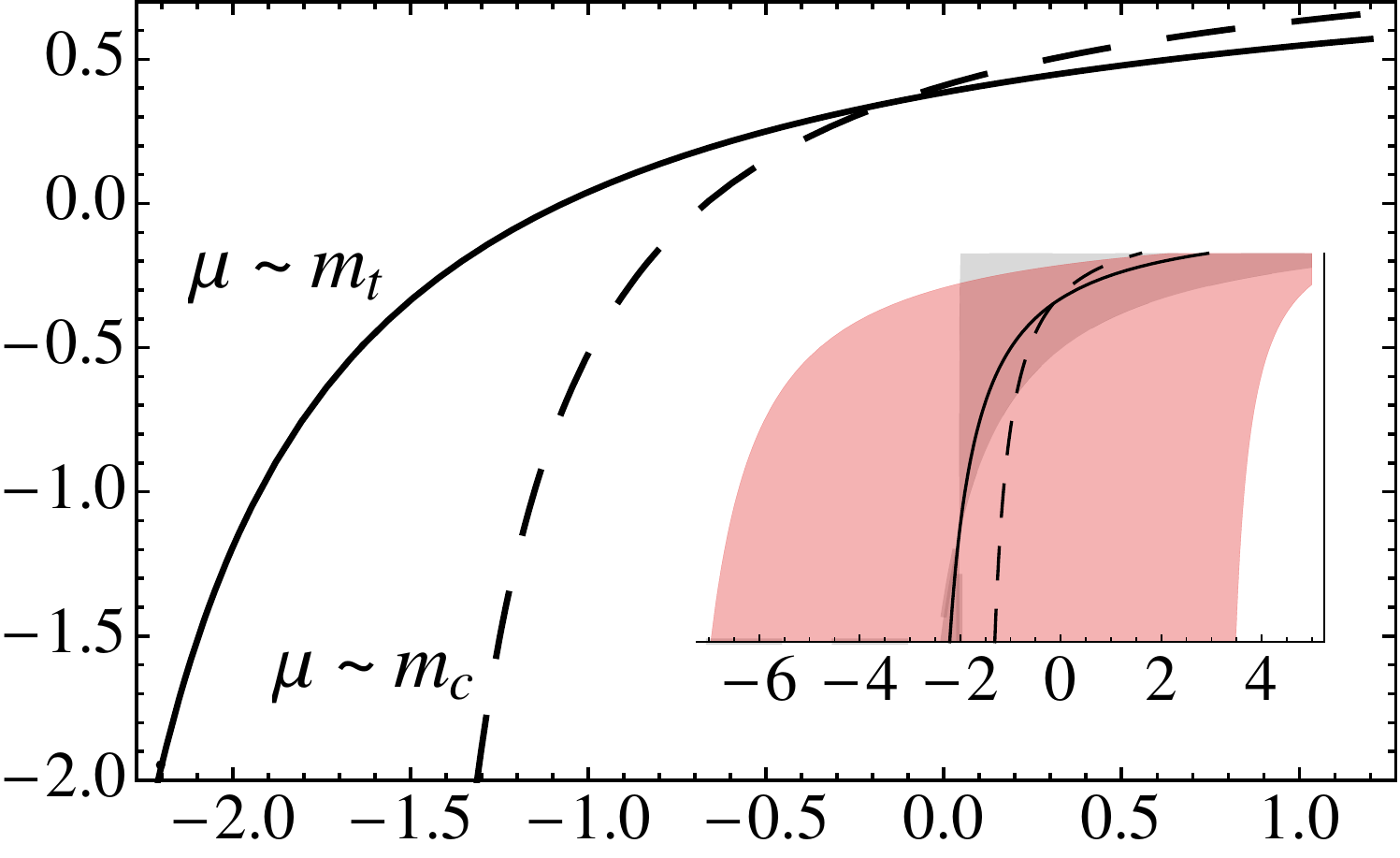}
\begin{minipage}{0cm}
\vspace*{1.3cm}\hspace*{-3.6cm}$b_g$\hspace*{-9.5cm}$-b_u/b_d$
\vspace*{0.5cm}
\end{minipage}
\begin{minipage}{0cm}
\vspace*{-5.05cm}\hspace*{-8.4cm}\rotatebox{90}{$f_n/f_p$}
\end{minipage}
\begin{minipage}{0cm}
\vspace*{-5cm}\hspace*{-16.9cm}\rotatebox{90}{$f_n/f_p$}
\end{minipage}
\caption{\label{fig:contact}
The ratio $f_n/f_p$ of the effective WIMP-neutron ($f_n$) and
WIMP-proton ($f_p$) couplings in terms of the
parameters $b_i$ in Eq.~(\ref{eq:IVDM}). For $b_g=0$ (left panel), $f_n/f_p$ is independent of $\Lambda$ and depends on only the ratio $b_u/b_d$. The uncertainty bands are from variation of the matrix element $\Sigma_-$ (gray) and the ratio $R_{ud}=m_u/m_d$ (red), with ranges given in (\ref{eq:sigmas}) and~(\ref{eq:massratios}). 
We illustrate the effect of non-zero $b_g$ in the right panel, with $b_d=-b_u=0.01$ and $\Lambda = 400 \, {\rm GeV}$. The solid (dashed) line is the
prediction assuming that the coefficients $b_i$ are defined at a high
(low) scale $\mu \sim m_t$ ($\mu \sim m_c$). The inset shows the
curves over the same vertical range, including uncertainty bands for
the solid line from variation of $\Sigma_-$ (gray) and $R_{ud}$
(red). In both cases the variation from $\Sigma_{\pi N}$ is subdominant.
}
\end{figure}

Consider the contact interactions between a Majorana fermion WIMP and
SM fields given in Eq.~(\ref{eq:fermionL}).  As a simple illustration,
let us focus on the set of operators
\be\label{eq:IVDM} {\cal L}_{\chi,SM} =  {1\over \Lambda^2}
\bar{\chi} \chi \bigg[ b_u \bar{u} u + b_d \bar{d} d + {b_g \over
\Lambda} (G^a_{\mu\nu})^2 \bigg]  \,, \ee
where coefficients $b_{u,d,g}$ may be constrained by collider
production bounds~\cite{Goodman:2010ku} or engineered to produce a
desired WIMP-nucleus scattering phenomenology~\cite{Feng:2011vu}.  An
observable of interest for the latter is the ratio $f_n/f_p$ of the
effective spin-independent WIMP-neutron ($f_n$) and WIMP-proton ($f_p$) couplings.%
\footnote{
In terms of the couplings in (\ref{eq:PT}), $f_{p}$ and $f_{n}$ are proportional to $d_2^{(p)}$ and $d_2^{(n)}$, 
respectively.   
} 

We show in Fig.~\ref{fig:contact} predictions for $f_n/f_p$ from the
model in Eq.~(\ref{eq:IVDM}), highlighting large effects from 
hadronic matrix element uncertainties and the choice of 
QCD renormalization scale. 
The left panel illustrates uncertainties from varying
the SM quantities $\Sigma_-$ and $R_{ud} = m_u/m_d$ given in~(\ref{eq:sigmas}) and~(\ref{eq:massratios}).%
\footnote{The point $-b_u/b_d = 1.08$ was highlighted in \cite{Feng:2011vu}.  Hadronic 
uncertainties are severe at this point.}
The right panel illustrates the uncertainty from not specifying the renormalization
scale at which the coefficients $b_i$ are defined. Meaningful
predictions for $f_n/f_p$ require both a precise knowledge of
hadronic inputs and a careful treatment of renormalization effects.
Similar considerations apply to other applications that relate constraints 
on contact interactions at the electroweak scale to low energy observables such as direct detection cross sections or annihilation rates for low mass WIMPs. 

\subsection{Heavy, electroweak-charged WIMPs}

We consider the heavy WIMP limit ($M \gg m_W$) for the cases of a self-conjugate 
electroweak triplet of hypercharge zero (``wino-like"), and an 
electroweak doublet of hypercharge $1/2$ (``higgsino-like''). 
For the latter, we assume mass perturbations that cause the mass eigenstates 
after EWSB to be self-conjugate combinations, thus forbidding a 
phenomenologically disfavored tree-level vector coupling 
between the lightest electrically neutral state and $Z^0$
(see Section 4 of Ref.~\cite{Hill:2014yka} for details).
The bare effective lagrangian at the weak-scale describing interactions 
of the lightest electrically neutral self-conjugate WIMP (of arbitrary spin) 
with low-energy SM degrees of freedom is given by
\begin{align}\label{eq:LWIMPSM}
{\cal L}_{\chi_v, \, {\rm SM}} &= 
 \bar{\chi}_v\chi_v  
\bigg\{ 
\sum_{q=u,d,s,c,b} \bigg[ 
c_{q}^{(0)} O_{q}^{(0)}   + c_{q}^{(2)} v_\mu v_\nu O_{q}^{(2)\mu\nu} 
\bigg]
+ c_{g}^{(0)} O_{g}^{(0)} + c_{g}^{(2)} v_\mu v_\nu O_{g}^{(2)\mu\nu} 
\bigg\}  + \dots \,,
\end{align}
where the scalar and $C$-even spin-two operators, $O^{(0)}_{q,g}$ and $O^{(2)}_{q,g}$, are given in Table~\ref{tab:QCDoperators}, and the coefficients are defined to include the mass suppression $1/m_W^3$. The bare matching coefficients for both wino-like and higgsino-like cases were computed explicitly in Ref.~\cite{Hill:2014yka} and are reproduced here for completeness:%
\footnote{
Spin-$0$ results were also obtained in \cite{Hisano:2011cs}. 
}
\begin{align}\label{eq:purebare}
c_{U}^{(0)} &= {\pi \Gamma(1+\epsilon) g_2^4 \over (4\pi)^{2-\epsilon} } \Bigg\{ 
- {m_W^{-3-2\epsilon}\over 2 x_h^2}\left[ {\cal C}_W + {{\cal C}_Z \over c_W^3}\right]  
+  {m_Z^{-3-2\epsilon} {\cal C}_Z \over 8 c_W^4} \big[ c_V^{(U)2} - c_A^{(U)2} \big]
+ \order(\epsilon) 
\Bigg\} 
\,,
\nl
c_{D}^{(0)} &= {\pi \Gamma(1+\epsilon) g_2^4 \over (4\pi)^{2-\epsilon} } \Bigg\{ 
- {m_W^{-3-2\epsilon}\over 2 x_h^2}\left[ {\cal C}_W + {{\cal C}_Z \over c_W^3}\right]  
+  {m_Z^{-3-2\epsilon} {\cal C}_Z \over 8 c_W^4} \big[ c_V^{(D)2} - c_A^{(D)2} \big]
\nl
&\quad
-  \delta_{Db} \, m_W^{-3-2\epsilon} {\cal C}_W    {x_t \over 8(x_t+1)^3}
+ \order(\epsilon) 
\Bigg\} 
\,,
\nl
c_{g}^{(0)} &= {\pi [\Gamma(1+\epsilon)]^2 g_2^4 g^2 \over (4\pi)^{4-2\epsilon}}\Bigg\{
{m_W^{-3-4\epsilon} \over 2} \Bigg[ 
{1\over 3 x_h^2}\left[ {\cal C}_W + { {\cal C}_Z \over  c_W^3}\right]   
+ {\cal C}_W \left[ \frac13 + {1 \over 6 (x_t+1)^2 }   \right]
\Bigg] 
\nl
&\quad
+  {m_Z^{-3-4\epsilon} {\cal C}_Z\over 64 c_W^4}
\Bigg[
 4 \big[ c_V^{(D)2} + c_A^{(D)2} \big] 
 + \big[ c_V^{(U)2} + c_A^{(U)2} \big]
\bigg[ \frac83
+  {32 y_t^6 (8 y_t^2 -7) \over (4y_t^2-1)^{7/2} }\arctan\big(\sqrt{\smash[b]{4y_t^2-1}} \big) 
\nl
&\quad
- \pi y_t 
+{4(48y_t^6-2y_t^4+9y_t^2-1) \over 3(4y_t^2-1)^3} 
\bigg] 
+ \big[ c_V^{(U)2} - c_A^{(U)2} \big]
\bigg[ 3\pi y_t 
- {4(144y_t^6 -70y_t^4+9y_t^2-2)\over 3(4y_t^2-1)^3} 
\nl
&\quad
- {32 y_t^4(24y_t^4-21 y_t^2+5) \over (4y_t^2-1)^{7/2} }\arctan\big(\sqrt{\smash[b]{4y_t^2-1}} \big)
\bigg] 
\Bigg]
+ \order(\epsilon) 
\Bigg\} \, ,
\nl
c_{U}^{(2)} &= {\pi \Gamma(1+\epsilon)  g_2^4 \over (4\pi)^{2-\epsilon} }  \Bigg\{
\bigg[ m_W^{-3-2\epsilon} {\cal C}_W + { m_Z^{-3-2\epsilon} {\cal C}_Z \over 2 c_W^4} 
\big[ c_V^{(U)2}+c_A^{(U)2} \big]  \bigg]  \bigg[ 
{1\over 3} + \left( {11\over 9} -{2\over 3}\log 2  \right) \epsilon 
\bigg]+ \order(\epsilon^2) 
\Bigg\} \, ,
\nl
c_{D}^{(2)} &= {\pi \Gamma(1+\epsilon)  g_2^4 \over (4\pi)^{2-\epsilon} }  \Bigg\{
\bigg[ m_W^{-3-2\epsilon} {\cal C}_W + { m_Z^{-3-2\epsilon} {\cal C}_Z \over 2 c_W^4} 
\big[ c_V^{(D)2}+c_A^{(D)2} \big]  \bigg]  \bigg[ 
{1\over 3} + \left( {11\over 9} -{2\over 3}\log 2  \right) \epsilon 
\bigg]
\nl
&\quad
+ \delta_{Db} \, { m_W^{-3-2\epsilon} {\cal C}_W \over 2}  
  \bigg[ {(3x_t +2) \over 3(x_t+1)^3} - \frac23
+ \bigg(
{ 2 x_t(7x_t^2 -3) \over 3(x_t^2-1)^3 }\log x_t 
- {2(3x_t+2)\over 3(x_t+1)^3}\log 2
\nl
&\quad 
- {2( 25x_t^2-2x_t -11) \over 9(x_t^2-1)^2(x_t+1) } 
- {22 \over 9} + {4 \over 3}\log 2
\bigg)\epsilon \bigg]  
+ \order(\epsilon^2) 
\Bigg\} \, ,
\nl
c_g^{(2)} &= 
{\pi [\Gamma(1+\epsilon)]^2 g_2^4 g^2 \over (4\pi)^{4-2\epsilon}}\Bigg\{
{m_W^{-3-4\epsilon} {\cal C}_W \over 2} 
\bigg[
 -{16 \over 9 \epsilon} -{284 \over 27} +{32 \over 9} \log2
- {2 (3x_t + 2)\over 9(x_t+1)^3} {1\over \epsilon} 
\nl
&\quad 
+ {8 ( 6 x_t^8 - 18 x_t^6 + 21 x_t^4 - 3x_t^2 -2 )\over 9 (x_t^2-1)^3}\log(x_t+1) 
+ {4 (3 x_t^4 -21 x_t^3 + 3 x_t^2 + 9x_t - 2)\over 9(x_t^2-1)^3}\log 2 
\nl
&\quad
- {4 ( 12 x_t^8 - 36 x_t^6 + 39 x_t^4 + 14 x_t^3- 9 x_t^2 - 6x_t - 2 )\over 9(x_t^2-1)^3}\log x_t 
\nl
&\quad
- {144 x_t^6 + 72 x_t^5 -312 x_t^4 - 105 x_t^3 - 40 x_t^2 + 47 x_t + 98 \over 27 (x_t^2-1)^2(x_t+1)} 
\bigg]
\nl
&\quad 
+  {m_Z^{-3-4\epsilon} {\cal C}_Z \over 64 c_W^4}
\Bigg[ 
\bigg[ 8 
\big[  c_V^{(U)2} + c_A^{(U)2} \big]
+ 12 \big[ c_V^{(D)2} + c_A^{(D)2} \big]
\bigg] 
\bigg[
-{16\over 9\epsilon} - {284\over 27} + {32\over 9}\log{2} 
\bigg]
\nl
&\quad 
+ \big[ c_V^{(U)2}+c_A^{(U)2}\big]  \bigg[ 
{ 128 ( 24 y_t^8 -21 y_t^6 -4y_t^4 + 5y_t^2 -1  )  \over 9 (4y_t^2-1)^{7/2} } \arctan\big(\sqrt{\smash[b]{4y_t^2-1}} \big)
- {4 \pi y_t \over 3}
\nl
&\quad 
+ {16 ( 48 y_t^6 +62 y_t^4 -47 y_t^2 +9 )\over 9(4y_t^2-1)^3 } 
 \bigg]
 + \big[ c_V^{(U)2}-c_A^{(U)2} \big] \bigg[ 
{16 y_t^2 ( 624 y_t^4 - 538 y_t^2 +103 ) \over 9 (4y_t^2-1)^3} 
- {52 \pi y_t \over 3} 
\nl
&\quad 
+{ 128 y_t^2 (104 y_t^6 -91 y_t^4 + 35 y_t^2 - 5) \over 3 (4y_t^2-1)^{7/2} }
\arctan\big(\sqrt{\smash[b]{4y_t^2-1}} \big)
\bigg] 
\Bigg]
+ \order(\epsilon) 
\Bigg\} \,,
\end{align}
where $x_i= m_i/m_W$, $y_i=m_i/m_Z$ and
\begin{align}\label{eq:quarkscvca}
c_V^{(U)} = 1-\frac83 s_W^2 \,, \quad
c_V^{(D)} = -1 + \frac43 s_W^2 \,, \quad
c_A^{(U)} = -1 \,, \quad
c_A^{(D)} = 1 \,.
\end{align}
We denote generic up- and down-type quarks by $U$ and $D$, respectively, and 
the Kronecker delta, $\delta_{Db}$, is equal to unity for $D=b$, and vanishes for $D=d,s$.
We have used CKM unitarity, $\sum_{D} |V_{UD}|^2=1$, to 
simplify the results; in practice it is sufficient to set $V_{tb}= 1$ for the numerical analysis.   
Beyond the specification of the WIMP electroweak quantum numbers $J$ and $Y$ through the constants
\be\label{eq:WZcoeff}
{\cal C}_W = [ J(J+1) - Y^2 ]\,,\quad {\cal C}_Z = Y^2,
\ee
the matching coefficients are completely given by SM parameters in the heavy WIMP limit.
The wino-like and higgsino-like results are obtained by setting ${\cal C}_W = 2\, , {\cal C}_Z =0$ and ${\cal C}_W = 1/2\, , {\cal C}_Z =1/4$, respectively.

Let us now consider the evolution down to low energies for these weak scale coefficients, and the subsequent evaluation of hadronic matrix elements to obtain the benchmark low-velocity single-nucleon scattering cross section. 

\subsubsection{Coefficient renormalization}\label{sec:renpure}

Let us employ $Z^{(0)}$ 
and $Z^{(2)}$ through $\order(\alpha_s)$ given in Table~\ref{tab:QCDZ} to derive the relation between bare and renormalized coefficients at first 
non-vanishing order. From the definition in (\ref{eq:Oren}),
the renormalized coefficients for the scalar operators are 
\begin{align} \label{eq:weakmatch0}
c_q^{(0)}(\mu) &= \sum_{q^\prime} Z^{(0)}_{q^\prime q}(\mu) c_{q^\prime}^{(0){\rm bare}} + Z^{(0)}_{gq}(\mu) c_g^{(0){\rm bare}} 
= c_q^{(0){\rm bare}} + \order(\alpha_s^2) \,,
\nl
c_g^{(0)}(\mu) &= \sum_{q^\prime} Z^{(0)}_{q^\prime g}(\mu) c_{q^\prime}^{(0){\rm bare}} + Z^{(0)}_{gg}(\mu) c_g^{(0){\rm bare}} 
= c_g^{(0){\rm bare}} + \order(\alpha_s^2) \,,
\end{align} 
while for the $C$-even spin-two operators, we find
\begin{align} \label{eq:weakmatch2}
c_q^{(2)}(\mu) &= \sum_{q^\prime} Z^{(2)}_{q^\prime q}(\mu) c_{q^\prime}^{(2){\rm bare}} 
+ Z^{(2)}_{gq}(\mu) c_g^{(2){\rm bare}} = c_q^{(2){\rm bare}} + \order(\alpha_s) \,,
\nl
c_g^{(2)}(\mu) &= \sum_{q^\prime} Z^{(2)}_{q^\prime g}(\mu) c_{q^\prime}^{(2){\rm bare}} + Z^{(2)}_{gg}(\mu) c_g^{(2){\rm bare}} 
= \sum_q {1\over \epsilon}{\alpha_s \over 6\pi}  c_q^{(2){\rm bare}} + c_g^{(2){\rm bare}} 
+ \order(\alpha_s^2) \,.
\end{align}
In particular, a nontrivial subtraction requiring
the $\order(\epsilon)$ part of the coefficients $c_{q}^{(2) {\rm bare}}$ is necessary to obtain the renormalized coefficient $c_g^{(2)}(\mu)$. Employing (\ref{eq:weakmatch0}) and~(\ref{eq:weakmatch2}), we find the renormalized coefficients
\begin{align}\label{eq:pureren}
c_{U}^{(0)}(\mu) &= {\pi \alpha_2^2 \over m_W^3 } \Bigg\{ 
- {1 \over 2 x_h^2}\left[ {\cal C}_W + {{\cal C}_Z \over c_W^3}\right]  
+  { {\cal C}_Z \over 8 c_W} \big[ c_V^{(U)2} - c_A^{(U)2} \big]
\Bigg\} 
\,,
\nl
c_{D}^{(0)}(\mu) &= {\pi \alpha_2^2 \over m_W^3 } \Bigg\{ 
- {1\over 2 x_h^2}\left[ {\cal C}_W + {{\cal C}_Z \over c_W^3}\right]  
+  { {\cal C}_Z \over 8 c_W} \big[ c_V^{(D)2} - c_A^{(D)2} \big]
-  \delta_{Db} \, {\cal C}_W    {x_t \over 8(x_t+1)^3}
\Bigg\} 
\,,
\nl
c_{g}^{(0)}(\mu) &= {\pi \alpha_2^2  \over m_W^3} { \alpha_s(\mu) \over 4 \pi }\Bigg\{
{1 \over 2} \Bigg[ 
{1\over 3 x_h^2}\left[ {\cal C}_W + { {\cal C}_Z \over  c_W^3}\right]   
+ {\cal C}_W \left[ \frac13 + {1 \over 6 (x_t+1)^2 }   \right]
\Bigg] 
\nl
&\quad
+  {{\cal C}_Z\over 64 c_W}
\Bigg[
 4 \big[ c_V^{(D)2} + c_A^{(D)2} \big] 
 + \big[ c_V^{(U)2} + c_A^{(U)2} \big]
\bigg[ \frac83
+  {32 y_t^6 (8 y_t^2 -7) \over (4y_t^2-1)^{7/2} }\arctan\big(\sqrt{\smash[b]{4y_t^2-1}} \big) 
\nl
&\quad
- \pi y_t 
+{4(48y_t^6-2y_t^4+9y_t^2-1) \over 3(4y_t^2-1)^3} 
\bigg] 
+ \big[ c_V^{(U)2} - c_A^{(U)2} \big]
\bigg[ 3\pi y_t 
- {4(144y_t^6 -70y_t^4+9y_t^2-2)\over 3(4y_t^2-1)^3} 
\nl
&\quad
- {32 y_t^4(24y_t^4-21 y_t^2+5) \over (4y_t^2-1)^{7/2} }\arctan\big(\sqrt{\smash[b]{4y_t^2-1}} \big)
\bigg] 
\Bigg]
\Bigg\} \, ,
\nl
c_{U}^{(2)}(\mu) &= {\pi \alpha_2^2 \over m_W^3 }  \Bigg\{
  {{\cal C}_W \over 3} + {  {\cal C}_Z \over 6 c_W} 
\big[ c_V^{(U)2}+c_A^{(U)2} \big]
\Bigg\} \, ,
\nl
c_{D}^{(2)}(\mu) &= {\pi \alpha_2^2 \over m_W^3 }  \Bigg\{
{ {\cal C}_W \over 3} + {  {\cal C}_Z \over 6 c_W} 
\big[ c_V^{(D)2}+c_A^{(D)2} \big]  
+ \delta_{Db} \, { {\cal C}_W \over 2}  
  \bigg[ {(3x_t +2) \over 3(x_t+1)^3} - \frac23
 \bigg]  
\Bigg\} \, ,
\nl
c_g^{(2)}(\mu) &= 
{\pi \alpha_2^2 \over m_W^3} { \alpha_s(\mu) \over 4 \pi }\Bigg\{  {\cal C}_W \Bigg[  -{ 2(8x_t^3 + 24x_t^2 + 27x_t +10) \over 9(x_t +1)^3 }\log {\mu \over m_W} - { 4x_t (7x_t^2 -3 )\over 9 (x_t^2 - 1)^3 } \log 2 
\nl
&\quad
- { 2(12x_t^5 -36x_t^4 +36x_t^3 -12x_t^2 +3x_t -2) \over 9 (x_t - 1)^3 } \log x_t 
\nl
&\quad
+ { 4( 6x_t^8 -18x_t^6 +21 x_t^4 -3x_t^2 -2 )  \over 9(x_t^2 - 1)^3  } \log(x_t+1)
\nl
&\quad
- { 48x_t^6 +60x_t^5 - 68x_t^4 -107 x_t^3 -52x_t^2 + 49x_t + 54 \over 18 (x_t^2-1)^2 (x_t +1) } \Bigg]
\nl
&\quad
+{ {\cal C}_Z \over c_W} \Bigg[  \bigg[  
  2 \big[ c_V^{(U)2}+c_A^{(U)2} \big] 
+ 3 \big[ c_V^{(D)2}+c_A^{(D)2} \big]
\bigg]  \bigg[ - \frac14 - \frac29 \log {\mu \over m_Z}  \bigg]
+ \big[ c_V^{(U)2}+c_A^{(U)2} \big] \bigg[ - { \pi y_t \over 48}  
\nl
&\quad
+ { 2 ( 24y_t^8 -21y_t^6 -4y_t^4 + 5y_t^2-1 ) \over 9  (4y_t^2-1)^{7/2} }\arctan\big(\sqrt{\smash[b]{4y_t^2-1}}\, \big)
+ { 48 y_t^6 +62 y_t^4 -47 y_t^2 + 9 \over 36 (4y_t^2-1)^3 }
\bigg]
\nl
&\quad
+ \big[ c_V^{(U)2}-c_A^{(U)2} \big] \bigg[ - { 13 \pi y_t \over 48}  
+ { 2 y_t^2 ( 104 y_t^6 - 91 y_t^4 + 35y_t^2-5) \over 3  (4y_t^2-1)^{7/2} }\arctan\big(\sqrt{\smash[b]{4y_t^2-1}}\, \big)
\nl
&\quad
+ { y_t^2 (624 y_t^4 - 538 y_t^2 + 103) \over 36 (4y_t^2-1)^3 }
\bigg]
\Bigg]
\Bigg\} \, .
\end{align}
We proceed to study the evolution of these coefficients down to low-energies.

\subsubsection{Coefficient evolution}

Let us illustrate the evolution of scalar and $C$-even spin-two coefficients from
a high scale down to a low scale, employing the solutions for RG running and threshold matching, $R$ and $M$, given in Tables~\ref{tab:QCDrunning} and~\ref{tab:HQmatching}. For definiteness, we consider the high scale
coefficients given by the renormalized coefficients in~(\ref{eq:pureren}) for an electroweak triplet (i.e., a pure
wino). Results for an electroweak doublet (i.e., a pure Higgsino) are
qualitatively similar. For illustration, we choose the default scale values $\mu_t = (m_t + m_W)/2 \approx 
126 \ {\rm GeV}$, $\mu_b = 4.75 \ {\rm GeV}$, $\mu_c = 1.4 \ {\rm
GeV}$ and $\mu_0 = 1.2 \ {\rm GeV}$. 

The results for scalar coefficients presented in the left panel of
Table~\ref{tab:coefficientsEVO} employ $R^{(0)}$ and
$M^{(0)}$ at NLO. The high-scale gluon coefficient is small,
having a factor of $\alpha_s(\mu_t)$, but increases at lower scales
due to running and heavy quark threshold effects. A large nucleon matrix element for the gluon makes it a dominant contribution to the scattering cross section. 

In the present example, mixing between the scalar quark
and gluon operators shift the quark coefficients by
$\order(5-10\%)$. For applications with only a gluon coefficient
$c_g^{(0)}(\mu_t)$ at the high scale, the mixing would induce nonzero
quark coefficients at the low scale, e.g., $c_q^{(0)}(\mu_b) = -2.8
c_g^{(0)}(\mu_t)$, and could be phenomenologically relevant.

The results for $C$-even spin-two coefficients presented in the right panel of
Table~\ref{tab:coefficientsEVO} employ $R^{(2)}$ and
$M^{(2)}$ at LO. The high-scale gluon coefficient is
$\order(10\%)$ of the $u,d,s,c$ quark coefficients, and contains a
large uncertainty of $\pm \order( 40\%)$ from scale variation of
$\mu_t$. 
Hence, the $C$-even spin-two gluon coefficient is
required for a robust estimate of perturbative uncertainties. In the next section, we will see that due to destructive interference between the scalar and $C$-even spin-two amplitudes, the gluon coefficient has a sizable impact on scattering cross sections of heavy WIMPs. 

\begin{table}[t]
\begin{center}
\begin{tabular}{c|c|c|c|c}
 & $c_{u,d,s}^{(0)}$ & $c_c^{(0)}$ & $c_b^{(0)}$ & $c_g^{(0)}$ \\
\hline
$\mu_t$  & -0.407 & -0.407 & -0.424 & 0.004\\
$\mu_b$ & -0.418 & -0.418 & -0.436 & 0.009 \\
$\mu_b$  & -0.418 & -0.418 & - & 0.012\\
$\mu_c$  & -0.443 & -0.443 & - & 0.022 \\
$\mu_c$  & -0.443 & - & - & 0.028 \\
$\mu_0$  &  -0.454 & - & - & 0.032 \\
\end{tabular}
\hspace{0.5cm}
\begin{tabular}{c|c|c|c}
$c_{u,d,s}^{(2)}$ & $c_c^{(2)}$ & $c_b^{(2)}$ & $c_g^{(2)}$ \\
\hline
 0.667 & 0.667 & 0.091 & -0.050\\
 0.498 & 0.498 & 0.073 & 0.080 \\
 0.498 & 0.498 & - & 0.080 \\
 0.418 & 0.418 & - & 0.140 \\
 0.418 & - & - & 0.140 \\
 0.405 & - & - & 0.147 \\
\end{tabular}
\end{center}
\caption{\label{tab:coefficientsEVO}
Evolution of scalar (left panel) and $C$-even spin-two (right panel) coefficients
for the pure triplet (with overall factors $\pi \alpha_2^2 /m_W^3$
extracted) from a high scale, $\mu_t$, down to a low scale,
$\mu_0$. The number of active quark flavors changes at heavy quark
thresholds $\mu_b$ and $\mu_c$ for the bottom and charm,
respectively. Isospin symmetry and $|V_{tb}|\approx 1$ lead to identical results for $u,d,s$.} 
\end{table}

\subsubsection{Amplitudes and cross section predictions}

\begin{figure}[top]
\centering
\hspace*{-0.45cm}
\includegraphics[scale=0.5]{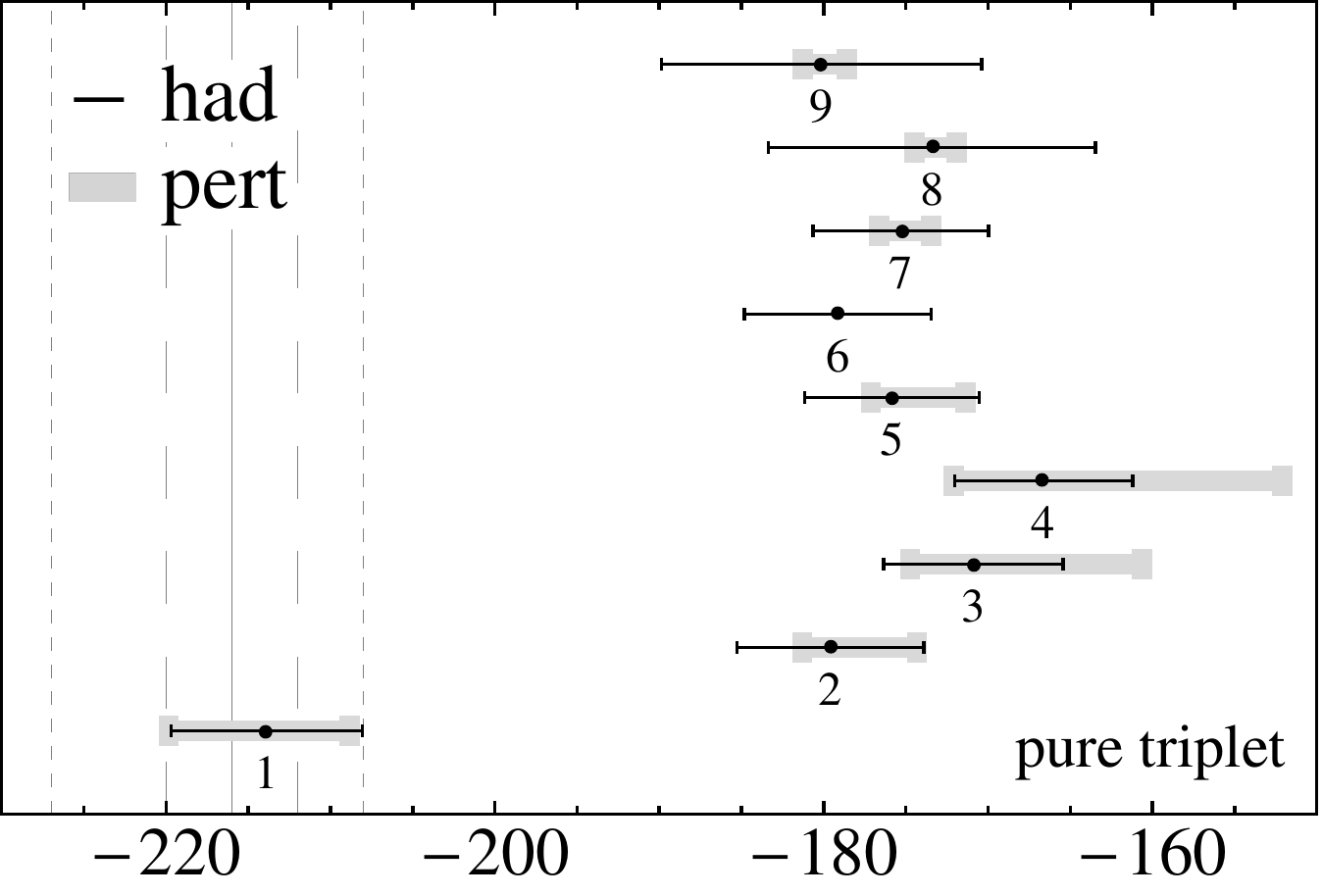}
\hspace*{12mm}\includegraphics[scale=0.5]{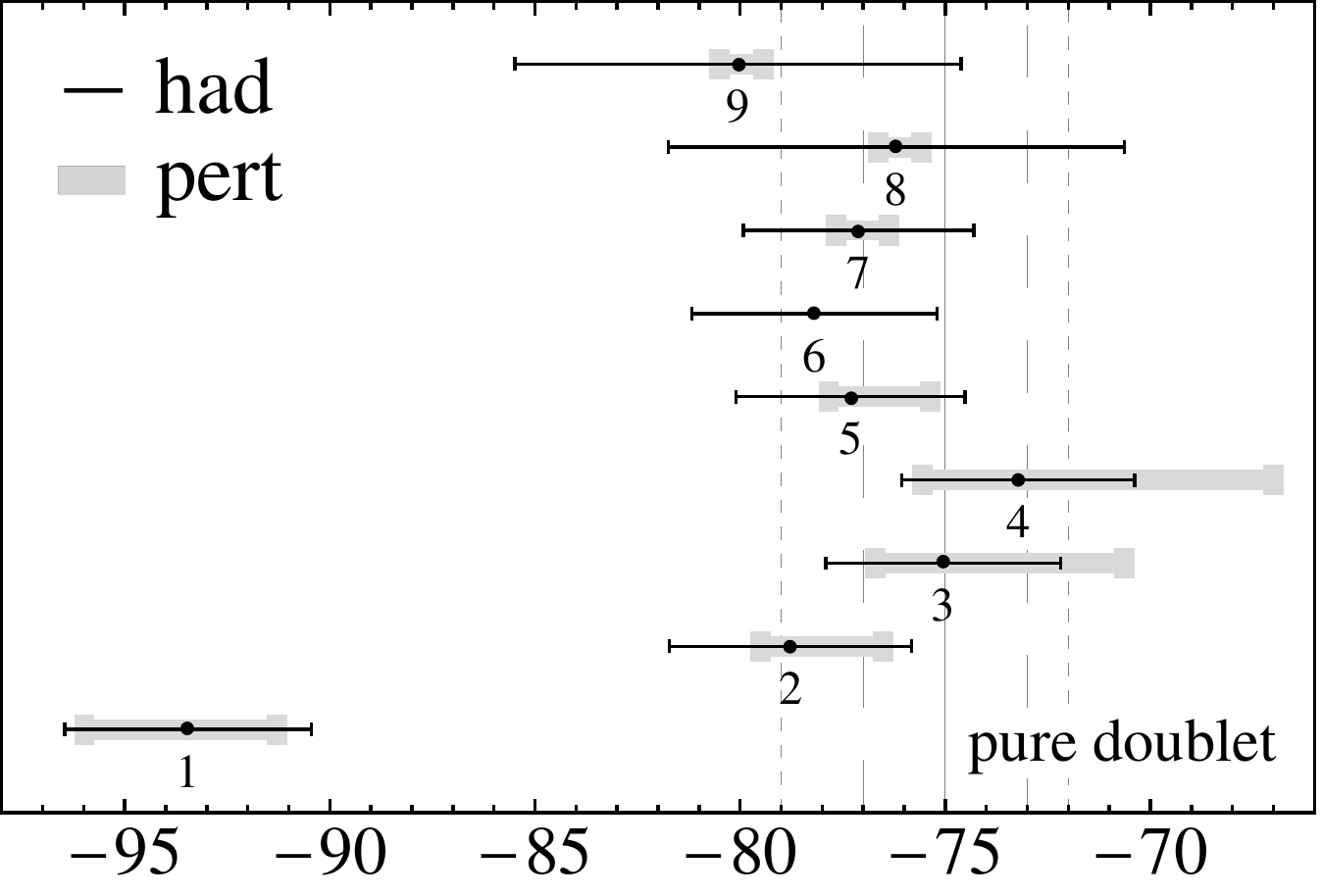}
\centering
\begin{minipage}{0cm}
\vspace*{1.3cm}\hspace*{-4.5cm}${\cal M}_p^{D (0)} \ ({\rm MeV})$\hspace*{-10.6cm}${\cal M}_p^{T (0)}\ ({\rm MeV})$
\vspace*{0.5cm}
\end{minipage}
\centering
\includegraphics[scale=0.5]{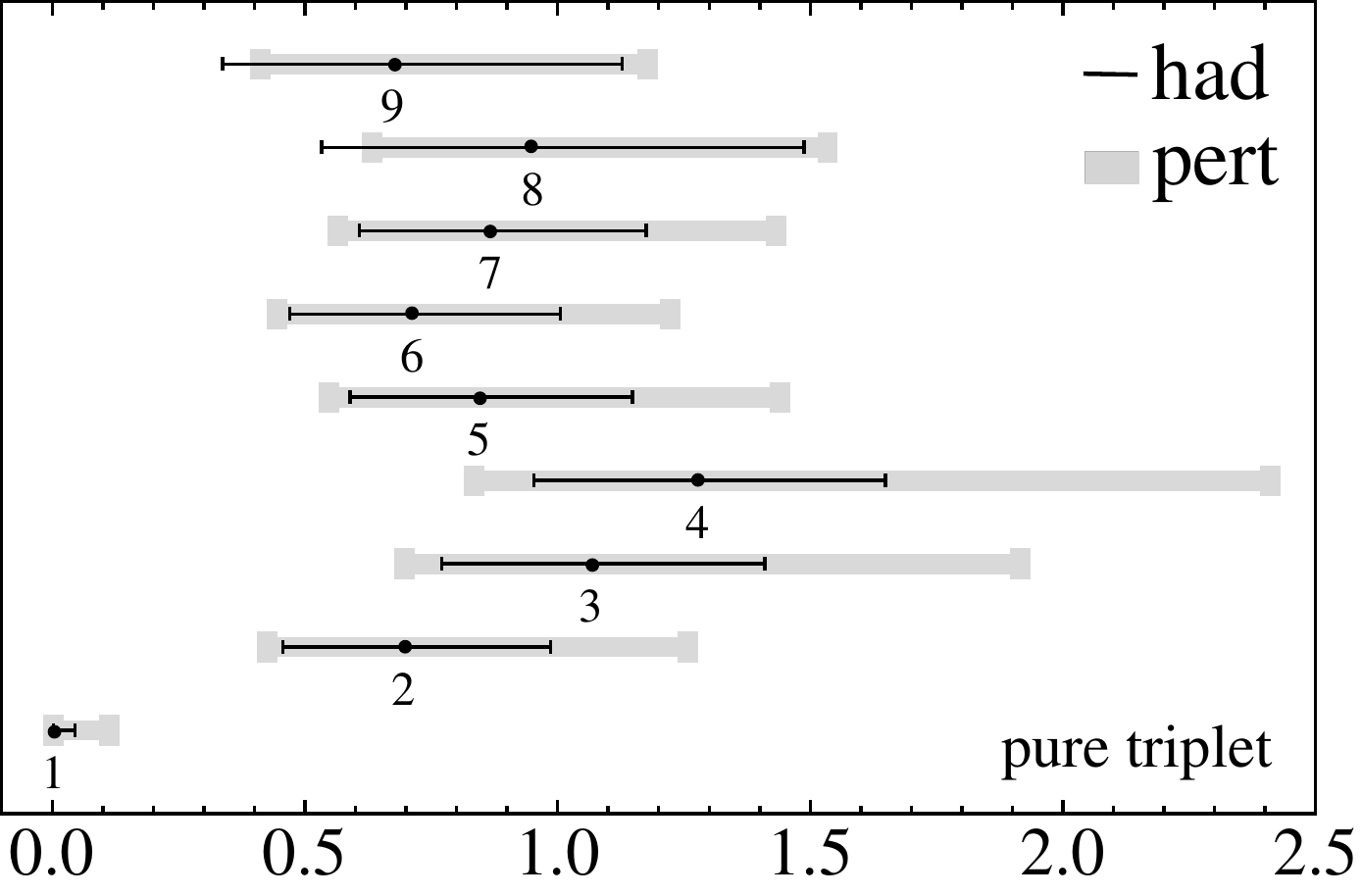}
\hspace*{12mm}\includegraphics[scale=0.5]{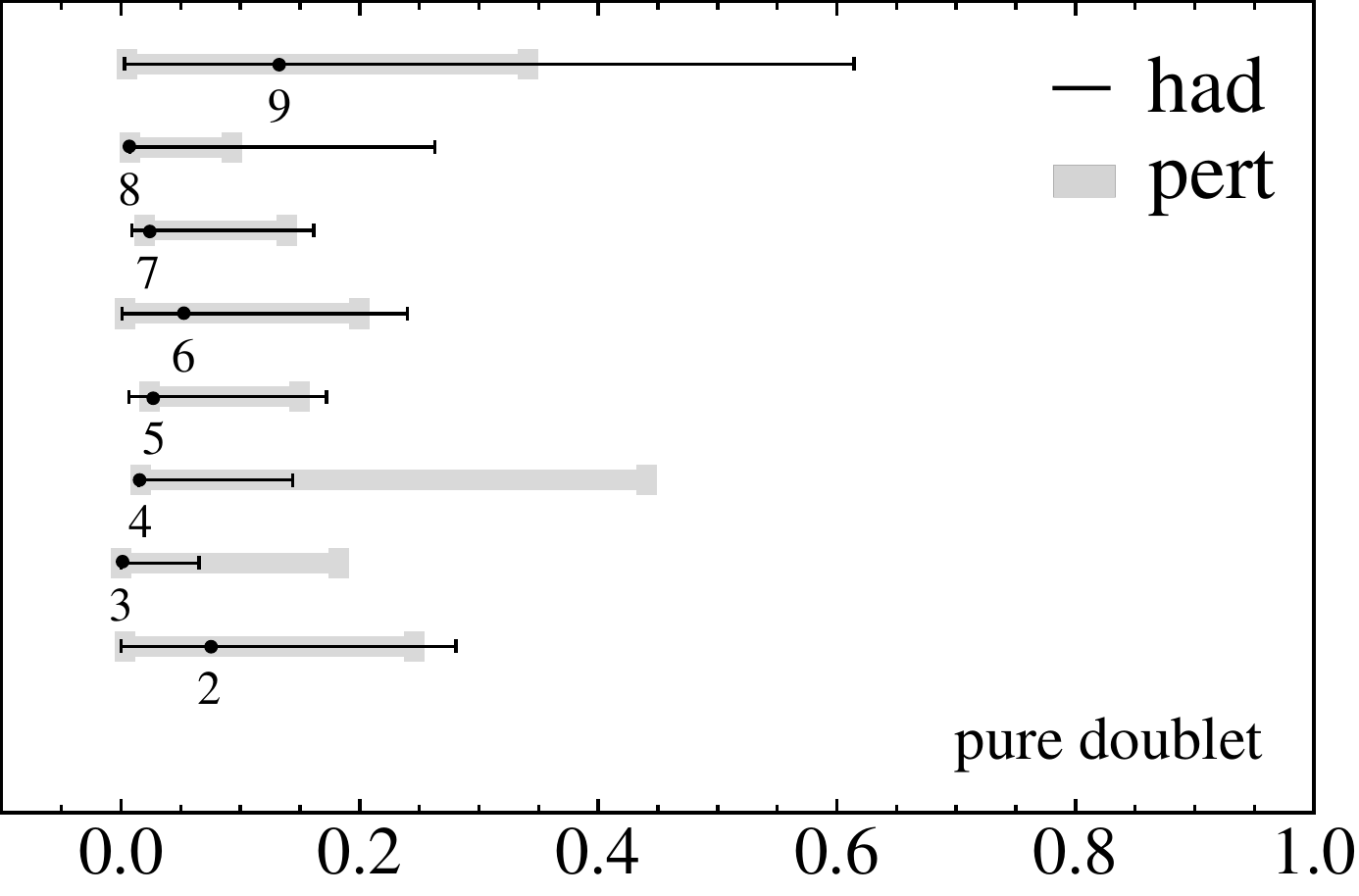}
\centering
\begin{minipage}{0cm}
\vspace*{1.3cm}\hspace*{-5cm}$\sigma_{\rm SI}^D \  (10^{-48} \ {\rm cm}^2)$\hspace*{-11cm}$\sigma_{\rm SI}^T \  (10^{-47} \ {\rm cm}^2)$
\vspace*{0.5cm}
\end{minipage}
\caption{\label{fig:pQCD} Scalar amplitudes (upper panels) and spin-independent cross sections (lower panels) for the pure triplet (left panel) and pure doublet (right panel) cases. The perturbative (hadronic) uncertainties are denoted by thick gray (thin black) lines, and we have extracted the factor $\pi \alpha_s^2 /m_W^3$ from the amplitudes. Vertical lines denote the magnitude of the $C$-even spin-two amplitude (solid) and its perturbative (short dash) and hadronic (long dash) uncertainties as given in Eq.~(\ref{eq:tensorampTD}). We describe each evaluation (labelled 1 through 9) in the text. For the pure doublet, the cross section corresponding to amplitude 1 is not shown.
}
\end{figure}

Let us evaluate the scalar and $C$-even spin-two amplitudes in the 3-flavor theory at a low scale $\mu_0$ to determine the scattering cross section for heavy electroweak charged dark matter.\footnote{As previously discussed, the $C$-even spin-two matrix elements are parametrized in terms of PDF moments and may thus be evaluated directly at the high scale.} 
From the coefficients of the previous section and the matrix elements discussed in Sec.~\ref{sec:hadron}, the amplitudes are given by
\be\label{eq:amplitudeGEN}
{\cal M}^{(S)}_N = \sum_{i=u,d,s, g} c_i^{(S)} (\mu_0) \langle N |  O_i^{(S)} (\mu_0) | N \rangle \, .
\ee
For the cases of pure triplet and
pure doublet scattering on a proton target, we find the $C$-even spin-two ($S=2$) amplitudes
\begin{align}\label{eq:tensorampTD}
  {\cal M}_p^{T(2)} =   {\pi \alpha_2^2 \over m_W^3  } \Bigg[ 216 \big( {}^{+11}_{-8} \big)  \big( {}^{+4}_{-4} \big) \ {\rm MeV} \Bigg] \, , \quad {\cal M}_p^{D(2)} = {\pi \alpha_2^2 \over m_W^3  } \Bigg[ 75  \big( {}^{+4}_{-3} \big)  \big( {}^{+2}_{-2} \big) \ {\rm MeV} \Bigg]\, ,
\end{align}
where the superscripts $T$ and $D$ denote triplet and doublet,
respectively. The first uncertainty is from scale variation, while the second is from PDF inputs. We neglect the uncertainties from variation of $\mu_b$,
$\mu_c$ and $\mu_0$ which are of $\order(1\%)$. The amplitudes for scattering on a neutron are numerically similar.

For the scalar amplitude ($S=0$) we present
several evaluations in the upper panels of Fig.~\ref{fig:pQCD} to illustrate the impact of perturbative QCD corrections. The amplitudes numbered 1 through 4 employ NLO solutions for the
running from $\mu_t$ to $\mu_c$ and for the matching at the bottom and
charm thresholds. Below the charm threshold, amplitudes 1, 2, 3 and 4
respectively employ LO, NLO, NNLO and NNNLO in the running to the low
scale $\mu_0$ and in the scalar gluon matrix element determined from
the sum rule.\footnote{The amplitudes 1, 2, 3, and 4 shown here correspond to the
cross sections labelled LO, NLO, NNLO and NNNLO in Figure 1 of
Ref.~\cite{Hill:2013hoa}.} For amplitudes 1, 2, 3 and 4, the scale $\mu_c$ dominates the perturbative
uncertainty. 

The increased uncertainty in amplitudes 3 and 4
reflects
$\alpha_s(\mu_c)$ corrections to the charm threshold matching and to the running from $\mu_b$ to $\mu_c$ beyond the included NLO corrections; i.e., reduction of the $\mu_c$ scale dependence requires a cancellation between $\alpha_s(\mu_c)$ corrections above, at and below the charm scale. The new corrections to threshold matching obtained in Sec.~\ref{sec:sumrule} 
provide
the missing ingredients required for such a higher order analysis. Including corrections through NNNLO to the running from $\mu_b$ to $\mu_c$ and to the charm threshold matching (employing~(\ref{eq:MsolutionsNNLO}),~(\ref{eq:MsolutionsNNNLO}) and
Ref.~\cite{Chetyrkin:1997un} for $M_{gQ}$ and $M_{qQ}$), we obtain amplitude 5 with reduced
perturbative uncertainty. 

The framework described in terms of the solutions $R$ and $M$ is equivalent to a perturbative determination of the scalar amplitude. The $\mu_0$ dependence cancels between the gluon matrix element and the running below the charm threshold, yielding a result depending on the
scales $\mu_t$, $\mu_b$ and $\mu_c$ only. Working through NLO, we obtain
the amplitude
\begin{align}\label{eq:scalarampNLO}
{ {\cal M}^{(0)}_N \over m_N} &= \sum_{q=u,d,s}  f_{q,N}^{(0)} c_q^{(0)} (\mu_t) + {2    \over  27 } (1-\lambda) c^{(0)}_c(\mu_t)  \Bigg\{1  + { \alpha^{(4)}_s(\mu_c) \over 4\pi} \Bigg[ {107 \over 9} - {8 \over 1-\lambda}\Bigg] \Bigg\} 
\nl
&\quad
+ {2 \over 27}  (1-\lambda) c_b^{(0)}(\mu_t) \Bigg\{1
 + { \alpha^{(4)}_s(\mu_c) \over 4\pi} \Bigg[ -{214 \over 225}  + {16 \over 25 (1-\lambda)} \Bigg]
+ { \alpha^{(5)}_s(\mu_b) \over 4\pi} \Bigg[ {321 \over 25} - {216 \over 25 (1-\lambda) }  \Bigg]
 \Bigg\} 
  \nl
&\quad
-{8 \pi \over  9 \alpha_s^{(6)}(\mu_t)} {  f_{N}^{(0)} } c_g(\mu_t) \Bigg\{ 1 + 
{ \alpha^{(4)}_s(\mu_c) \over 4\pi} \Bigg[ -{214 \over 225}  + {16 \over 25 (1-\lambda )}  \Bigg] + { \alpha^{(5)}_s(\mu_b) \over 4\pi} \Bigg[ -{642 \over 575}  + {432 \over 575 (1-\lambda) }  \Bigg]  
 \nl
&\quad
+ { \alpha^{(6)}_s(\mu_t) \over 4\pi} \Bigg[ {68 \over 23}  - {216 \over 23 (1-\lambda) }  +{4 \over 3} \log {\mu_t \over m_t} \Bigg]
\Bigg\} + \order(\alpha_s^2, 1/m_c, 1/m_b,1/m_N)  \, ,
\end{align}
where $\lambda = \sum_{q=u,d,s}  f_{q,N}^{(0)}$, and the leading-order
$\alpha_s$ result is well-known from Ref.~\cite{Shifman:1978zn}. Since
the quark matrix elements are scale independent and, neglecting power corrections, are not corrected at heavy quark thresholds, the result in
Eq.~(\ref{eq:scalarampNLO}) is the same whether obtained in a 4- or
5-flavor theory employing the NLO solution given in~(\ref{eq:heavyME}) for the charm and bottom matrix elements. Given the ingredients in Sec.~\ref{sec:sumrule}, it is straightforward to extend this result to NNNLO. For illustration, we include in
Fig.~\ref{fig:pQCD} the amplitudes corresponding to the LO and NLO result in Eq.~(\ref{eq:scalarampNLO}), labelled 6 and 7, respectively. The LO result has no scale variation,
while the
NLO result gives an estimate of perturbative corrections that is
consistent with amplitudes 2 and 5, albeit smaller.\footnote{For amplitudes 2 and 5, we employ $R$ and $M$ matrices expanded order by order in $\alpha_s$, and the residual scale uncertainty can be traced to spurious terms appearing in the product of these matrices.}

The amplitudes 8 and 9 in Fig.~\ref{fig:pQCD} are evaluated in the
4-flavor theory, employing the charm matrix elements given in~(\ref{eq:fcharmlattice}). The large hadronic uncertainty reflects those of the lattice measurements, while the scale uncertainty is small, having
avoided a perturbative treatment of the scale $\mu_c$. 

The cross section for scattering on a nucleon target is obtained from the amplitudes as
\be
\sigma_{\rm SI} = {m_N^2 \over \pi} | {\cal M}^{(0)}_N + {\cal M}^{(2)}_N |^2 \, .
\ee
In the case of heavy electroweak-charged WIMPs, opposite signs of
the scalar and 
$C$-even spin-two
amplitudes lead to destructive interference. There is a large cancellation for scalar
amplitudes near the vertical lines denoting the magnitude of the
$C$-even spin-two
amplitude in the upper panels of Fig.~\ref{fig:pQCD}. Cross section predictions are shown in the lower panels with labels corresponding to the scalar
amplitude employed.

For the triplet, the cross section prediction given in
Ref.~\cite{Hill:2013hoa} corresponds to amplitude 4 in
Fig.~\ref{fig:pQCD}, and gives a conservative estimate of scale
uncertainty. An improved estimate with reduced scale uncertainty is given by the cross section corresponding to amplitude 5,
\be
\sigma_{\rm SI}^T = 8  {}^{+6}_{-3} {}^{+3}_{-3}  \times 10^{-48} \ {\rm cm}^2 \, ,
\ee
where the first (second) uncertainty is from scale variation (hadronic inputs).
The remaining scale uncertainty is dominated by $\mu_t$ variation in the 
$C$-even spin-two
amplitude, and its reduction requires higher-order matching at the weak-scale. 

For the doublet case, the improved estimates lead to the same conclusion in Ref.~\cite{Hill:2013hoa},
\be
\sigma_{\rm SI}^D \lesssim 10^{-48} \ {\rm cm}^2 \quad (95\% \ {\rm C.L.}) \, .
\ee
In the case of strong destructive interference, the cross
section prediction and its fractional uncertainty become highly
sensitive to perturbative corrections and changes in parameter inputs.

\section{Summary and discussion\label{sec:summary}}

The analysis of WIMP dark matter scattering on an atomic nucleus is a challenging field theory 
problem involving multiple energy scales, ranging from mass scales of SM extensions ($\gtrsim{\rm TeV}$), 
to the electroweak scale ($\sim 100\,{\rm GeV}$), heavy quark thresholds ($\sim 5\,{\rm GeV}$), 
QCD and pion mass scales ($\sim 100\,{\rm MeV}$), nuclear excitation scales ($\sim\,{\rm MeV}$),
and finally recoil energies in direct detection experiments ($\sim~{\rm keV}$).    
A sequence of effective theories capitalizes on these scale separations, permitting a systematic expansion 
in small ratios such as $m_W/M_{\rm DM}$, $m_b/m_W$ and $\Lambda_{\rm QCD}/m_c$. 
A corresponding sequence of matching computations and renormalization group evolution provides the
connection between the parameters of high scale physics models and low energy observables.  
The preceding paper~\cite{Hill:2014yka} of this series treated the problem of
weak scale matching, which may proceed from a specified UV completion, or employ 
the heavy WIMP expansion to compute matching coefficients 
independent of the detailed UV completion. 
The remaining steps in the sequence are independent of the origin of the weak scale matching coefficients, and 
in this paper we have treated the problem of relating the resulting theory renormalized at the weak scale to
an effective theory renormalized at low scales ($\lesssim m_c$) where hadronic matrix elements are evaluated.    
We discussed some aspects of the further evaluation of nuclear matrix elements, and presented either the $n_f=3$ flavor QCD theory, or the single nucleon theory discussed in Section~\ref{sec:nucleon}, as a natural handoff point to detailed 
nuclear modeling.

Section~\ref{sec:eft} presented the basis of lagrangian interactions between scalar or fermion WIMPs 
and SM fields; for fermionic WIMPs, we considered photon interactions through dimension five, and quark or gluon interactions through dimension seven.
These capture the leading interactions for either complex (Dirac) or self-conjugate (Majorana) WIMPs.  
A sample matching calculation onto this basis from a gauge-singlet WIMP UV completion was performed in \ref{sec:weakmatch}; the case of electroweak charged 
dark matter was discussed in \cite{Hill:2014yka}. 
Seemingly dramatic effects can emerge when passing from high to low scales, generically involving processes that
are naively absent but not forbidden by symmetry.   Examples include the chiral rotation to mass eigenstates
that induces an operator mediating spin-independent scattering, as considered in Section~\ref{sec:weakmatch}. 
Similarly, dipole interactions of WIMPs with the electromagnetic field 
($c_{\chi 1}$ and $c_{\chi 2}$) can be induced by heavy quark loops 
from a theory which at some renormalization scale contains only contact interactions with quarks~\cite{Haisch:2013uaa}.  
These examples highlight the importance of working in a low energy basis that is closed under renormalization, 
and that contains all operators not forbidden by symmetry. 
For self-conjugate WIMPs of mass $M\gtrsim m_W$,  
the $n_f+1$ spin zero operators involving $O^{(0)}_{q,g}$ and 
$n_f+1$ spin two operators involving $O^{(2)}_{q,g}$ in (\ref{eq:LWIMPSM}) 
determine spin independent interactions with nuclei.   
Remaining agnostic regarding UV completion, one could investigate direct detection constraints on 
these 12 coefficients ($n_f=5$).  Large redundancies in the parameters would appear since the effects of heavy quarks 
are degenerate with those of light quarks and gluons; passing to $n_f=3$ leaves 8 coefficients that could be constrained
in principle by a suite of direct detection observables.   The spin zero operators $m_q \bar{q}q$ and $(G^a_{\mu\nu})^2$ 
have received most attention, but equally large contributions are obtained in many cases from spin 
two operators~\cite{Hill:2013hoa}. 

Section~\ref{sec:weak} considered the seven classes of QCD operators appearing in 
the basis for WIMP interactions with quarks and gluons.  Each class is separately closed 
under QCD renormalization.  Leading operator renormalization factors, anomalous dimensions, 
and threshold matching corrections were presented.   Special attention was paid to 
the dimension four scalar operators, since this sector drives the final cross section 
uncertainty in many WIMP models.  In particular, poor convergence of perturbation theory
at the charm mass scale implies sensitivity of scattering observables to high orders in the 
$\alpha_s$ expansion.  We performed a new analysis using sum rule
constraints to derive the heavy quark threshold matching corrections for light quark and gluon 
interactions induced in the presence of a high scale gluon operator.  
To our knowledge, the expressions (\ref{eq:MsolutionsNNNLO}) are new. 
The solutions to RG evolution were obtained; combined with threshold matching corrections, these
results provide the mapping of weak scale coefficients onto the low-energy theory containing
the WIMP and $n_f=3$ flavor QCD. 

While a complete analysis of general nuclear matrix elements is beyond the scope of this work, 
in order to compute benchmark single nucleon cross sections, and make contact with 
nuclear models, Section~\ref{sec:hadron} considers the nucleon matrix elements for each 
of the seven classes of relevant QCD operators. 
Again, special attention is paid to the scalar operator matrix elements.   We provide
an updated value for the perturbative prediction of the charm scalar matrix element 
in terms of $n_f=3$ flavor QCD quark matrix elements. 
We surveyed current knowledge concerning the remaining nucleon matrix elements, providing
a guide to the level of uncertainty in each case.  Constraints on these matrix elements
arise from a wide range of techniques and approximations: 
elastic and inelastic electron and neutrino scattering; $SU(3)$ baryon spectroscopy and 
chiral perturbation theory; lattice QCD; and sum rule and anomaly constraints to relate 
gluon and quark matrix elements.  

The significance of the remaining uncertainties depends on the WIMP model under investigation. 
Several matrix elements are also of relevance to nucleon electric dipole moment
searches~\cite{Ellis:2008zy,Engel:2013lsa} and remain a target for further improvement 
from lattice studies, or potentially (as concerns $F_A^{(p,0)}(0)$ in (\ref{eq:FA380})) 
neutrino scattering~\cite{Adams:2013qkq,Miceli:2014hva}.   Let us single out several 
hadronic quantities that can be traced directly to significant (sometimes dramatic) 
uncertainties in WIMP models.   The strange scalar matrix element is a well-known 
source of uncertainty in spin independent WIMP-nucleon scattering~\cite{Ellis:2008hf}, as illustrated in Fig.~\ref{fig:pQCD}. 
The quark mass ratio $m_u/m_d$ and isovector scalar matrix element $\Sigma_-$ in (\ref{eq:sigmas}) drive an 
$O(1)$ uncertainty in connecting $f_n/f_p$ to underlying quark-gluon operators in 
some well-studied scenarios (cf. Fig.~\ref{fig:contact}).    
A charm scalar matrix significant different from the OPE (in $1/m_c$) prediction would 
significantly alter predictions for spin-independent scattering (cf. Fig.~3 of \cite{Hill:2013hoa}). 
Clearly there is further room for significant, albeit model-dependent, impact from lattice 
QCD on dark matter direct detection. 

For scenarios including light force mediators in the dark sector, we considered the 
heavy particle effective lagrangian for nucleons and WIMPs in Section~\ref{sec:nucleon}. 
As discussed in Section~\ref{sec:lorentz}, our basis differs 
(apart from notation) 
from a basis constrained by Galilean invariance~\cite{Fitzpatrick:2012ix}
by the inclusion of operators forbidden by Galilean but not Lorentz invariance, and 
corresponding coefficient relations.   

As phenomenological illustrations we considered two examples.   The first involved
relating contact interactions specified at the weak scale to the low-energy effective 
theory at hadronic scales.   The chosen example exhibited a strong and previously 
unappreciated sensitivity of the ratio $f_n/f_p$ (spin-independent WIMP nucleon couplings)
to both hadronic matrix element uncertainties and QCD scale choice.  
The formalism presented here may be used to systematically relate more general 
classes of weak-scale contact interactions~\cite{Beltran:2010ww,Goodman:2010ku} (and other) models
to low energy observables of direct detection, or annihilation of low-mass WIMPs. 
The second phenomenological example provided details of the first complete calculation of the leading spin-independent WIMP-nucleon 
cross section in 
SM extensions consisting of one or two heavy ($M\gg m_W$) 
electroweak $SU(2)_W \times  U(1)_Y$ multiplets.
Inclusion of higher orders in perturbation theory for charm threshold corrections improves upon but does
not significantly alter the conclusions previously reported in \cite{Hill:2011be,Hill:2013hoa}. 
While further analysis of power corrections and nuclear modeling is warranted, it is likely
that such WIMP candidates will remain an elusive target for next generation direct detection
searches~\cite{Aprile:2012nq,Akerib:2013tjd}.   

QCD corrections have an important impact on many WIMP models, and must be systematically incorporated in order to meaningfully compare theory and observation.
A number of extensions can be readily considered, e.g., including new gauge interactions beyond the SM, 
higher spin DM, and inelastic scattering.  

\vskip 0.2in
\noindent
{\bf Acknowledgements}
\vskip 0.1in
\noindent
We thank Andreas Kronfeld for comments on the manuscript. This work was supported by the United States Department of Energy under
Grant No. DE-FG02-13ER41958. MS acknowledges
support from a Bloomenthal Fellowship at the University of Chicago, and
from the  Office of Science, Office of High Energy Physics, of the
U.S.\ Department of Energy under contract DE-AC02-05CH11231.

\appendix

\section{Renormalization constants \label{sec:constants}}
\subsection{Finite corrections to the axial-vector and pseudoscalar renormalization constants}
The renormalization constants given in Table~\ref{tab:QCDZ} for the axial-vector currents and pseudoscalar operators include a finite correction in addition to the ${\overline {\rm MS}}$ scheme~\cite{Larin:1993tq},
\begin{align}\label{eq:ZAP}
Z_A^{\rm singlet} = \big( Z_5^{\rm s} \big)^{-1} \big( Z_{\overline{\rm MS} }^{\rm s} \big)^{-1} \, , \quad Z_A^{\rm non-singlet} = \big( Z_5^{\rm ns} \big)^{-1} \big( Z_{\overline{\rm MS} }^{\rm ns} \big)^{-1}\, , \quad Z^{(0)}_{5qq} = Z_m   \big( Z_5^{\rm P} \big)^{-1} \big( Z_{\overline{\rm MS} }^{\rm P} \big)^{-1} \, ,
\end{align}
where
\begin{align}\label{eq:ZMSZ5}
Z^{\rm ns}_{\overline{{\rm MS}}} &= 1 
+ \left(\alpha_s\over 4\pi \right)^2 
{1\over \epsilon} 
\left( {88\over 3} - {16\over 9} n_f \right)
+ \order(\alpha_s^3) \,, \nl
Z^{\rm ns}_{5} &= 1 + {\alpha_s \over 4\pi} 
\left( -{16\over 3} \right) + \order(\alpha_s^2)\,, \nl
Z^{\rm s}_{\overline{{\rm MS}}} &= 1 + 
\left({\alpha_s\over 4\pi}\right)^2 {1\over \epsilon} \left( 
{88 \over 3} + {20 \over 9} n_f 
\right) + \order(\alpha_s^3) \,, \nl
Z^{\rm s}_{5} &= 1 + {\alpha_s\over 4\pi} 
\left( -{16\over 3} \right) 
+ \order(\alpha_s^2) 
\,, \nl
Z_{\overline{{\rm MS}}}^p &= 1 
+ {\alpha_s\over 4\pi}
\left(-4\over \epsilon\right) 
+ 
\left(\alpha_s\over 4\pi\right)^2 
\left[ \left(30-\frac43 n_f\right){1\over\epsilon^2}
  + \left(25-{22\over 9} n_f \right) {1\over \epsilon} 
\right] + \order(\alpha_s^3) \,,
\nl
Z_{5}^p &= 1 + {\alpha_s\over 4\pi} 
\left( -{32\over 3} \right) 
+ \order(\alpha_s^2) \,.
\end{align} 
The mass renormalization constant $Z_m$, also appearing in the renormalization constant $Z_T$ of the antisymmetric tensor current $T_q$, is given by
\begin{align}\label{eq:Zm}
Z_m &= 1 + 
{\alpha_s\over 4\pi}{1\over \epsilon}
\left( - 4 \right)
  + \left(\alpha_s\over 4\pi\right)^2
\bigg[ {1\over\epsilon^2}
\left( 30 - \frac43 n_f \right) 
+ {1\over \epsilon}
\left( -{101\over 3} + {10\over 9} n_f \right) 
\bigg] + \order(\alpha_s^3) \,.
\end{align}
Terms contributing to one-loop matching and two-loop anomalous dimension are retained in the renormalization constants in (\ref{eq:ZMSZ5}) and in $Z^{(0)}_{5gq}$ and $Z^{(0)}_{5gg}$ given in Table~\ref{tab:QCDZ}.\footnote{
In the notation of Ref.~\cite{Larin:1993tq}, where a different operator 
basis involving $J \equiv \sum_q \partial_\mu A_q^\mu$ was chosen, we have 
$Z^{(0)}_{5gq} = \big( Z_{G\tilde{G},{\overline{\rm MS}}} \big)^{-1} Z_{GJ,{\overline{\rm MS}}} Z_A^{\rm singlet}$ and 
$Z^{(0)}_{5gg} = \big( Z_{G\tilde{G},{\overline{\rm MS}}} \big)^{-1}\bigg[ 1 - {g^2 \over 32\pi^2 } n_f Z_{GJ,{\overline{\rm MS}}} Z_A^{\rm singlet} \bigg]$,
where $Z_A^{\rm singlet}$ is given in (\ref{eq:ZAP}) and $Z_{G\tilde{G},{\overline{\rm MS}}}$, $Z_{GJ,{\overline{\rm MS}}}$ are 
the quantities denoted by $Z_{G\tilde{G}}$, $Z_{GJ}$ in 
Ref.~\cite{Larin:1993tq}. 
}

\subsection{QCD beta function and quark anomalous dimension}
The renormalization constant for the scalar operators is given in terms of the QCD beta function $\beta$ and the quark mass anomalous dimension $\gamma_m$. We define these as
\begin{align}
{\beta \over g} &= { d \log g \over d \log \mu} = -\beta_0 \left({\alpha_s \over 4\pi}\right) - \beta_1 \left({\alpha_s \over 4\pi}\right)^2 - \beta_2 \left({\alpha_s \over 4\pi}\right)^3 - \beta_3 \left({\alpha_s \over 4\pi}\right)^4  + \dots \,,
\nl
\gamma_m &= { d \log m_q \over d \log \mu} =  - \gamma_0 \left({\alpha_s \over 4\pi}\right) - \gamma_1 \left({\alpha_s \over 4\pi}\right)^2 - \gamma_2 \left({\alpha_s \over 4\pi}\right)^3 - \gamma_3 \left({\alpha_s \over 4\pi}\right)^4  \dots\, ,
\end{align}
where the ellipses denote terms higher order in $\alpha_s$, and the required functions are
\begin{align}
\beta_0 &= 11 - \frac23 n_f \, , \nl
\beta_1 &= 102 - {38 \over 3} n_f \, , \nl 
\beta_2 &= {2857 \over 2} -{5033 \over 18} n_f + {325 \over 54} n_f^2 \, , \nl
\beta_3 &= {149753 \over 6} + 3564 \zeta(3) -\left( {1078361 \over 162} + {6508 \over 27} \zeta(3) \right) n_f + \left( {50065 \over 162} + {6472 \over 81} \zeta(3) \right) n_f^2 + {1093 \over 729} n_f^3 \, , 
\end{align}
and
\begin{align}
\gamma_0 &= 8 \, , \nl
\gamma_1 &= {404 \over 3} - {40 \over 9} n_f \, , \nl
\gamma_2 &= 2498 - \left( {4432 \over 27} +{320 \over 3} \zeta(3) \right) n_f - {280 \over 81} n_f^2 \, , \nl
\gamma_3 &= {4603055 \over 81} + {271360 \over 27} \zeta(3) - 17600\zeta(5) +\left( - {183446 \over 27} - {68384 \over 9} \zeta(3) + 1760 \zeta(4) + {36800 \over 9} \zeta(5) \right) n_f 
\nl
&\quad 
+ \left( {10484 \over 243} +{1600 \over 9} \zeta(3) -{320 \over 3} \zeta(4) \right) n_f^2 + \left( -{664 \over 243} +{128 \over 27} \zeta(3) \right) n_f^3 \, .
\end{align}

\section{Nucleon matrix elements}\label{sec:MEappendix}

\subsection{Corrections to zero momentum transfer}
The corrections to zero momentum transfer are severely suppressed in the 
nonrelativistic regime of typical WIMP-nucleon scattering processes. 
To gauge the impact of these corrections in general models, 
we summarize current knowledge, identifying 
uncertainties that should be revisited if observables 
are found to be sensitive to these parameters.

The $q^2$ dependence of vector form factors may be investigated using the definition of the nucleon charge radii,
\begin{align}
{d\over dq^2} F_1^{(N)}\bigg|_{q^2=0} & \equiv \frac16 [r_E^{(N)}]^2 - {a_N \over 4 m_N^2} \,, 
\end{align} 
with $[r_E^{(p)}]^2 = 0.70\,{\rm fm}^2 - 0.77\,{\rm fm}^2$~\cite{Mohr:2012tt} and 
$[r_E^{(n)}]^2 = -0.1161(22) \,{\rm fm}^2$~\cite{Nakamura:2010zzi}.%
\footnote{
This definition is motivated by considering the Sachs electric form factor 
$G_E(q^2) = F_1(q^2) + (q^2/m_N^2) F_2(q) \approx G_E(0) + \frac16 r_E^2 q^2 + \order(q^4)$. 
The proton charge radius is the subject of significant debate, but discrepancies are 
at a level of precision 
far beyond what is currently required for dark matter applications. 
}   
Together with estimates for the numerically small strange contribution, 
\be
{d\over dq^2} F_1^{(p,s)}\bigg|_{q^2=0} \equiv \frac16 r_s^2 \,, \quad
r_s^2 = 
\begin{array}{lc} 
0.021 \pm 0.063 \,{\rm fm}^2  & \cite{Leinweber:2006ug}
\\
\end{array}
\,,
\ee
we may solve for the leading Taylor expansion of the quark vector current matrix elements, 
\begin{align}
{d\over dq^2} F_1^{(p,u)}\bigg|_{q^2=0} &= 
2\left( \frac16 [r_E^{(p)}]^2 - {a_p \over 4 m_p^2} \right) 
+ \left( \frac16 [r_E^{(n)}]^2 - {a_n \over 4 m_n^2} \right) 
+ \frac16 r_s^2 \,,
\nl
{d\over dq^2} F_1^{(p,d)}\bigg|_{q^2=0} &= 
\left( \frac16 [r_E^{(p)}]^2 - {a_p \over 4 m_p^2} \right) 
+ 2\left( \frac16 [r_E^{(n)}]^2 - {a_n \over 4 m_n^2} \right) 
+ \frac16 r_s^2 \,. 
\end{align} 
Again, we approximate the neutron form factors by the corresponding proton form factors using approximate isospin symmetry expressed in (\ref{eq:PNiso}).

The $q^2$ dependence of axial-vector form factors may be investigated, writing
\be\label{eq:mA}
F_A^{(p,a)}(q^2) = F_A^{(p,a)}(0)\big[ 1 + 2 q^2/m_A^{(a)2} + \order(q^4)\big] \,,
\ee
with $m_A^{(3)} \approx m_A^{(8)} \approx m_A^{(0)} \approx 1.0\,{\rm GeV}$ denoting  
an ``axial mass'' scale. 
There remains considerable uncertainty on the $q^2$ dependence of the 
isovector axial-vector form factor, with a conservative treatment of
shape uncertainty yielding~\cite{Bhattacharya:2011ah} 
$m_A^{(3)} = 0.85(22)(8) \, {\rm GeV}$ from neutrino scattering 
and $m_A^{(3)} = 0.92(13)(8) \,{\rm GeV}$ from pion electroproduction. 
The octet and flavor-singlet cases are less constrained.  For definiteness we 
take also $m_A^{(8)}=m_A^{(0)}=1.0(3)\,{\rm GeV}$ as default values, which should 
be revisited if observables are found to be sensitive to these parameters.  

For the induced pseudoscalar form factors, 
the isovector component is best determined~\cite{Bernard:2001rs}, 
\be
F_{P^\prime}^{(p,3)}(q^2) = \frac12\left[ {4 m_p g_{\pi N} f_\pi \over m_\pi^2 - q^2} -\frac23 g_A m_p^2 r_A^2 + \order(q^2,m_\pi^2) 
\right] \,,
\ee
with $f_\pi=93\,{\rm MeV}$, $m_\pi$ and $m_p$ denoting the charged pion and proton masses 
respectively, $g_A = 1.267(4)$ the axial coupling constant, 
$g_{\pi N}=13.1(4)$ the pion-nucleon coupling and 
$r_A^2/6 = 2/m_A^{(3)2}$ giving 
the slope of the form factor as in (\ref{eq:mA}).   In particular, in the chiral limit, 
\be
F_{P^\prime}^{(p,3)}(0) \sim  {g_A\over 2} {4 m_p^2\over m_\pi^2} \,,
\ee
diverging as $m_\pi\to 0$.    
For the octet contribution, a similar enhancement behaving as $m_p^2/m_\eta^2$ emerges, 
\be\label{eq:FP8}
F_{P^\prime}^{(p,8)}(0) \sim {2m_p^2 \over m_\eta^2} \,. 
\ee
while for the flavor singlet case the absence of a pseudo Nambu Goldstone boson coupling
to the current implies the absence of such an enhancement,
\be
F_{P^\prime}^{(p,0)}(0) \sim \order(1) \,. 
\ee
Large $SU(3)$ breaking corrections modify relations such as (\ref{eq:FP8}), which should be 
accounted for by allowing significant variation in the assumed form factors. 

\subsection{Heavy quark scalar matrix elements}
The $\order(\alpha_s^3)$ term in (\ref{eq:heavyME}) is given by
\begin{align}
\langle O_Q^{\prime (0)} \rangle_4 &= 
\frac{6628017}{8} \log ^2{ \mu_Q \over m_Q }+\frac{3325355}{8} \log { \mu_Q \over m_Q }-\frac{37326201}{8} \lambda \log ^2{ \mu_Q \over m_Q }-\frac{32191115}{8} \lambda \log { \mu_Q \over m_Q }
\nl
&\quad
-\frac{2387203071  \zeta (3)}{512} \lambda +\frac{3852721945 }{768}\lambda+\frac{3399188991 \zeta (3)}{512}-\frac{6064085209}{768}
\nl
&\quad
+ n_f \Bigg\{ 25234 \log ^2{ \mu_Q \over m_Q }-\frac{244697}{8} \log { \mu_Q \over m_Q }+1167584 \lambda \log ^2{ \mu_Q \over m_Q }+\frac{6596273}{8} \lambda \log { \mu_Q \over m_Q }
\nl
&\quad
+\frac{181905471  \zeta (3)}{128} \lambda -\frac{36859013 }{32} \lambda-\frac{227904831 \zeta (3)}{128}+\frac{195412223}{96}\Bigg\}
\nl
&\quad
n_f^2 \Bigg\{ -\frac{116243}{3}  \log ^2{\mu_Q \over m_Q}+\frac{433163}{48} \log {\mu_Q \over m_Q}-\frac{328597}{3} \lambda \log ^2{\mu_Q \over m_Q}-\frac{3504107}{48} \lambda \log {\mu_Q \over m_Q}
\nl
&\quad
-147631 \lambda \zeta (3)+\frac{51820165 }{576}\lambda+169411 \zeta (3)-\frac{120231661}{576}\Bigg\}
\nl
&\quad
n_f^3 \Bigg\{ \frac{41840}{9} \log ^2{\mu_Q \over m_Q}-\frac{340895}{216} \log {\mu_Q \over m_Q}+\frac{13696}{3} \lambda \log ^2{\mu_Q \over m_Q}+\frac{820223}{216} \lambda \log {\mu_Q \over m_Q}
\nl
&\quad 
+\frac{1839305  \zeta (3)}{288} \lambda -\frac{1088479 }{324}\lambda -\frac{1966025 \zeta (3)}{288}+\frac{3637345}{324}\Bigg\}
\nl
&\quad
n_f^4 \Bigg\{ -\frac{1934}{9}  \log ^2{\mu_Q \over m_Q}+\frac{9421}{108} \log {\mu_Q \over m_Q}-\frac{214}{3} \lambda \log ^2{\mu_Q \over m_Q}-\frac{12397}{108} \lambda \log {\mu_Q \over m_Q}-\frac{28297  \zeta (3)}{288}\lambda
\nl
&\quad 
+\frac{7519 }{162}\lambda+\frac{28297 \zeta (3)}{288}-\frac{886}{3}\Bigg\}
\nl
&\quad 
+ n_f^5 \Bigg\{ \frac{32}{9} \log ^2{\mu_Q \over m_Q}-\frac{77}{54} \log {\mu_Q \over m_Q}+\frac{77}{54} \lambda \log {\mu_Q \over m_Q}+\frac{5 }{54}\lambda+\frac{481}{162}\Bigg\} \, ,
\end{align}
where $m_Q$ is the $\overline{\rm MS}$ quark mass. Scheme dependence enters at this order and we have checked that the result in terms of the pole mass, employing the relevant functions in Ref.~\cite{Chetyrkin:1997un}, is consistent with the relation between $\overline{\rm MS}$ and pole masses given in Ref.~\cite{Chetyrkin:2000yt}. This result is employed in the determination of the charm quark matrix element in~(\ref{eq:fcharm}) which, in terms of an $\alpha_s$ expansion, is given by
\begin{align}\label{eq:fcharmalpha}
{\tilde f}^{(0)}_{c,N} &= 0.074(1-\lambda) 
+ { \alpha^{(4)}(\mu_c) \over \pi} \Bigg\{ 0.072-0.220 \lambda \Bigg\}
+ \Bigg( { \alpha^{(4)}(\mu_c) \over \pi} \Bigg)^2 \Bigg\{ 0.100-0.528 \lambda
\nl
&\quad
 +\left[  0.300 -0.917 \lambda \right] \log{\mu_c \over m_c} \Bigg\} 
+  \Bigg( { \alpha^{(4)}(\mu_c) \over \pi} \Bigg)^3 \Bigg\{ 
-0.391 + 0.761 \lambda \nl
&\quad
+ \left[ 0.694778  - 3.977 \lambda \right] \log {\mu_c \over m_c}  
+ \left[ 1.25029  -  3.8223 \lambda \right] \log^2{\mu_c \over m_c}
\Bigg\} + \order(\alpha_s^4) \, .
\end{align}


\end{document}